\def\huawei{0}

\if\huawei1
\documentclass[acmsmall,authordraft,nonacm]{acmart}
\else
\documentclass[acmsmall]{acmart}
\fi

\usepackage{longtable}
\usepackage{amsfonts} 

\usepackage{array}
\usepackage{lipsum}

\usepackage{xspace}
\usepackage{algorithm}
\usepackage[noend]{algpseudocode}
\usepackage{amsmath}
\usepackage{subcaption}
\usepackage{relsize}
\usepackage{xcolor}

\newcolumntype{L}[1]{>{\raggedright\let\newline\\\arraybackslash\hspace{0pt}}m{#1}}
\newcolumntype{C}[1]{>{\centering\let\newline\\\arraybackslash\hspace{0pt}}m{#1}}
\newcolumntype{R}[1]{>{\raggedleft\let\newline\\\arraybackslash\hspace{0pt}}m{#1}}
\usepackage{multirow}
\usepackage{array, caption, tabularx,  ragged2e,  booktabs}

\newcommand{\APPR}{SAFE\xspace} 

\newcommand{\CHANGED}[1]{\textcolor{black}{#1}}

\newcommand{\ASE}[1]{\textcolor{black}{#1}}
\newcommand{\ASEnew}[1]{\textcolor{black}{#1}}
\newcommand{\ASEnn}[1]{\textcolor{black}{#1}}

\definecolor{mygreen}{rgb}{0.0, 0.2, 0.13}
\newcommand{\ASEnnn}[1]{\textcolor{black}{#1}}

\usepackage{marginnote}
\usepackage[backgroundcolor=white,bordercolor=blue,linecolor=blue]{todonotes}

\newcommand{\IEE}{IEE\xspace}

\newcommand{\CloseDNN}{OC\xspace}

\newcommand{\OC}{OC\xspace}

\newcommand{\GD}{GD\xspace}
\newcommand{\HPD}{HPD\xspace}
\newcommand{\FLD}{FLD\xspace}

\newcommand{\TrafficDNN}{TS\xspace}
\newcommand{\ODDNN}{OD\xspace}

\newcommand{\MAJOR}[2]{#2}

\newcommand{\MAJORSTARTS}[0]{}
\newcommand{\MAJORENDS}[0]{}

\newcommand{\MINOR}[2]{#2}

\usepackage{pifont}
\newcommand{\cmark}{\ding{51}}%
\newcommand{\xmark}{\ding{55}}%

\begin{document}

\title{Black-box Safety Analysis and Retraining of DNNs based on Feature Extraction and Clustering}



\author{Mohammed~Oualid~Attaoui}
\affiliation{%
  \institution{SnT Centre, University of Luxembourg}
  \streetaddress{JFK 29}
  \city{Luxembourg}
  \country{Luxembourg}}
\email{mohammed.attaoui@uni.lu}

\author{Hazem Fahmy}
\affiliation{%
  \institution{SnT Centre, University of Luxembourg}
  \streetaddress{JFK 29}
  \city{Luxembourg}
  \country{Luxembourg}}
\email{hazem.fahmy@uni.lu}

\author{Fabrizio Pastore}
\affiliation{%
  \institution{SnT Centre, University of Luxembourg}
  \streetaddress{JFK 29}
  \city{Luxembourg}
  \country{Luxembourg}}
\email{fabrizio.pastore@uni.lu}

\author{Lionel Briand}
\affiliation{%
  \institution{SnT Centre, University of Luxembourg}
  \streetaddress{JFK 29}
  \city{Luxembourg}
  \country{Luxembourg}}
  \affiliation{%
  \institution{School of EECS, University of Ottawa}
  \city{Ottawa}
  \country{Canada}}
\email{lionel.briand@uni.lu}



\begin{abstract}

Deep neural networks (DNNs) have demonstrated superior performance over classical machine learning to support many features in safety-critical systems. Although DNNs are now widely used in such systems (e.g., self driving cars), there is limited progress regarding automated support for functional safety analysis in DNN-based systems. For example, the identification of root causes of errors, to enable both risk analysis and DNN retraining, remains an open problem. In this paper, we propose \APPR, a black-box approach to automatically characterize the root causes of DNN errors. \APPR relies on a transfer learning model pre-trained on ImageNet to extract the features from error-inducing images. It then applies a density-based clustering algorithm to detect arbitrary shaped clusters of images modeling plausible causes of error. Last, clusters are used to effectively retrain and improve the DNN.
The black-box nature of \APPR is motivated by our objective not to require changes or even access to the DNN internals to facilitate adoption. 

Experimental results show the superior ability of \APPR in identifying different root causes of DNN errors based on case studies in the automotive domain. It also yields significant improvements in DNN accuracy after retraining, while saving significant execution time and memory when compared to alternatives.

\end{abstract}




\maketitle

\section{Introduction}
Deep neural networks (DNN) have become an essential computational tool inside many cyber-physical systems. This success is partly due to their capacity to automate complex tasks that are typically performed by humans and are difficult to program. It is also driven by the high performance they have achieved in many important fields regarding perception and decision-making tasks in smart grids \cite{xu2021stealthy, kabir2021detection}, networked surveillance \cite{vallathan2021suspicious, fu2021dvqshare}, medical imaging \cite{dif2021transfer, sahaai2021brain}, and  autonomous vehicles \cite{tian2018deeptest, li2021testing}. A good example of the latter is  IEE \cite{IEE}, our industry partner in this research, who is extending its portfolio of in-vehicle monitoring systems with DNN-based products.


A DNN model is often regarded as a black-box. Despite their high performance, such models cannot easily provide meaningful explanations on how a specific prediction (decision) is made. Without such explanations to enhance the transparency of DNN models, it remains challenging to build up trust and credibility among end-users, especially in the context of safety-critical systems that need to be certified. 

Such trustworthiness, for a DNN model, can be addressed predominantly by two processes: a certification process and an explanation process \cite{huang2020survey}. The certification process is held before the deployment of the product to ensure that it is reliable and safe.  The explanation process is performed whenever needed during the lifetime of the product. Explanation is required in the safety-critical context to support safety analysis. The safety standards, such as ISO26262 \cite{ISO26262}, and ISO/PAS 21448 \cite{ISO24765}, enforce the identification of the situations in which the system might be unsafe (i.e., provide erroneous and unsafe outputs) and the design of countermeasures to put in place (e.g., integrating different types of sensors).

Explanation methods aim at making neural networks decisions trustworthy \cite{gilpin2018explaining}.  It is usually defined as a visual aid accompanying a prediction to provide insights into the underlying reasons for the model output. Existing works in the literature have provided alternative techniques to explain DNNs \cite{jeyakumar2020can}, which focus on different model elements, e.g., the training dataset or the learned feature representations. 

When DNN-based systems are used in a safety-critical context, root cause analysis is required to support safety analysis \cite{fahmysupporting}. A root cause is a source of a failure, which is in our context an incorrect DNN prediction or classification. 

One example root cause may be that certain classes are harder to distinguish. For example, in CIFAR-10 \cite{cifar}, dog and cat classes tend to confuse the DNN model since they share many standard semantic features. For multi-label classification, one root cause may be that two classes frequently appear together. For example, in the COCO dataset \cite{LinMBHPRDZ14}, mouse and laptop appear in the same image frequently, making it hard for the DNN model to distinguish between them  \cite{tian2021detect}.

In general, there are two categories of root cause analysis methods based on machine learning \cite{gomez2015automatic}: supervised and unsupervised. Supervised methods perform well in the systems where the different classes of problems are known a priori. Several supervised learning methods have been used for root cause analysis, for instance, SVM \cite{he2016big} and Bayesian models \cite{alaeddini2011using}.  

Unlike supervised methods, unsupervised methods do not require any training labels but automatically cluster failures and mine common features in each cluster. Several unsupervised root cause analysis approaches have been proposed in the literature, \MAJOR{R3.10}{for instance, 
Sparse Filtering \cite{lei2016intelligent}, and Frequent Pattern Mining \cite{pan2021unsupervised}.}  

Though supervised root-cause analysis is widely used, it is not adequate for all scenarios, especially when labels are missing or not numerous enough. Further, in modern architectures, engineers cannot guess all potential failure causes. Moreover, new root causes can appear when changing configurations and settings. Thus, more dynamic and less human-dependent, unsupervised root-cause-analysis methods have been proposed \cite{pan2021unsupervised, fahmysupporting, gomez2015automatic}. These unsupervised methods automatically cluster failures according to common causes, without expert's involvement. The main limitation of this approach is in the clustering evaluation. In this step, the authors use external validation measures to evaluate clustering quality. Such measures require the data to be labeled, for example Normalized Mutual Information \cite{strehl2002cluster} and Adjusted Rand Index \cite{hubert1985comparing}. However, in most case studies, the ground truth is not available and we have no other choice but to use internal measures like the Silhouette Index \cite{rousseeuw1987silhouettes} and the Davies Bouldin Index \cite{davies1979cluster}.

The current paper proposes \APPR (Safety Analysis based on Feature Extraction), a new automated approach for root cause analysis. The foremost objective is to provide a black-box solution that does not rely on internal information about the DNN or its modification, thus facilitating its adoption in practice. Indeed, engineers do not have access to such information in many contexts or are not sufficiently skilled to modify the DNN, whose development is often outsourced.  \MINOR{R1.1}{Our approach targets DNNs that process images since they are the most common form of inputs for many DNN-based components in the automotive and other safety-critical domains (e.g., manufacturing robots). \APPR can be be extended to deal with other types of inputs by choosing a feature extraction method adapted to this kind of data (e.g., BERT~\cite{devlin2018bert} for text data).}

\APPR makes use of transfer learning-based feature extraction, dimensionality reduction, and unsupervised learning. This approach is an improvement over HUDD \cite{fahmysupporting} to avoid reliance on heatmap-based distance, which requires access to the DNN's internal information, and to improve the quality of the root cause clusters' identification. Transfer Learning transfers the knowledge from a generic domain to another specific domain using a pre-trained model. Besides a large amount of time saved by using these methods, it has been shown that starting from a pre-trained model may perform better than training from scratch even on a different problem \cite{loey2021hybrid, wan2021review}. In our approach, we propose to extract the features from our error-inducing images based on convolutional layers in a pre-trained model instead of relying on heatmaps.

We conducted an empirical evaluation on six DNNs. Our empirical results show the cost-effectiveness of \APPR in identifying plausible root causes with a reasonable human effort and its efficiency in memory usage and computation time. \APPR also achieved significant improvements in the retraining of DNNs (up to 35\% improvement over the original models) and overall better results than alternatives (e.g., HUDD).

The rest of the paper is organized as follows. In Section \ref{sec:background}, we present HUDD and its main features and limitations. In Section \ref{sec:approach}, we describe our proposed approach and its expected advantages. In Section \ref{sec:empirical}, we present the experiment questions, design, and results, including a comparison with HUDD. In Section \ref{sec:related}, we discuss and compare related work. Finally we conclude this paper in Section \ref{sec:conclusion}.



\section{Background}
\label{sec:background}

This Section introduces the body of work on which we build our approach, with a focus on DNN explanation and transfer learning-based feature extraction. We also describe our previous approach,  HUDD~\cite{fahmysupporting} \MAJOR{R1}{(Heatmap-based Unsupervised Debugging of DNNs)}, which is used as a baseline of comparison.  
\subsection{DNN Explanation and HUDD}
 \label{sec:background:hudd}
Approaches that aim to explain DNN results have been developed in recent years~\cite{GARCIA2018}. 
Most of these concern the generation of heatmaps that capture the importance of pixels in image predictions. They include black-box~\cite{Petsiuk2018rise,Dabkowski17} and white-box
approaches~\cite{Montavon2019,Selvaraju17,Zeiler14,DB15a,Zhou16}. 
Black-box approaches generate heatmaps for the input layer and do not provide insights regarding internal DNN layers.
White box approaches rely on the backpropagation of the relevance score computed by the DNN~\cite{Montavon2019,Selvaraju17,Zeiler14,DB15a,Zhou16}.

\begin{figure}[b]
\includegraphics[width=0.6\textwidth]{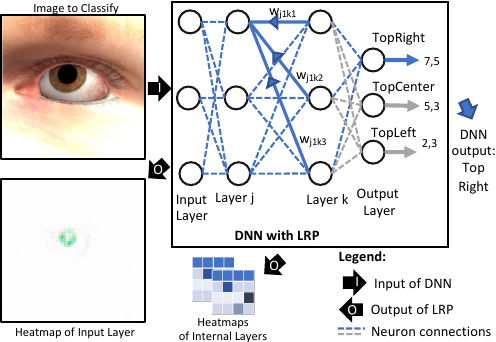}
\caption{Layer-Wise Relevance Propagation.}
\label{fig:LRP}
\end{figure}

For example, Layer-Wise Relevance Propagation (LRP)~\cite{Montavon2019} redistributes the relevance scores of neurons in a higher layer to those of the lower layer. 
Figure~\ref{fig:LRP} illustrates the execution of LRP on a fully connected network used to classify inputs. 
In the forward pass, the DNN receives an input and generates an output (e.g., classifies the gaze direction as TopLeft) while keeping trace of the activations of each neuron.
The heatmap is generated in a backward pass. 
The heatmap in Figure~\ref{fig:LRP}  shows that the result computed by the DNN was mostly influenced by the pupil and part of the eyelid, which are the non-white parts in the heatmap.
In his backward pass, LRP generates \emph{internal heatmaps}. An internal heatmap for a DNN layer $k$ consists of a matrix with the relevance scores computed for all the neurons of layer $k$.

\begin{figure}[b]
\includegraphics[width=0.8\textwidth]{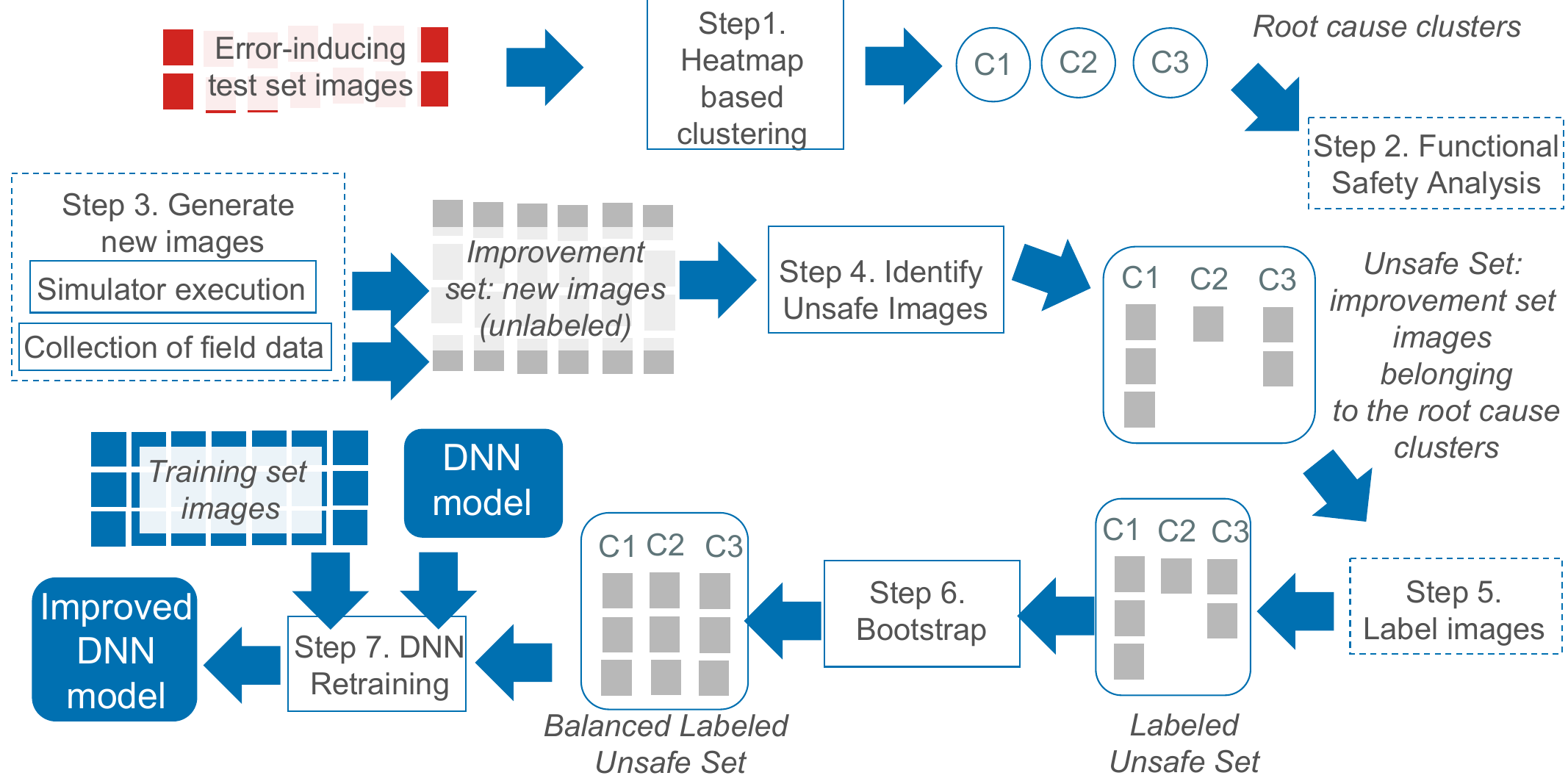}
\caption{Overview of HUDD.}
\label{fig:HUDD}
\end{figure}

Although heatmaps may provide useful information to determine the characteristics of an image that led to an erronous result from the DNN, they are of limited applicability because, to determine the cause of all DNN errors observed in the test set, engineers may need to visually inspect all the error-inducing images, which is practically infeasible. To overcome such limitation, we recently developed HUDD~\cite{fahmysupporting}, a technique that facilitates the explanation and removal of the DNN errors observed in a test set. HUDD generates clusters of images that led to a DNN error because of a same root cause.
The root cause is determined by the engineer who visualizes a subset of the images belonging to each cluster and identifies the commonality across each image (e.g., for a Gaze detection DNN, all the images present a closed eye).
To further support DNN debugging, HUDD automatically retrains the DNN by selecting from a pool of unlabeled images a subset that will likely lead to DNN errors because of the same root causes observed in the test set. 

HUDD consists of seven steps, shown in Figure~\ref{fig:HUDD}.
In Step 1, HUDD performs heatmap-based clustering, which consists of three activities: (1) generate heatmaps for the error-inducing test set images, (2) compute distances between every pair of images using the euclidean distance applied to their heatmaps, and (3) execute hierarchical agglomerative clustering to group images based on the computed distances. 
Heatmaps enable HUDD to determine similarities based on a characteristic that actually caused the erroneous DNN result.
Step 1 leads to the identification of root cause clusters, i.e., clusters of images with a common root cause for the observed DNN errors. 

In Step 2, engineers inspect the root cause clusters (typically a small number of representative images) to identify unsafe conditions, as required by functional safety analysis. The inspection of root cause clusters is an activity performed to gain a better understanding of the limitations of the DNN and thus introduce countermeasures for safety purposes, if needed.

In Step 3, engineers select a new set of unlabeled real-world images to retrain the DNN, referred to as the \emph{improvement set}.

\ASE{In Step 4, HUDD \emph{automatically} identifies the subset of images belonging to the improvement set that are likely to lead to DNN errors, referred to as the  \emph{unsafe set}. It is obtained by assigning 
the images of the improvement set to the root cause clusters according to their heatmap-based distance.} 

In Step 5, engineers manually label the images belonging to the unsafe set. Different from traditional practice, HUDD requires that engineers label only a small subset of the improvement set.

In Step 6, to improve the accuracy of the DNN for every root cause observed,  regardless of their frequency of occurrence in the training set, HUDD balances the labeled unsafe set using a bootstrap resampling approach \MAJOR{R3.11}{(i.e., replicating samples in the unsafe set) in order to have a sufficiently large number of unsafe images to improve the DNN.}

In Step 7, the DNN model is retrained by relying on a training set that consists of the union of the original training set and the balanced labeled unsafe set.

Although effective, HUDD presents a number of limitations.
First, it can only analyze DNN implementations extended to compute LRP. 
Although LRP implementations for multiple DNN architectures relying on the tensorflow framework are available~\cite{iNNvestigate}, it might be particularly complex for engineers to integrate LRP into a different DNN architecture. 
Indeed, the relevance computation formula to be adopted for each layer depends on the layer type (e.g., input, normalization, spatial pooling, internal layer) and the presence of recursion~\cite{Montavon2019}. 


Also, companies often acquire off-the-shelf DNNs which cannot be modified, thus preventing the computation of LRP and the application of HUDD.
Moreover, computing a heatmap-based euclidean distance might become particularly expensive when layers are made of thousands of neurons and hundreds of error-inducing images need to be processed.
Finally, given that the neurons relevant for a specific DNN error (i.e., the neurons with high relevance scores) might represent a small proportion of the neurons in a DNN layer, computing the euclidean distance considering all the items in a heatmap may potentially lead to imprecise clusters caused by noise (i.e., the sum of many differences that are almost zero).

For all the reasons above, although the safety analysis and improvement of DNNs through the automated identification of root cause clusters has demonstrated to be effective, achieving a wider adoption requires a black-box substitute for HUDD.

%
%
%

\subsection{Transfer Learning and Feature Extraction}
\label{sec:back:FeatureExtraction}
To maximize the accuracy of DNNs in a cost-effective way, engineers often rely on the transfer learning approach, which consists of transferring knowledge from a generic domain, usually ImageNet~\cite{IMAGENET}, to another specific domain, (e.g., Safety Analysis, in our case). In other terms, we try to exploit what has been learned in one task and improve generalization in another task. Researchers have demonstrated the efficiency of transfer learning from ImageNet to other domains \cite{talo2019automated}. The hierarchical nature of convolutional neural networks (CNNs) encouraged the computer vision community to use this technique with distant datasets due to the similarities between the features extracted by the first CNN layers. Transfer learning saves training time, gives better performance in most cases, and reduces the need for a large dataset. 

Transfer learning-based \textit{Feature Extraction} is an efficient method to transform unstructured data into structured raw data to be exploited by any machine learning algorithm. In this method, the features are extracted based on a pre-trained CNN model~\cite{dif2021transfer}. 


The standard  CNN architecture \MAJOR{R3.12}{\cite{albawi2017understanding, zhang2021improved, sony2021systematic}} comprises three types of layers: convolutional layers, pooling layers, and fully connected layers.
	The convolutional layer is considered the primary building block of a CNN. This layer extracts relevant features from input images during training. Convolutional and pooling layers are stacked to form a hierarchical feature extraction module. The model captures the resulting feature map from the last pooling layer as a 3D matrix of size ($N$, $N$, $N_c$), where $N$ is its width and height, and $N_c$ is its depth. For features extraction, this feature map is flattened to form a vector of size (1, $N \times N $). In summary, the CNN model receives an input image of size $(224,224,3)$. This image is then passed through the network's layers to generate a vector of features. The feature extraction process from all images generates raw data represented by a 2D matrix (denoted as X) formalized below:  
	{
		\begin{equation}
			X = \begin{bmatrix}
				x_{11} & x_{12}  & ...  &   x_{1m} & l_2\\ 
				x_{21} &  x_{22}  & ...  & x_{2m} & l_c \\ 
				... & ...   & ...   & ...  & ... \\ 
				x_{k1} & x_{k2}  & ...  & x_{km} & l_1 \\
			\end{bmatrix}, l_i \in \left \{ l_1,l_2,...,l_{c}  \right \} 
		\end{equation}
	}
	where $l_i$ represent the class labels, c is the number of categories, $m = N \times N $ is the number of features, and k is the size of the dataset. \MAJOR{R3.28a}{In our case, the class categories will not be used since our approach is unsupervised. They are useful if the user is working on a supervised problem or if fine-tuning is required.}

There are several pre-trained models to extract features based on transfer learning:
InceptionV3 \cite{szegedy2015going}, VGGNet \cite{simonyan2014very}, ResNet50 \cite{ he2016deep}, and MobileNet \cite{howard2017mobilenets}.
The extracted features are related to the used architecture, where, in our context, InceptionV3, VGGNet-16, VGGNet-19, ResNet50, and MobileNet generated 2048, 512, 512, 2048, and 1024 features, respectively. We notice that the VGGNet architectures extract the least amount of features. We rely on VGGNet-16 instead of VGGNet-19 because the latter is more costly in execution time (19 layers instead of 16 for VGGNet-16). We describe in the following the VGGNet architecture.

VGGNet \cite{simonyan2014very} is a CNN characterized by a high number of layers (11 to 19 layers). The purpose of this architecture is to minimize the number of trainable parameters.  Controlling the number of parameters helps to reduce overfitting issues. To this end,  VGGNet proposes to increase the network's depth and to decrease the size of filters from $7 \times 7$ and $5 \times 5$ to $3 \times 3$. The comparative study between the number of parameters in 3 stacked convolutional layers associated with $3 \times 3$ filters and a single convolutional layer associated with $7 \times 7$   filters demonstrated that small filters reduce the total number of parameters. 
%
	Also, it enhances non-linearity through ReLu activation functions in intermediate layers.

	VGGNet proposes six different configurations: A, A-LRN, B, C, D, and E, where the depth varies from 11 to 19 layers. In this contribution, we exploited the configurations D and E, which are composed of 16 and 19 layers, respectively. 

\section{The \APPR Approach}
\label{sec:approach}



In this Section, we present \APPR, a solution to overcome the previously mentioned limitations of HUDD. \APPR relies on a new black-box approach to extract features and compute root cause clusters. This black-box approach is based on transfer learning and dimensionality reduction. \MAJOR{R3.15}{\APPR also allows the detection of non-convex clusters~\cite{kriegel2011density}. A cluster is convex if, for every pair of points within this cluster, every point on the straight line segment that joins them is also within the cluster \cite{kriegel2011density}. This kind of clusters is usually represented by its centroid. But in many practical cases, the data leads to clusters with arbitrary, non-convex shapes. Such clusters, however, cannot be properly detected by a centroid-based algorithm or even a hierarchical clustering algorithm, as they are not designed for arbitrary-shaped clusters.}

The presented method has the following merits compared to HUDD:

\begin{itemize}
    \item It avoids the need for extending the DNN under analysis to compute LRP, which is achieved by relying on a transfer learning model that extracts features from the images.
    \item It applies a feature-based distance instead of a heatmap-based distance, thus saving training time and memory.
    \item It applies a density-based clustering algorithm to detect non-convex clusters, modeling the DNN errors' root causes.
    \item It relies on the non-convex root cause clusters to select an unsafe set for the retraining of the DNN.
\end{itemize}

\begin{figure}[b]
\includegraphics[width=0.8\textwidth]{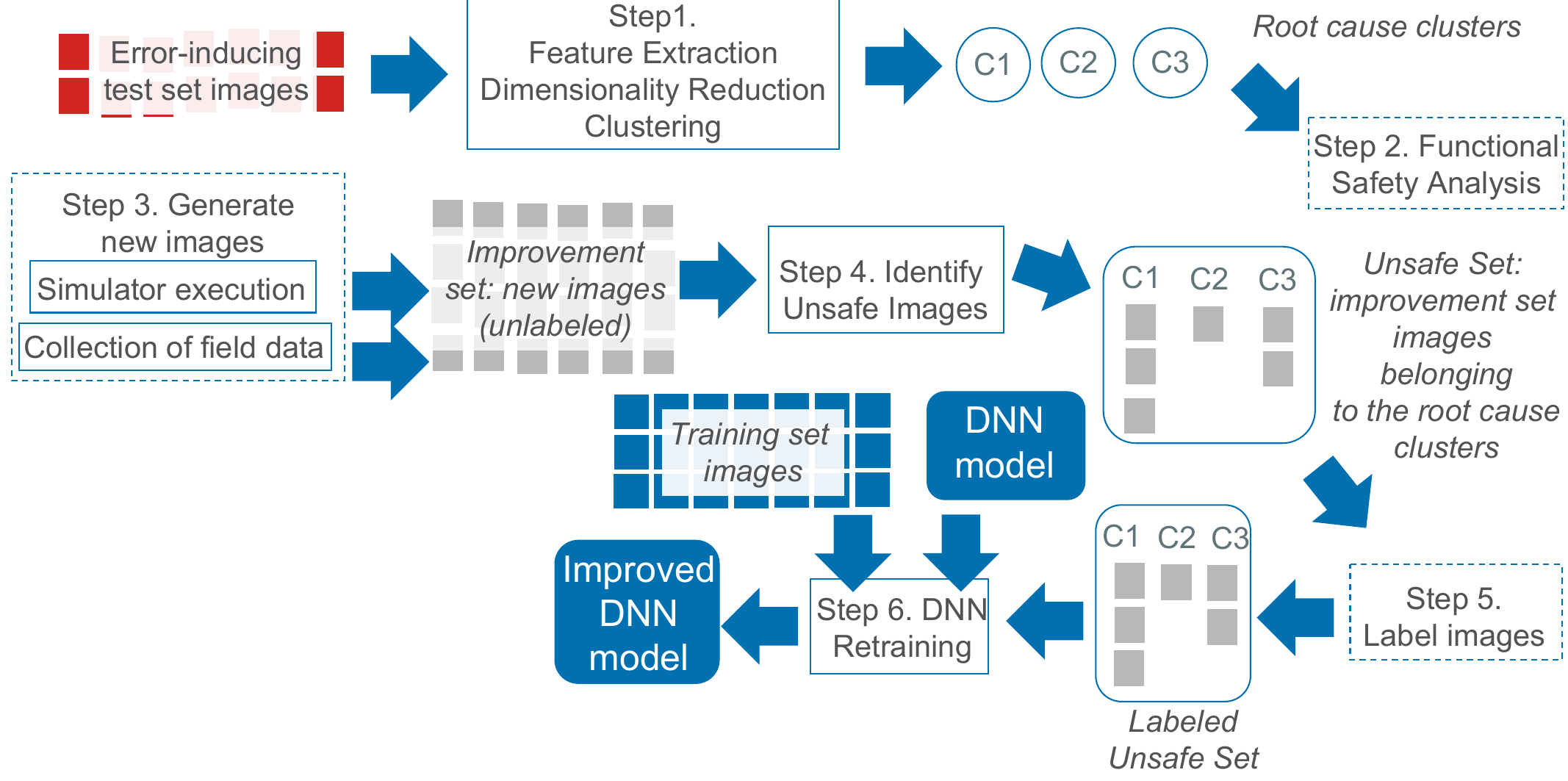}
\caption{Overview of \APPR}
\label{fig:Approach}
\end{figure}

Figure~\ref{fig:Approach} presents an overview of \APPR.
It consists of six steps. In Step 1, root cause clusters are identified by relying on feature extraction-based clustering. Step 2 involves a visual inspection performed by engineers as required by functional safety analysis \cite{ISO24765, ISO26262}. In Step 3, a new set of images, referred to as the \textit{improvement set}, is provided by the engineers to retrain the model. In Step 4, \APPR automatically selects a subset of images from the improvement set called the \textit{unsafe set}. The engineers label the images in the unsafe set in Step 5. Finally, in Step 6, \APPR automatically retrains the model to enhance its prediction accuracy.

The main contributions of \APPR, when compared to HUDD, are in Step 1 and Step 4. Step 1 consists of four stages, namely (i) data acquisition and preprocessing, (ii) features extraction, (iii) dimensionality reduction, and (iv) clustering. Step 4 relies on a new method for the identification of the unsafe set that fits the clustering solution integrated in \APPR.

 \MAJOR{R1.2}{Similar to HUDD, SAFE includes some manual activities: Step 2 (visual inspection of images for functional safety analysis), Step 3 (generate new images), and Step 5 (label images). Such activities are, however, required also by state-of-the-practice approaches. Indeed, to debug and improve DNNs, engineers usually visually inspect all error-inducing images, select an improvement set, and manually label the images to be reused for retraining the DNN. 
 However, both HUDD and \APPR significantly reduce the costs associated with these activities. Indeed, by inspecting a few representative images for each root cause clusters, instead of the whole set of error-inducing images, engineers can save substantial effort; further, since clusters group similar images together and each cluster is considered, it is less likely for engineers to overlook characteristics  and associated causes appearing in a small subset of the images. Also, Step 2 is required only for functional safety analysis (e.g., to determine the unsafe cases to be discussed in safety analysis documents); it is not necessary if engineers aim at automated DNN improvement only.
The effort required for the acquisition of the improvement set is the same for both HUDD, \APPR, and common practice (e.g., purchase an additional stock of real-world pictures); however, HUDD and \APPR require only the unsafe images to be labelled, thus reducing development costs. Finally, by relying on feature-extraction rather than heatmaps, \APPR eliminates the effort required to integrate LRP-based heatmap generation into the DNN under analysis, which is required by HUDD.} 

\begin{figure}[b]
\includegraphics[width=0.8\textwidth]{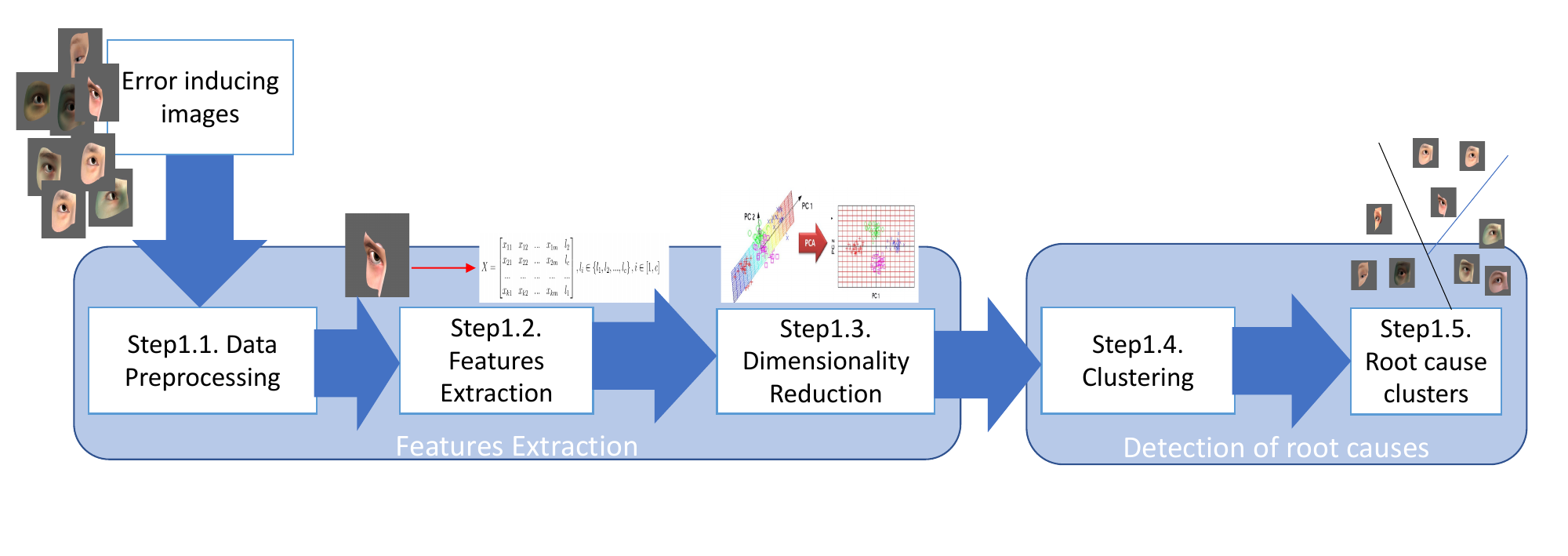}
\caption{Generation of root cause clusters with \APPR}
\label{fig:step1}
\end{figure}

Figure \ref{fig:step1} depicts the stages composing Step 1. We provide further details about each stage in the following subsections. Steps 2 and 4 are described last. Steps 3 and 5 are not further described because they are standard practice.

\subsection{Step1.1: Data Preprocessing}
\label{sec:preprocess}
This step aims to downsample the image sizes to the size required by the transfer learning models. Since we rely on a VGG16 model (see Section~\ref{sec:fe}), which requires an input size of $224 \times 224$ pixels, the images have to be resampled to match this requirement. 
\MAJOR{R3.29}{To downsample the size of each image we rely on the Python Numpy reshape function\footnote{https://sparrow.dev/numpy-reshape/}, which changes the shape of an array so that it has a new, compatible  shape\footnote{The reshape function just changes the shape of the array containing the data (not the image itself) to match the VGG requirement. It does not change the data. The new shape must include the same total number of elements as the original shape.}. In our experiments, we downsampled images from ($640 \times 480$) to ($224 \times 224$).}


\subsection{Step1.2: Feature extraction}
\label{sec:fe}
Feature extraction aims to transform unstructured data into structured data for their exploitation by clustering algorithms \cite{gosztolya2017dnn}. 
	

As discussed in Section~\ref{sec:back:FeatureExtraction}, we rely on transfer learning-based feature extraction. More precisely, we rely on VGGNet-16 models pre-trained on the ImageNet database.



\subsection{Step1.3: Dimensionality reduction}
\label{sec:pca}
Dimensionality reduction aims at approximating data in high-dimensional vector spaces \cite{gorban2010principal}. 

This can be achieved using projections on hyperplanes. These methods, referred to as linear dimensionality reduction, include the well-known Principal Component Analysis (PCA) \cite{pearson1901liii, shlens2014tutorial}. 
Principal component analysis (PCA) is used for dimensionality reduction by projecting each data point onto  the first few principal components, i.e., eigenvectors of the data covariance matrix, to obtain lower-dimensional data while preserving as much of the data variation as possible. The first principal component can equivalently be defined as a direction that maximizes the variance of the projected data \cite{pearson1901liii, shlens2014tutorial}.

There are other methods to reduce dimensionality, including UMAP~\cite{mcinnes2020umap}, LDA~\cite{yang2003can} and T-SNE~\cite{wattenberg2016use}. We compared these methods with respect to clustering quality as measured by the Silhouette Index~\cite{rousseeuw1987silhouettes}. Two methods appeared to fare better, with comparable clustering quality: UMAP and PCA, the latter showing a much shorter execution time. In addition, PCA removes correlated features in contrast to UMAP.

\subsection{Step1.4: Clustering}
\label{sec:dbscanAlgorithm}
Clustering is an unsupervised learning technique that divides a set of objects into clusters: (a) objects in the same cluster must be as similar as possible, (b) objects in different clusters must be as different as possible. 


Since the root cause clusters of images may have any shape and their number $K$ cannot be determined a priori, we rely on an automatic $K$-determination clustering algorithm that can find clusters of arbitrary shapes. We use DBSCAN (density-based spatial clustering of applications with noise) \cite{ester1996density}, which is the most used density-based clustering algorithm \cite{kriegel2011density, mcinnes2017hdbscan}. Intuitively, regions with high density are considered clusters and  points with the most neighbors are referred to as the "core" of the clusters. The points with fewer neighbors are considered noise. The DBSCAN algorithm relies on two concepts: Reachability and Connectivity.
\begin{itemize}
    \item Reachability: A point is reachable from another if the distance between them is inferior to a threshold $\epsilon$.
    \item Connectivity: If two points $p$ and $q$ are connected \MAJOR{R3.19}{(i.e., $p$ is in the neighborhood of $q$ based on $\epsilon$)} they belong to the same cluster. 
\end{itemize}
DBSCAN introduces two parameters to apply these concepts: $MinPts$ and $\epsilon$.

\begin{itemize}
    \item $MinPts$: The minimum number of points that a region (a hypersphere with a diameter equal to $\epsilon$) should have to be considered dense.
    \item Threshold $ \epsilon $: A threshold to determine if a point belongs to another point's neighborhood.
\end{itemize} 
\APPR uses a common technique to choose optimal values of $\epsilon$ and $MinPts$. To select $\epsilon$, \APPR relies on the elbow method \cite{bholowalia2014ebk}. For this step, unlike that to find an optimal $MinPts$, \APPR does not require the execution of DBSCAN. 

More specifically, the optimal value for $\epsilon$ is selected as follows:
\begin{itemize}
\item First, we calculate the euclidean distance from each point to its closest neighbor.
\item Then, \MAJOR{R3.5}{we compute the average distance of every point to its closest neighbor and plot these distances in ascending order.}
\item Finally, we find the plot's elbow point \cite{rahmah2016determination}, which is a point where there is a sharp change in the distance plot, which serves as a threshold. This point corresponds to the optimal $\epsilon$ value.
\end{itemize}

To find the optimal value for $MinPts$, we run DBSCAN with $\epsilon$ equal to the optimal value found above, and with different values for $MinPts$. We then select the clustering configuration with the highest Silhouette Index value \cite{rousseeuw1987silhouettes}. The Silhouette index computes the compactness and the separateness of clusters. For a data point $x_i$ assigned to cluster $C_i$, the Silhouette index is calculated as follow:
\begin{equation}
    SI(i) = \frac{(b(i) - a(i))}{Max(b(i) - a(i))}
\end{equation}
where $a(i)$ is the average distance between $x_i$ and all the data points assigned to cluster $C_i$. $b(i)$ is the minimum average distance between $x_i$ and the data points assigned to one of the other clusters $C_j$ where $j = 1,..,K; j \neq i$.
Based on the concepts described above, DBSCAN defines three types of points:
\begin{itemize}
    \item Core point:  It has at least $MinPts$ points within a distance of $\epsilon$.
    \item Border point: It is not a core point, but it belongs to at least one cluster. That means that it lies within a distance $\epsilon$ from a core point.
    \item Noise point: It is a point that is neither a core point nor a border point.
\end{itemize}
 
The DBSCAN algorithm proceeds by sampling points randomly from the dataset until all points are selected. For each point, it determines if it is a Core, Border or Noise point based on the parameters $\epsilon$ and $MinPts$. The core points are considered representative of clusters. The clusters are then expanded by recursively repeating the neighborhood calculation for each point within the region.  
\MAJOR{R3.3}{In DBSCAN, randomness affects only the selection of the point to be treated next; however, since  each point ends up being assigned to the cluster containing the closest core point, randomness affects results only when border points are reachable from more than one cluster, which is unlikely~\cite{DBSCAN:revisited}. 
If we run the algorithm several times with the same parameter settings, the same clusters will likely be obtained each time.}

\begin{figure}[t]
\includegraphics[width=0.6\textwidth]{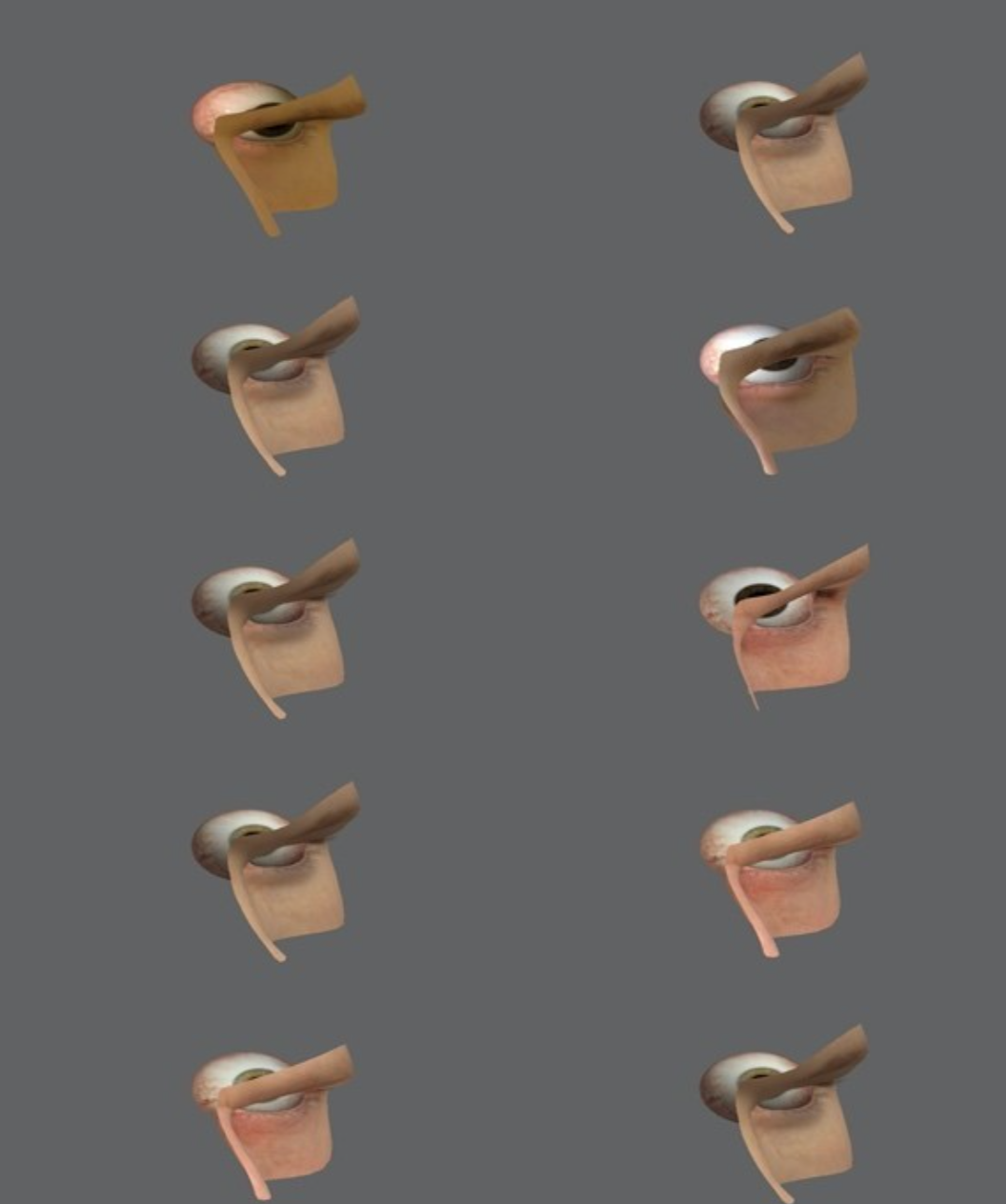}
\caption{Example of a root cause cluster with rare cases}
\label{fig:RCC_rare}
\end{figure}

\MAJOR{R2.1}{DBSCAN also helps identifying rare cases in the set of error-inducing images. Indeed, if there are enough rare cases to form a cluster (i.e., there are at least \emph{MinPts} data points within a hypersphere with a diameter equal
to $\epsilon$), they are grouped together. If rare cases cannot form a cluster, most of them are considered noise and excluded from the set of clusters returned to the end-user. Since we select \emph{MinPts} and
$\epsilon$ based on the analysis of the distribution of the error-inducing images, rare cases are considered noise only if they occur in regions of low density (i.e., the region contain less images than $MinPts$).} \MINOR{2.4}{Figure~\ref{fig:RCC_rare} shows an example of a cluster containing images representing a rare case (the eye is in an unusual position). In this case, there are ten error-inducing images with such characteristic. When $MinPts <= 10$, then all the ten images form a cluster. However, if $MinPts > 10$, the images will be considered noise. In practice, $MinPts$ is unlikely to be above 10 because its value is automatically derived considering all the dense regions, including the area with these rare cases.
Further, if rare cases that are considered noise share similarities with images in other clusters, they will be assigned to the cluster with the most similar features.
However, this case is unlikely to happen since such rare cases appear in sparse areas while clusters only appear in dense areas. In practice, a rare case assigned to a cluster might actually share the same root cause of the other images belonging to the cluster.}

\subsection{Step1.5: Functional safety analysis through root cause clusters visualization}
To analyze the clusters, safety engineers can use the same approach as in HUDD \cite{fahmysupporting}. For each cluster, a subset of elements is visually inspected to determine the associated unsafe conditions, as functional safety analysis requires. Similarities among images within each cluster may suggest the cause of failures of the DNN. In other words, engineers attempt to identify the root cause of each cluster to gain a better understanding of the DNN behavior.

\begin{figure}[t]
\includegraphics[width=0.8\textwidth]{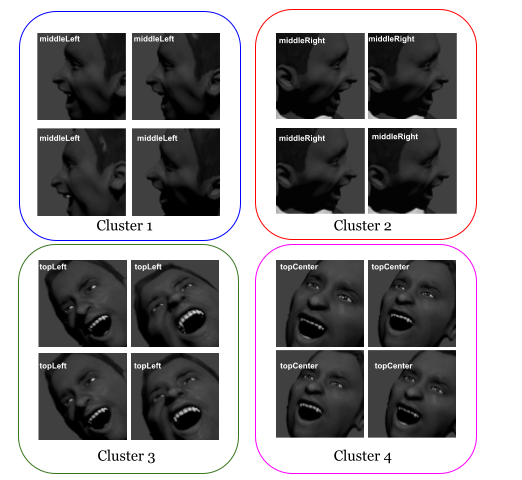}
\caption{Examples of root cause clusters for Head Pose detection}
\label{fig:RCC_examples}
\end{figure}

Figure \ref{fig:RCC_examples} shows examples of root cause clusters identified by \APPR for the Head Pose Detection case study subject considered in our empirical evaluation (see Section~\ref{sec:empirical}). We notice that in cluster 1 the hidden eye seems to confuse the DNN and is a plausible reason for failure. The same observation can be made for Cluster 2, where, because the head is turned to the right, the right eye is not visible. Clusters 3 and 4 identify common causes of error due to an incomplete training set, i.e., the absence of images with a head pose close to a borderline.

\MAJOR{R2.2}{In general, \APPR generates clusters that differ for at least one common characteristic in the images. For example, the root causes presented by Cluster 1 and Cluster 2 are not the same. Indeed, Cluster 1 concerns the right eye while Cluster 2 concerns the the left eye. Engineers might be interested in knowing if the training set is lacking images with only the left eye being hidden or both; for this reason, we believe that separating such clusters is beneficial. Empirical results are reported in Section~\ref{sec:exp:distinctErrorCauses}.}

\begin{figure}[t]
\includegraphics[width=0.8\textwidth]{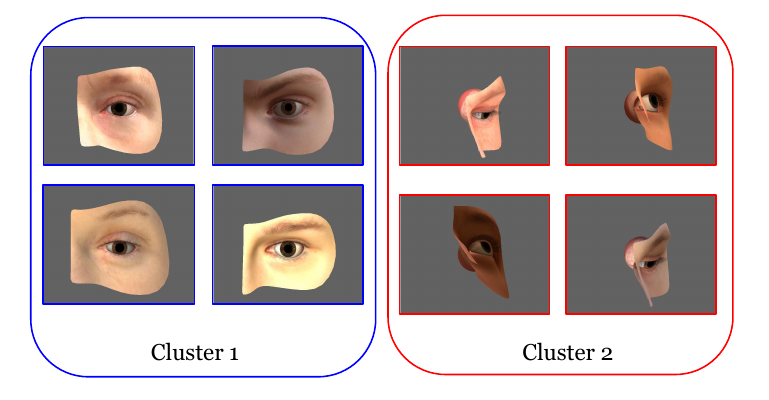}
\caption{Examples of root cause clusters for Gaze detection}
\label{fig:RCC_examples:two}
\end{figure}

\MAJOR{R3.7}{Similar to HUDD, \APPR can identify different root causes
of errors, including (1) borderline cases (e.g., the gaze and
head pose angle detected by Cluster 1 in Figure~\ref{fig:RCC_examples}), (2) an incomplete
training set (e.g., Clusters 3 and 4 in Figure~\ref{fig:RCC_examples}), (3) an incomplete definition of the predicted classes
(e.g., Cluster 1 in Figure~\ref{fig:RCC_examples:two} shows a cluster generated from our GD case study in  Section~\ref{sec:exp:distinctErrorCauses}, with eyes looking middle center, a class missing from our configuration) and (4) limitations in our capacity
to control the simulator (e.g., unlikely face positions detected by
Cluster 2 in Figure~\ref{fig:RCC_examples:two}). In general, \APPR identifies commonalities (i.e., root causes) across images leading to failures. 
Based on a recent taxonomy~\cite{humbatova:2019}, the cases described above concern training data quality; in general, \APPR can help engineers to discover any fault that affects the correctness of the DNN output (e.g., a missing model layer). However, we do not integrate mechanisms to automatically determine the underlying cause for the observed failures. In general, we suggest engineers to first inspect root cause clusters to determine major pitfalls (e.g., missing output class) then proceed with automated retraining (i.e., Steps 3-6); if the DNN accuracy is not sufficiently improved then it is necessary to modify the DNN (e.g., add layers or change architecture).}

\subsection{Step 4: Identify unsafe images}
\label{sec:step4}
\hfill\\

Since the manual labeling of images is expensive, it is necessary to automatically identify an unsafe set of reduced size containing images that can improve the DNN accuracy to achieve a cost-effective retraining process. An unsafe set is a subset of unlabelled images from the improvement set, to be labeled and used for retraining the model. Unlike HUDD, \APPR relies on a new clustering-based selection method to automatically select this unsafe set. We identify images from the improvement set that belong to a root cause cluster and select images that, within the same cluster, are representative of the different types of images in the cluster (e.g., different face shapes with a gaze angle that confuses a gaze classifier). Therefore, we assume that the selected images are representative of a cluster in the sense that they contain sufficient information to replace the other cluster members \cite{marc20072} and can be effectively used for retraining.

The key difference between \APPR and HUDD, in this step, is that the latter relies on a Hierarchical Clustering Algorithm. This method only finds spherical clusters. In addition,  HUDD depends on the cluster medoids and the cluster ratio to determine if improvement set images belong to the root cause clusters. Unfortunately, such a method might be suboptimal if clusters are not spherical. With DBSCAN, the identification of core points enables the determination of representative cluster members even in non-spherical clusters. 
It does not rely on cluster centroids but core points in each cluster, assuming that, taken together, these points represent the entire cluster. A point close to any core point will be assigned to the cluster represented by this core point.

In \APPR, the selection is performed according to Algorithm \ref{alg:unsafe_set_selection}, which is detailed below.
After selecting the unsafe set, we follow the same retraining process as HUDD. We retrain the DNN model by relying on a training set consisting of the union of the original training set and the manually labeled unsafe set. The original training set is reused to prevent reducing the accuracy of the DNN for parts of the input space that are safe. The retraining process is thus expected to lead to an improved DNN model.

\begin{algorithm}[ht!]
\caption{\APPR Unsafe Set Selection Algorithm}
 \hspace*{\algorithmicindent} \textbf{Input:} improvement set images $\mathcal{X}$, core points detected for each cluster, a set of clusters $\mathcal{C}$. \\
 \hspace*{\algorithmicindent} \textbf{Output:} unsafe set 
\begin{algorithmic}[1]
  \For{$x$ in $\mathcal{X}$}
        \State Assign $x$ to the cluster corresponding to the closest core point.
\EndFor

\State Compute $N$, the number of selected images for the unsafe set based on Equation \ref{eq:unsafe_set_selection}.
\State Compute the number of selected images $r_i$ from each cluster $c_i$, computed as $N$ over the proportion of images in the cluster.
 \For{$c_i$ in $\mathcal{K}$}
        \State Sort the images in ascending distance to their respective closest core point.
        \State From the sorted images, take the $r_i$ first images and add them to the unsafe set.
\EndFor

\end{algorithmic}
\label{alg:unsafe_set_selection}
\end{algorithm}


 
 To minimize the retraining effort, and most particularly that related to labeling, we look for representatives images in each cluster. For that, we rely on the core points (see Section~\ref{sec:dbscanAlgorithm}).
 The choice of the core points is motivated by the fact that they represent better the shape of a  particular cluster than a centroid. As described in Section \ref{sec:dbscanAlgorithm}, a root cause cluster is typically represented by several core points. The border points that define the cluster's shape are localized around the core points. We assume that the points close to the core points contain enough information to replace the other border points. These points usually take approximately the same shape as the cluster \cite{kriegel2011density}. Recall that core points, unlike centroids, can represent a non-convex cluster with arbitrary shapes.

Algorithm~\ref{alg:unsafe_set_selection} show the steps for the selection of the unsafe set for retraining. The algorithm requires the root cause clusters and their core points. It also requires an improvement set.

The algorithm starts by assigning every image in the improvement set to the closest core point in terms of euclidean distance (line 1, Algorithm~\ref{alg:unsafe_set_selection}).

On lines 2 and 3, \APPR computes $N$, the number of images to be selected for the unsafe set and the number of images to be selected from each root cause cluster $i$, denoted as $r_i$, where $r_i = N \times \frac{C_i}{C}$. $C$ is the number of images in the improvement set and $C_i$ is the number of such images assigned to cluster $i$. Unsafe set images across root cause clusters therefore preserve the distribution of the improvement set across such clusters.

As in HUDD, we assume that the distribution of error-inducing images across clusters is similar in the improvement and test sets. We determine the number of images $N$ selected from the improvement set to include in the unsafe set as follows: 
 \begin{equation}
 N= (| \mathit{TestSet} | \times \mathit{sf} ) \times (1 - \mathit{TestSetAcc} ) 
 \label{eq:unsafe_set_selection}
\end{equation}
 $|\mathit{TestSet}|$ is the size of the test set, while $sf$ is a selection factor in the range $[0-1]$ (we use $0.3$ in our experiments, \MAJOR{R3.6}{same as HUDD to make the comparison fair}). $\mathit{TestSetAcc}$ represent the accuracy of the original model on the test set of the case study subject. The term $(1 - \mathit{TestSetAcc})$ indicates the proportion of error-inducing images that are observed in the improvement set. $(|\mathit{TestSet}| \times \mathit{sf} ) \times (1 - \mathit{TestSetAcc})$ estimates the number of error-inducing images that should be selected from the improvement set. The term $(|\mathit{TestSet} | \times \mathit{sf} )$ provides an upper bound for the unsafe set size as a proportion of the test set size. 

In line 4, the images are sorted according to ascending distance to their respective closest core point. Finally, in line 5, for each cluster $i$, we select the first $r_i$  sorted images to include in the unsafe set. 

 \begin{figure*}[t]
\includegraphics[width=\textwidth]{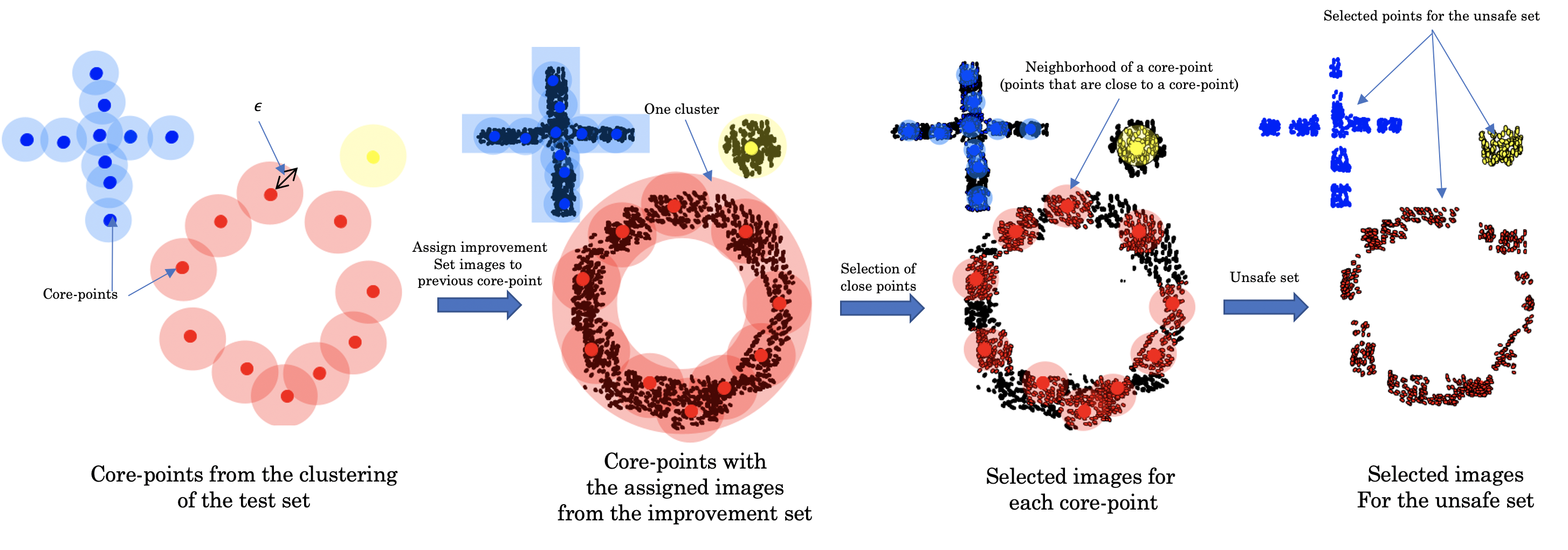}
\caption{Unsafe set selection}
\label{fig:unsafe_set_selection}
\end{figure*}

\MAJOR{R3.23}{We also illustrate the algorithm steps in Figure \ref{fig:unsafe_set_selection}. The first image represents the core points obtained from the clustering of the error-inducing images (the dots represent the core points). In the second step, we assign the images in the improvement set to their closest core point, respectively. Then, we select a subset of points from the neighborhood of each core point based on Algorithm \ref{alg:unsafe_set_selection}. The last image represents the selected unsafe set.}

\MAJOR{R2.5}{Our algorithm excludes from the unsafe set the images leading to DNN errors due to root causes not observed in the test set; indeed, such images will be distant from clusters' core points. Furthermore, such images will not help improve the DNN performance on the test set, which is our objective since the test set is assumed to be representative of real-world scenarios. Finally, when the improvement set does not include any image belonging to a root cause cluster then \APPR does not assign any image to the cluster; this is not the case for HUDD which, different from \APPR, will not prevent engineers from labeling useless images.}

\subsection{\APPR running example}

\begin{figure}[t]
\includegraphics[width=0.7\textwidth]{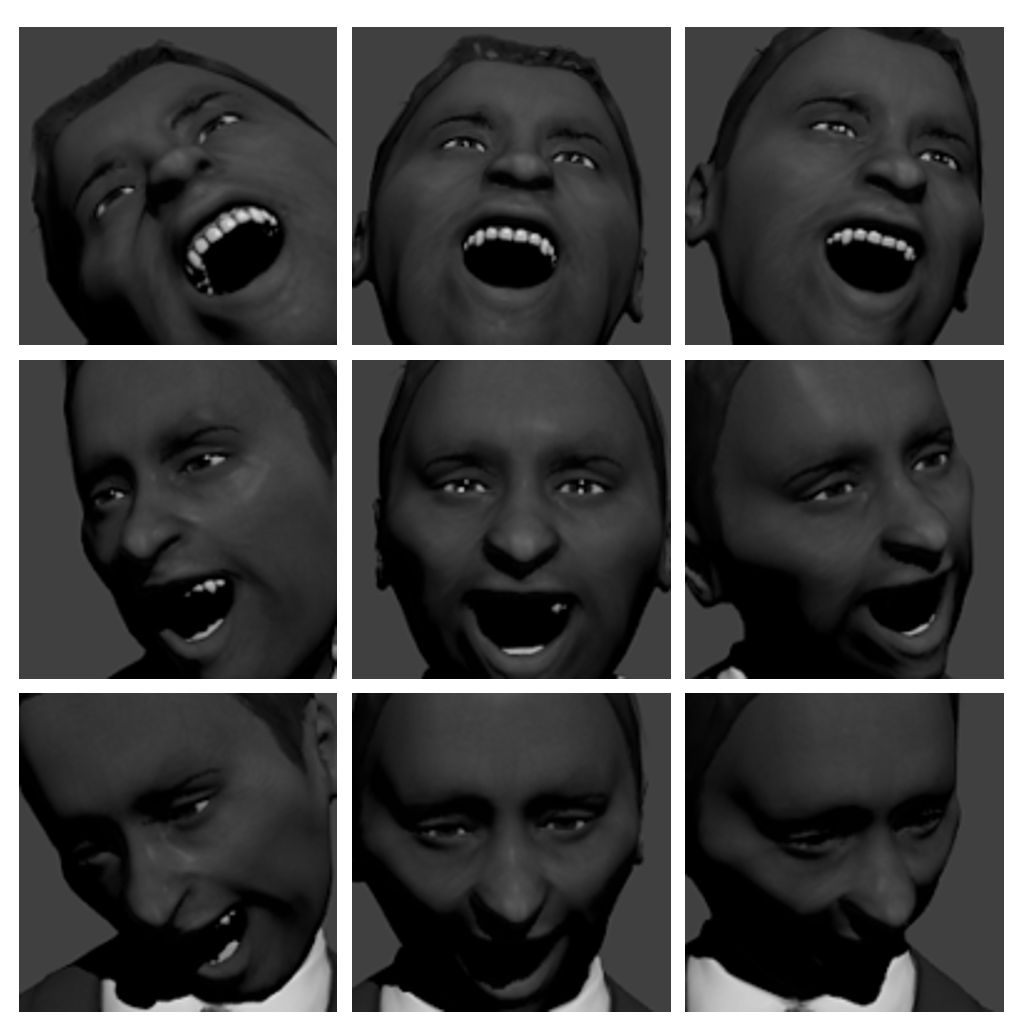}
\caption{Example images from the HPD dataset}
\label{fig:HPD_example}
\end{figure}

\MAJOR{R3.31 and R1.2b}{This section presents an example of \APPR usage. It is based on the headpose detection DNN (HPD) considered in our empirical evaluation.  HPD receives as input a picture taken from a camera positioned inside a car; the picture is automatically cropped to a size of $640 \times 640$ pixels.
HPD classifies the head position according to 
nine classes: straight, turned bottom-left, turned left, turned top-left, turned bottom-right, turned
right, turned top-right, reclined, looking up. Figure~\ref{fig:HPD_example} provides an example of an image classified by \APPR.}

\MAJORSTARTS{}

To reduce development costs, HPD has been trained and tested using images generated by a simulator capable of generating pictures of human heads. The training and test sets consists of 16013 and 2825 images, respectively, both generated by randomly selecting simulator parameters' values. The training and the test set could also have included real-world images. After 30 epochs, we obtained an accuracy of 88.03\%. From the test set, 1580 images were misclassified, they represent the error-inducing images that should be investigated to determine the root causes of errors.

We implemented the \APPR pipeline as a Jupyter Notebook\footnote{http://jupyter-notebook.readthedocs.io}. 

The \APPR pipeline starts by pre-processing the error-inducing images to match our model's input requirement (\APPR Step 1.1, in Figure~\ref{fig:step1}). It automatically converts the images into a NumPy\footnote{http://numpy.org} array and downsample them as explained in Section \ref{sec:preprocess}.

After preprocessing, the images are ready for feature extraction (\APPR Step 1.2, in Figure~\ref{fig:step1}). The \APPR pipeline automatically extracts the features by relying on a pre-trained VGG model loaded by the Notebook. 
Precisely, for each image, \APPR stores the output of the second-last fully connected layer of the VGG model, which leads to an array of 512 features for each image.

The pipeline continues by applying the PCA dimensionality reduction method to reduce the number of features from 512 to 256 (\APPR Step 1.3, in Figure~\ref{fig:step1}). The output of the PCA method is an $1580 \times 256$ array, where $1508$ is the number of error-inducing images and $256$ is the number of features. The number of features (i.e., $256$) has been empirically determined in preliminary experiments (see Section~\ref{sec:pca}) and is not supposed to be further modified by end-users. This array is passed to DBSCAN as an input. We use DBSCAN from the SciKitLearn library\footnote{http://scikit-learn.org}. 

Before performing clustering (\APPR Step 1.4), we first need to choose the optimal parameter settings. We apply the method explained in Section \ref{sec:emp:clusters} to obtain $\epsilon = 0.9$ and $minPts = 9$. Using these parameters, we run our algorithm to generate the final root cause clusters, which are $20$ in this case.
The next step of the pipeline (i.e., \APPR Step 1.5) includes a procedure that, from the clusters generated by DBSCAN, generates several folders, each containing the images belonging to the cluster. It also generates an animated gif image (similar to a video) with the images belonging to each cluster, to help the user visualize it.
The end-user is then expected to visualize a portion of the images appearing in each cluster (e.g., five according to our experimental results in Section~\ref{sec:images:number}) to perform the functional safety analysis step of \APPR (i.e., Step 2 in Figure~\ref{fig:Approach}). Example images are reported in Figure~\ref{fig:RCC_examples}.

Next, the end-user should provide an improvement set (\APPR Step 3). 
For our experiments with HPD, we rely on an improvement set of $4103$ images generated with additional executions of the simulator. 
We could have also included real-world images collected by our industry partner in the field but they might have prevented replicability (e.g., they cannot be publicly shared because of privacy agreements).
In our execution, \APPR selected $154$ images as an unsafe set for  retraining. The number of selected images is computed automatically based on the selection factor ($\mathit{sf}$) configured by the end-user (see Section \ref{sec:step4})
These images need to be labelled by the end-user (\APPR Step 5) but for our case study subject, labels are automatically derived from simulator parameters. In case the improvement set includes real-world images, labelling is performed manually.

After obtaining the unsafe set, the end-user simply combines it with the training set and retrains the model according to the specific procedure of the DNN under analysis. 

\MAJORENDS{}

\hfill \\

\section{Empirical Evaluation}
\label{sec:empirical} 

In this section, we aim to evaluate our approach. \APPR is expected to perform better than HUDD in terms of the quality of the clustering and DNN accuracy after retraining.  \MAJOR{R1.5}{We choose HUDD for comparison not only because \APPR aims to improve over it but because they are the only approaches in the literature that aim to support safety analysis, which is achieved through the identification of root cause clusters and the selection of images for retraining based on these clusters.} 
To investigate whether if such expectations hold and \APPR is useful, we compare these two approaches following the experimental procedures described below addressing the following research questions.

\textbf{RQ1}. Does \APPR enable engineers to identify the root causes of DNN errors?
The clusters produced by \APPR should provide useful information to identify plausible causes of DNN errors in a form that is amenable to practical analysis.

\textbf{RQ2} Does \APPR enable engineers to more effectively and efficiently retrain a DNN when compared with HUDD and baselines approaches? We expect \APPR to lead to a higher model accuracy after retraining.

\textbf{RQ3} Does \APPR provides time and memory savings compared to HUDD?
\APPR's black-box nature should provide significant time and memory savings, compared to HUDD, which is a white-box approach.

To perform our empirical evaluation, we have implemented \APPR as a toolset that relies on the PyTorch~\cite{PyTorch} and SciPy~\cite{SciPy} libraries.
Our toolset, case study subjects, and results are available for download~\cite{REPLICABILITY}.
In our experiments, steps 1 to 5  were carried out on an Intel Core i9 processor running macOS with 32 GB RAM. Step 6 (retraining) was conducted on the HPC facilities of the University of Luxembourg (see http://hpc.uni.lu). We relied on a Dual Intel Xeon Skylake CPU (28 cores) and 128 GB of RAM.

\subsection{Subjects of the study}
\label{sec:subj}
We rely on images generated using simulators as it allows us to associate each generated image to values of the simulator's configuration parameters. These parameters capture information about the characteristics of the elements in the image and can thus be used to objectively identify the likely root causes of DNN errors. Such simulators are increasingly common, and of higher fidelity in many domains \cite{mukherjee2018neuraldrop}, including automotive and aerospace.

We consider the same DNNs as the HUDD paper, which
support gaze detection, drowsiness detection, headpose detection, and face landmarks detection systems under development at \IEE Sensing, our industry partner. 


Eye gaze detection systems (\GD) use DNNs to perform eye tracking. Gaze tracking is typically employed to determine a person's focus and attention. It classifies the gaze direction into eight classes (i.e., TopLeft, TopCenter, TopRight, MiddleLeft, MiddleRight, BottomLeft, BottomCenter, and BottomRight). The drowsiness detection system (\CloseDNN{}) features the same architecture as the gaze detection system, except that the DNN predicts whether eyes are opened or closed.

The headpose detection system (\HPD) is an important cue for scene interpretation and remote computer control like driver assistance systems. It determines a head pose in an image according to nine classes: straight, turned bottom-left, turned left, turned top-left, turned bottom-right, turned right, turned top-right, reclined, looking up.

\CHANGED{The face landmark detection system (\FLD) determines the location of the pixels corresponding to 27 face landmarks delimiting seven face elements: nose ridge, left eye, right eye, left brow, right brow, nose, mouth. Several face landmarks match each face element.}

\GD, \CloseDNN{}, and \HPD follow the AlexNet architecture~\cite{AlexNet} which is commonly used for image classification. \FLD, which addresses a regression problem, relies on an Hourglass-like architecture~\cite{Hourglass}. 

Since \APPR can be applied to DNNs trained using either a simulator or real images, we also considered additional DNNs trained using real-world images.  To do so, we selected the same DNNs included in the HUDD paper, which target traffic sign recognition (\TrafficDNN) and object detection (\ODDNN), and are typical features in automotive, DNN-based systems.

\TrafficDNN recognizes traffic signs in pictures whereas 
\ODDNN determines if a person wears eyeglasses. The latter has been selected to compare results with MODE, a state-of-the-art retraining approach whose implementation is not available, but with an objective that is close to that of \APPR. 
Both \ASEnew{\TrafficDNN and  \ODDNN follow the AlexNet architecture~\cite{AlexNet}.}

We further describe in Table~\ref{tab:dnns} the case study subjects used to evaluate \APPR. We indicate either the simulator used to generate the data or the real-world image dataset. The data is then randomly split into training and test sets whose sizes are reported. \MAJOR{R3.1}{Further, we report the number of error-inducing images, which are the images from the test set leading to a result different than the ground truth. For classifier DNNs (GD, OC, HPD, OD, TS), DNN errors are inconsistent predicted and expected classes. For FLD, we determine a DNN error when the average distance of the predicted keypoints is above four pixels, as suggested by IEE engineers.} 


\MAJOR{R3.3}{Since DNN errors and, consequently, clustering results, depend on the initial training of the DNN under analysis, to deal with such randomness we repeated the initial training ten times for the case study subjects GD, OC, and HPD; each training execution relies on a different split of the training and the test data sets. Unfortunately, we could not repeat the training for the FLD DNN because it was provided by our industry partner along with the error-inducing images. 
Further, we could not repeat the execution of HUDD ten times for each case study DNN because of the large amount of time required to compute distance matrices based on heatmaps (see Section~\ref{sec:emp:executionTime}). 
However, to discuss the statistical significance of the differences between HUDD and \APPR, for each of the metrics selected to address our research questions, we relied on a one-sample Wilcoxon signed rank test.
The one-sample Wilcoxon signed rank test is a non-parametric statistical hypothesis test used to determine whether the median of a population (here, the \APPR median) is greater than a reference value (here, the HUDD result). It enables us to test the null hypothesis: \emph{the \APPR median is equal to the HUDD result}.} 



\MAJOR{R3.4}{Finally, to determine the number of components to be selected by PCA, which is an input parameter for \APPR, we conducted a set of experiments. Precisely, we considered a number of features between 2 and 256, considering all powers of 2; then we applied DBSCAN and measured the quality of its result based on the Silhouette Index~\cite{rousseeuw1987silhouettes}. We performed the analysis on all the case studies and concluded that 256 is the number of features that provides the best results.} 


\begin{table}[tb]
\centering
\caption{Case Study Systems}
\footnotesize
\begin{tabular}{
|@{\hspace{1pt}}p{6mm}
|@{\hspace{1pt}}p{1.8cm}
|@{\hspace{1pt}}p{13mm}
|@{\hspace{1pt}}p{12mm}
|@{\hspace{1pt}}p{12mm}
|@{\hspace{1pt}}p{15mm}|}
\hline
\textbf{DNN}&		 \textbf{Data}	&\textbf{Training}& \textbf{Test} & \textbf{DNN} & \textbf{number of error} \\
&		 \textbf{Source}& \textbf{Set Size} & \textbf{Set Size} & \textbf{Accuracy} & \textbf{inducing images}	\\
\hline
\GD&	 UnityEyes & 61,063 &  132,630 & 95.95\% & 5371\\
\CloseDNN&      UnityEyes &1,704	  &	4,232 & 88.03\% & 506 \\
\HPD&      Blender &16,013  &	2,825 & 44.07\% & 1580 \\
\FLD&      Blender &16,013  &	2,825 & 44.99\% & 1554\\
\ODDNN& CelebA~\cite{Liu:15}&7916&5276 & 84.12\% & 838\\
\TrafficDNN& TrafficSigns~\cite{TRAFFICdataset}&29,416&12,631 & 81.65\% & 2317\\
\hline
\end{tabular}

\label{tab:dnns}
\end{table}%


\subsection{Experimental Results}
\label{sec:emp:clusters}

\ASEnew{We refine RQ1 into four complementary subquestions (RQ1.1, RQ1.2, RQ1.3, RQ1.4, and RQ1.5), which are described in the following, along with their corresponding experiment design and results.}

\subsubsection{RQ1.1} \emph{Is the number of generated clusters small enough for enabling visual inspection?}
\label{sec:evaluation:RQ1.1}

\CHANGED{\emph{Design and measurements.} Though this is to some extent subjective and context-dependent, we discuss whether \APPR finds a number of root cause clusters that is amenable to inspection by experts.}
\CHANGED{To respond to this research question, we assume that experts visually inspect five images per root cause cluster to be able to make a decision. \MAJOR{R2.6}{This assumption is supported by an experiment we conducted, as presented in Section \ref{sec:images:number}.} Under this assumption, we compare \APPR and HUDD in terms of the number of generated clusters and based on the ratio of error-inducing images that should be visually inspected when relying on each method. This ratio is calculated as follows:}
\begin{equation}
    ratio = \frac{(k \times 5) \times 100}{n} 
\end{equation}
where $k$ is the number of root cause clusters, and $n$ is the number of error-inducing images.

\emph{Results.} Table~\ref{table:RQ1.0:results} shows, for each case study subject, the total number of error-inducing images belonging to the test set, the number of root cause clusters generated by \APPR and HUDD, and the ratio of error-inducing images that should be visually inspected when using \APPR or HUDD.

For the respective DNNs, \APPR identifies 25 (\GD), 26 (\OC), 24 (\HPD), 64 (\FLD), 2 (\ODDNN), 9 (\TrafficDNN) root cause clusters (for GD, OC, HPD, OD, and TS we refer to median of the ten runs executed). In contrast, HUDD identifies 16 (\GD), 14 (\OC), 17 (\HPD), 71 (\FLD), 14 (\ODDNN), and 20 (\TrafficDNN) root cause clusters.

We notice that in 50\% of the case study subjects (see bold values in Table~\ref{table:RQ1.0:results}), \APPR yields a lower number of clusters. This experiment shows 
both \APPR and HUDD find an acceptable number of clusters for visual inspection. Indeed, the ratios of error-inducing images to be inspected are low (between 1.07 and 26.15, with a median of 5.12). These results suggest that using \APPR can save significant effort compared to the manual inspection of the entire set of images.


\begin{table*}[t!]

\footnotesize
\caption{Root cause clusters generated by \APPR.}
\centering
\makebox[0.8\textwidth]{\begin{tabular}{|c|ccccc|ccC{1cm}|}
\hline
    & \multicolumn{5}{c|}{SAFE}                                                                                                                                                                                                                              & \multicolumn{3}{c|}{HUDD}                                                                                                                                                                          \\ \hline
    & \multicolumn{1}{l|}{\begin{tabular}[l]{@{}l@{}}\# error inducing\\  images\end{tabular}} & \multicolumn{2}{c|}{\# of clusters}                                & \multicolumn{2}{c|}{\begin{tabular}[c]{@{}c@{}}\% of inspected \\ images\end{tabular}} & \multicolumn{1}{L{1cm}|}{\begin{tabular}[l]{@{}l@{}l@{}}\# error\\ inducing\\  images\end{tabular}} & \multicolumn{1}{L{1cm}|}{\# of clusters} & \begin{tabular}[l]{@{}l@{}l@{}}\% of\\ inspected\\  images\end{tabular} \\ \hline
    & \multicolumn{1}{c|}{min/max/med}                                                      & \multicolumn{1}{c|}{min/max/med} & \multicolumn{1}{c|}{p-value} & \multicolumn{1}{c|}{min/max/med}                       & p-value                    & \multicolumn{1}{c|}{}                                                                    & \multicolumn{1}{c|}{}               &                                                                   \\ \hline
GD  & \multicolumn{1}{c|}{4967/6290/5602}                                                      & \multicolumn{1}{c|}{14/31/25}       & \multicolumn{1}{c|}{0.004}   & \multicolumn{1}{c|}{1.41/2.46/2.24}                       & 0.004                      & \multicolumn{1}{c|}{5371}                                                                & \multicolumn{1}{c|}{16}             & 1.49                                                              \\ \hline
OC  & \multicolumn{1}{c|}{409/557/492}                                                         & \multicolumn{1}{c|}{21/33/26}       & \multicolumn{1}{c|}{0.002}   & \multicolumn{1}{c|}{25.67/29.62/26.15}                    & 0.002                      & \multicolumn{1}{c|}{506}                                                                 & \multicolumn{1}{c|}{14}             & 13.83                                                             \\ \hline
HPD & \multicolumn{1}{c|}{1371/2089/1519}                                                      & \multicolumn{1}{c|}{20/30/24}       & \multicolumn{1}{c|}{0.002}   & \multicolumn{1}{c|}{7.29/7.18/7.99}                       & 0.002                      & \multicolumn{1}{c|}{1580}                                                                & \multicolumn{1}{c|}{17}             & 5.38                                                              \\ \hline
FLD & \multicolumn{1}{c|}{1554}                                                                & \multicolumn{1}{c|}{64}             & \multicolumn{1}{c|}{/}       & \multicolumn{1}{c|}{20.5}                                 & /                          & \multicolumn{1}{c|}{1554}                                                                & \multicolumn{1}{c|}{71}             & 22.84                                                             \\ \hline
OD  & \multicolumn{1}{c|}{758/933/822}                                                                 & \multicolumn{1}{c|}{2/2/2}              & \multicolumn{1}{c|}{0.002}       & \multicolumn{1}{c|}{1.07/1.32/1.22}                                 & 0.004                         & \multicolumn{1}{c|}{838}                                                                 & \multicolumn{1}{c|}{14}             & 8.35                                                              \\ \hline
TS  & \multicolumn{1}{c|}{2239/2698/2450}                                                                & \multicolumn{1}{c|}{7/12/9}              & \multicolumn{1}{c|}{0.004}       & \multicolumn{1}{c|}{1.45/2.63/1.93}                                 & 0.004                          & \multicolumn{1}{c|}{2317}                                                                & \multicolumn{1}{c|}{20}             & 4.31                                                              \\ \hline
\end{tabular}}
\label{table:RQ1.0:results}
\end{table*}

\subsubsection{RQ1.2} \emph{Does \APPR generate root cause clusters with a significant reduction in variance for simulator parameters?}

\vspace{1mm}
\emph{Design and measurements.} This research question investigates if \APPR achieves high within-cluster similarity concerning at least one simulator parameter. 
\CHANGED{Indeed, since we rely on DNNs that are trained and tested with simulators, images assigned to the same cluster should present similar values for a subset of the simulator parameters.}
Within each cluster, the variance of these parameters should be significantly smaller than that computed on the entire test set. 
For a cluster $C_{i}$, the rate of reduction in variance for a parameter $p$ can be computed as follows:

\begin{math}
RR^{p}_{C_{i}}= 1 - \frac { \mathit{variance}\ \mathit{of} p\ \mathit{for}\ \mathit{the}\ \mathit{images}\ \mathit{in}\ C_{i} } {\mathit{variance}\  \mathit{of}\  p\  \mathit{for}\ \mathit{the}\ \mathit{entire}\ \mathit{error-inducing}\ \mathit{set}}
\end{math}
\newline
\newline
\ASE{Positive values  for $RR_{C_{i}}^p$ indicate reduced variance.}

%

Table~\ref{tab:parameters} provides the list of parameters considered in our evaluation.

\begin{table}[t]
\caption{Image parameters considered to address RQ1.2}
\footnotesize
\begin{tabular}{
|@{\hspace{1pt}}p{7mm}
|@{\hspace{1pt}}p{2cm}
|@{\hspace{1pt}}p{10cm}|}
\hline
\textbf{DNN}&\textbf{Parameter}&\textbf{Description}\\
\hline
\multirow{14}{*}{GD/OC}&Gaze Angle&Gaze angle in degrees.\\
\cline{2-3}
&Openness&Distance between top and bottom eyelid in pixels.\\
\cline{2-3}
&H\_Headpose&Horizontal position of the head (degrees)\\
\cline{2-3}
&V\_Headpose&Vertical position of the head (degrees)\\
\cline{2-3}
&Iris Size&Size of the iris.\\
\cline{2-3}
&Pupil Size&Size of the pupil.\\
\cline{2-3}
&PupilToBottom&Distance between the pupil bottom and the bottom eyelid margin. 
\\
\cline{2-3}
&PupilToTop&Distance between the pupil top and the top eyelid margin. 
\\
\cline{2-3}
&DistToCenter&Distance between the pupil center of the iris center. When the eye is looking middle center, this distance is below 11.5 pixels.\\
\cline{2-3}
&Sky Exposure&Captures the degree of exposure of the panoramic photographs reflected in the eye cornea.\\
\cline{2-3}
&Sky Rotation&Captures the degree of rotation of the panoramic photographs reflected in the eye cornea.\\
\cline{2-3}
&Light&Captures the degree of intensity of the main source of illumination.\\
\cline{2-3}
&Ambient&Captures the degree of intensity of the ambient illumination.\\
\hline
\multirow{6}{*}{HPD}&
Camera Location& Location of the camera, in X-Y-Z coordinate system.\\
\cline{2-3}
&Camera Direction& Direction of the camera (X-Y-Z coordinates).\\
\cline{2-3}
&Lamp Color& RGB color of the light used to illuminate the scene.\\
\cline{2-3}
&Lamp Direction& Direction of the illuminating light (X-Y-Z coordinates).\\
\cline{2-3}
&Lamp Location& Location of the source of light (X-Y-Z coordinates).\\
\cline{2-3}
&Headpose& Position of the head of the person (X-Y-Z coordinates). It is used to derive the ground truth.\\
\hline
\multirow{2}{*}{FLD}&
X coordinate of landmark& Value of the horizontal axis coordinate for the pixel corresponding to the $i^{th}$ landmark.\\
\cline{2-3}
&Y coordinate of landmark& Value of the vertical axis coordinate for the pixel corresponding to the $i^{th}$ landmark.\\
\hline
\end{tabular}

\label{tab:parameters}
\end{table}%


\CHANGED{In the case of \GD and \OC,} we rely on the parameters given by the simulator, except for the ones that capture coordinates of single points used to draw pictures (e.g., eye landmarks) since these coordinates alone are not informative about the elements in the image.
We also rely on parameters that are derived from the coordinates mentioned above and capture characteristics that are potentially related to error-inducing images: PupilToBottom, PupilToTop, DistToCenter, Openness. 

\CHANGED{For \HPD, we also considered the parameters provided by the simulator, omitting the landmark coordinates. Parameters expressed with X-Y-Z coordinates are considered as three separate parameters (e.g.,  Headpose).} 

\CHANGED{As for \FLD, since a DNN error may depend on the specific shape and expression of the processed face (i.e., the particular position of a landmark), we considered the coordinates of the 27 landmarks on the horizontal and vertical axes as distinct parameters (54 parameters in total).}

\MAJOR{R1.4}{Note that simulator parameters are only used to objectively evaluate the approach; they are not involved in the practical application of the approach. Section~\ref{sec:rq2} addresses the application of \APPR to case study subjects  for which a simulator is not available (i.e., TS and OD).}

Since the number of parameters that capture common error causes is not known a priori, we consider a significant variance reduction in at least one parameter to be enough for the cluster to be indicative of root causes. Therefore, we compute the percentage of clusters showing such a variance reduction for at least one of the parameters. 

Consistent with HUDD, we compute the percentage of clusters with a variance reduction between 0\% and 90\%, with incremental steps of 10\%.
To answer our research question positively, a high percentage of the clusters should reduce variance for at least one of the parameters. We compare our results to those of HUDD.





\begin{figure*}
\centering
\includegraphics[width=\textwidth]{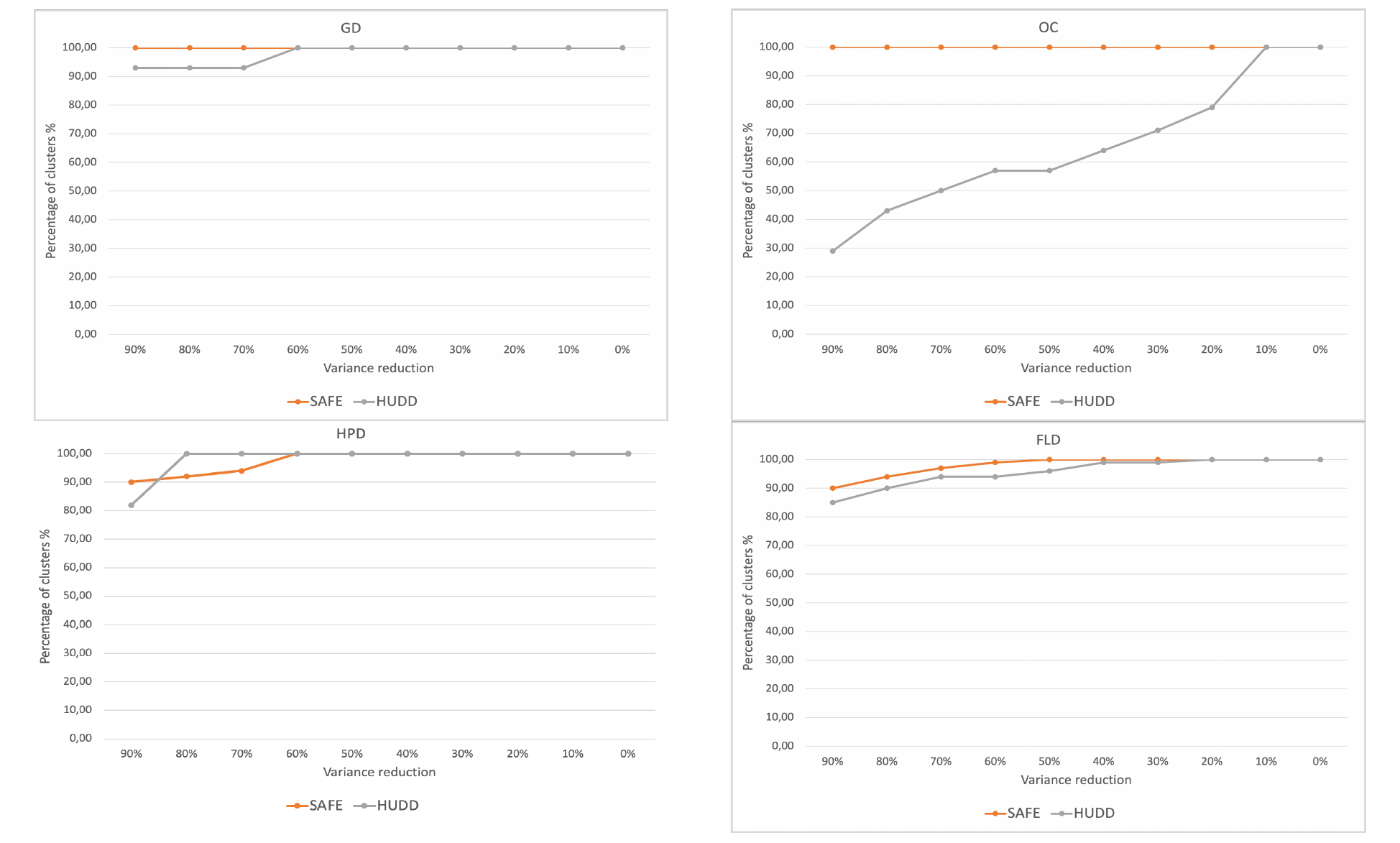}
\caption{RQ1.2: median percentage of clusters with at least one parameter showing a reduction rate above a threshold in the range [0\% - 90\%].}
\label{fig:RQ1.22}
\end{figure*}

\emph{Results.} \MAJOR{R3.3}{We report in Table \ref{tab:statistical_rq122} the maximum, minimum, and median percentage of clusters having at least one parameter showing a reduction rate above a threshold in the range [0\% - 90\%], compared to the percentage obtained by HUDD. We also report the p-values resulting from performing a one-sample Wilcoxon signed rank test (see Section~\ref{sec:subj}). We notice that, when the median obtained with \APPR is higher than the value obtained with HUDD (i.e., \APPR performs better), the p-values are always below 0.05. This implies that, in these cases, the median percentage of clusters with a reduced variance obtained with \APPR is significantly larger than the one obtained with HUDD.}

To enable visual comparison, Figure \ref{fig:RQ1.22} plots the median percentage of clusters with variance reduction for at least one of the simulator parameters, with different reduction rates, for both \APPR and HUDD.
\MAJOR{R3.3}{Results show that \APPR yields a higher percentage for three out of the four case study subjects (i.e., GD, OC, FLD). For HPD, based on the p-values reported in Table~\ref{tab:statistical_rq122}, the differences between HUDD and \APPR are not significant (i.e., HUDD does not perform better than \APPR).}

\begin{table}[H]
\caption{Minimum, Maximum and Median percentage of clusters with at least one parameter showing a reduction rate above a threshold in the range [10\% - 90\%], with the percentage obtained by HUDD and the p-value when comparing SAFE to HUDD.}
\footnotesize
\centering
\begin{tabular}{|c|c|c|c|c|c|}
\hline
            Threshold          &                     & GD                & OC                & HPD               & FLD   \\ \hline
\multirow{3}{*}{90\%} & SAFE min/max/median & 100\%/100\%/100\% & 100\%/100\%/100\% & 85\%/100\%/90\%   & 90\%  \\ \cline{2-6} 
                      & HUDD                & 93\%              & 29\%              & 82\%              & 85\%  \\ \cline{2-6} 
                      & p-value             & 0.001            & 0.001           & 0.004            &       \\ \hline
\multirow{3}{*}{80\%} & SAFE min/max/median & 100\%/100\%/100\% & 100\%/100\%/100\% & 85\%/100\%/92\%   & 94\%  \\ \cline{2-6} 
                      & HUDD                & 93\%              & 43\%              & 100\%             & 90\%  \\ \cline{2-6} 
                      & p-value             & 0.001            & 0.001            & 0.99              &       \\ \hline
\multirow{3}{*}{70\%} & SAFE min/max/median & 100\%/100\%/100\% & 100\%/100\%/100\% & 87\%/100\%/94\%   & 97\%  \\ \cline{2-6} 
                      & HUDD                & 93\%              & 50\%              & 100\%             & 94\%  \\ \cline{2-6} 
                      & p-value             & 0.001           & 0.001            & 0.99              &       \\ \hline
\multirow{3}{*}{60\%} & SAFE min/max/median & 100\%/100\%/100\% & 100\%/100\%/100\% & 100\%/100\%/100\% & 99\%  \\ \cline{2-6} 
                      & HUDD                & 100\%             & 57\%              & 100\%             & 94\%  \\ \cline{2-6} 
                      & p-value             & 1                 & 0.001           & 1                 &       \\ \hline
\multirow{3}{*}{50\%} & SAFE min/max/median & 100\%/100\%/100\% & 100\%/100\%/100\% & 100\%/100\%/100\% & 100\% \\ \cline{2-6} 
                      & HUDD                & 100\%             & 57\%              & 100\%             & 96\%  \\ \cline{2-6} 
                      & p-value             & 1                 & 0.001            & 1                 &       \\ \hline
\multirow{3}{*}{40\%} & SAFE min/max/median & 100\%/100\%/100\% & 100\%/100\%/100\% & 100\%/100\%/100\% & 100\% \\ \cline{2-6} 
                      & HUDD                & 100\%             & 64\%              & 100\%             & 99\%  \\ \cline{2-6} 
                      & p-value             & 1                 & 0.001           & 1                 &       \\ \hline
\multirow{3}{*}{30\%} & SAFE min/max/median & 100\%/100\%/100\% & 100\%/100\%/100\% & 100\%/100\%/100\% & 100\% \\ \cline{2-6} 
                      & HUDD                & 100\%             & 71\%              & 100\%             & 99\%  \\ \cline{2-6} 
                      & p-value             & 1                 & 0.001            & 1                 &       \\ \hline
\multirow{3}{*}{20\%} & SAFE min/max/median & 100\%/100\%/100\% & 100\%/100\%/100\% & 100\%/100\%/100\% & 100\% \\ \cline{2-6} 
                      & HUDD                & 100\%             & 79\%              & 100\%             & 100\% \\ \cline{2-6} 
                      & p-value             & 1                 & 0.001           & 1                 &       \\ \hline
\multirow{3}{*}{10\%} & SAFE min/max/median & 100\%/100\%/100\% & 100\%/100\%/100\% & 100\%/100\%/100\% & 100\% \\ \cline{2-6} 
                      & HUDD                & 100\%             & 100\%             & 100\%             & 100\% \\ \cline{2-6} 
                      & p-value             & 1                 & 1                 & 1                 &       \\ \hline

\end{tabular}
\label{tab:statistical_rq122}
\end{table}

\begin{table}[htp]
\centering
\caption{Number of core-points and number of clusters for each case study subject}
\footnotesize
\makebox[\textwidth]{\begin{tabular}{|C{2cm}|C{3cm}|C{3cm}|C{4cm}|}

\hline
Case Study Subject & \# of core points (min/max/median) & \# of clusters (min/max/median)& \% of images considered core points (min/max/median)  \\ \hline
GD         & 2389/5808/5080                  & 14/31/26  &   46\%/92\%/88\%           \\ \hline
OC         & 366/495/441                   & 21/33/26  &   89\%/97\%/90\%       \\ \hline
HPD        & 259/623/316                   & 20/30/24  &   18\%/30\%/22\%            \\ \hline
FLD        & 622                   & 64  &   40\%         \\ \hline
\end{tabular}}
\label{tab:core_points_size}
\end{table}

We report in Table \ref{tab:core_points_size} the minimum, maximum and median number of core points found by the clustering algorithm for each case study subject and the ratio of error-inducing images being considered as core points. 
We recall that a cluster consists of several core points and that the core points determine the cluster shape.
The ratio of error-inducing images being considered as core points is therefore an indicator of the complexity of the cluster shape. The larger this ratio, the more complex the cluster shape and the less likely it is to be convex. 
\MAJOR{R15}{Since non-convex clusters cannot be properly modeled by a centroid-based algorithm or even a hierarchical clustering algorithm (see Section~\ref{sec:approach}), this partly explains the results presented in Figure \ref{fig:RQ1.22}.} 

For GD and FLD, \APPR performs slightly better than HUDD. With GD, out of ten executions, \APPR yields a median of 100\% of the clusters with a variance reduction above 90\% compared to 93\% for HUDD. As for FLD, it obtained 90\% of the clusters with variance reduction above 90\%, compared to 85\% for HUDD. These values are close because the number of clusters detected by both methods is similar for GD and FLD. Detecting the optimal number of clusters is crucial as it leads to root cause clusters grouping very similar images with less noise. Consequently, identifying the right root cause clusters will result in higher variance reduction. Nevertheless, the slight superiority shown by \APPR is explained by the fact that the latter finds root cause clusters with arbitrary shapes, compared to convex shapes found by HUDD. This is important since arbitrary-shaped clusters can find more homogeneous clusters (i.e., clusters with higher within-cluster similarity) with very similar images. In contrast, a convex cluster tends to be less dense and can group rather dissimilar images. 
In the case of OC, the percentage of clusters with a given variance reduction obtained by \APPR is much higher than that obtained by HUDD. \APPR yielded 100\% of the clusters (median of ten executions) with variance reduction above 90\%, in contrast to 29\% for HUDD. This is explained by the fact that \APPR found a much higher number of clusters than HUDD (26 for \APPR compared to 14 for HUDD). Also, 90\% of the error-inducing images are considered core points by \APPR (median), thus indicating complex cluster shapes, which may, in turn, explain the detection of more clusters. A larger number of clusters leads to root cause clusters with a lower number of images (15 images per cluster on average for \APPR with OC), which have higher chances to contain similar images.



For the HPD case study subject, HUDD and \APPR results are very close. \APPR yields 90\% of the clusters (median of ten runs) with variance reduction above 90\% compared to 82\% for HUDD. At the same time, HUDD shows 100\% of the clusters with variance reduction above 80\%. Both methods yield 100\% of the clusters with variance reduction above 60\%. We notice that the number of clusters detected by both methods is pretty close (24 for \APPR compared to 17 for HUDD), thus explaining these results. Also, we observe that only 22\% of the error-inducing images are considered core points, hence indicating that clusters do not have complex shapes and are closer to being convex than in the OC case, for example. 

In general, when the cluster shapes obtained by \APPR are relatively simple, the results of the two approaches can be expected to be similar. 
In contrast, when clusters have arbitrarily complex shapes, there is a clear advantage in using \APPR, as illustrated by the \OC results and to a lesser extent the \GD results.

In general, in spite of being a black-box approach, \APPR tends to find a high percentage of clusters with at least one reduced parameter. For GD and OC, 100\% of the clusters show parameters with a variance reduction above 90\%. HPD and FLD yield both 90\%. Based on these results, we can positively answer RQ1.2 since all clusters present at least one parameter with a positive, significant reduction rate ($>50\%$ in Figure \ref{fig:RQ1.22}).



\subsubsection{RQ1.3} \emph{Do parameters with high variance reduction represent a plausible cause for DNN errors?}
\label{sec:evaluation:RQ1.3}

\emph{Design and measurements.} This research question investigates whether \APPR helps engineers understand the root causes for each error.

\CHANGED{We assume that DNN errors occur in specific areas of the simulator parameter space. Under this assumption, we identify a set of unsafe parameters and corresponding unsafe values around which a DNN error is susceptible to occur.}
Table~\ref{tab:boundary2} provides the list of unsafe parameters, along with the unsafe values identified.
For example, for the Gaze Angle parameter, unsafe values consist of the boundary values distinguishing labels pertaining to different gaze directions. 
\MAJOR{R1.3, R2.2b, R3.2}{These unsafe parameters were selected in a systematic manner for all the case study subjects. Precisely, we report as unsafe values all the values used to label different classes (e.g., eyes openness above/below 20 pixels) and values for borderline cases (i.e., cases in which portions of the face are hidden). 
Also, we have identified additional unsafe parameters that can cause a DNN error because they lead to masked elements in images (e.g., distance between the pupil center and the iris center, distance between the pupil bottom and the bottom eyelid margin).} Note that determining unsafe parameters and values is only required for experimental purposes here, as detailed below, and not for applying \APPR in practice.

\MAJOR{R2.3}{Face images generated with such unsafe values may lead to DNN errors because the DNN either cannot distinguish two classes or because part of the human face is not present in the image. Therefore, we expect all the error-inducing images having such characteristics to belong to an appropriate cluster; precisely, they should belong to a cluster having (a) high variance reduction for the unsafe parameter and (b) an average value close to the identified unsafe value.}

\begin{table}[tb]
\caption{Safety parameters considered to address RQ1.3}
\footnotesize
\begin{tabular}{
|@{\hspace{1pt}}p{7mm}
|@{\hspace{1pt}}p{3cm}
|@{\hspace{1pt}}p{8cm}|}
\hline
\textbf{DNN}&\textbf{Parameter}&\textbf{Unsafe values}\\
\hline
\multirow{7}{*}{\GD,\OC}&Gaze Angle&Values used to label the gaze angle in eight classes (i.e., 22.5$^{\circ}$, 67.5$^{\circ}$, 112.5$^{\circ}$, 157.5$^{\circ}$, 202.5$^{\circ}$, 247.5$^{\circ}$, 292.5$^{\circ}$, 337.5$^{\circ}$).\\
&Openness&Value used to label the gaze openness in two classes (i.e., 20 pixels) or an eye abnormally open (i.e., 64 pixels).\\
&H\_Headpose&Values indicating a head turned completely left or right (i.e., 160$^{\circ}$, 220$^{\circ}$)\\
&V\_Headpose&Values indicating a head looking at the very top/bottom (i.e., 20$^{\circ}$, 340$^{\circ}$)\\
&DistToCenter&Value below which the eye is looking middle center (i.e., 11.5 pixels).\\
&PupilToBottom&Value below which the pupil is mostly under the eyelid (i.e., -16 pixels).\\
&PupilToTop&Value below which the pupil is mostly above the eyelid (i.e., -16 pixels).\\
\hline
\multirow{2}{*}{\HPD}&Headpose-X&Boundary cases (i.e.,-28.88$^{\circ}$,21.35$^{\circ}$), values used to label the headpose in nine classes (-10$^{\circ}$,10$^{\circ}$), and middle position (i.e., 0$^{\circ}$).\\
\cline{2-3}
&Headpose-Y&Boundary cases (i.e.,-88.10$^{\circ}$,74.17$^{\circ}$), values used to label the headpose in nine classes (-10$^{\circ}$,10$^{\circ}$), and middle position (i.e., 0$^{\circ}$).\\
\hline
\end{tabular}

\label{tab:boundary2}
\end{table}%


For our experiment, we consider that a root cause cluster is explanatory in terms of root causes if it satisfies two requirements: (1) It should have at least one unsafe parameter with a variance reduction above 50\%, (2) the cluster average should be close to one unsafe value. For Gaze Angle, Openness, Headpose-X, and Headpose-Y, an average value is considered close to an unsafe value if the difference between them is below 25\% of the length of the subrange including the average value. For DistToCenter, PupilToBottom, and PupilToTop, an average value is considered close to an unsafe value if it is below or equal to it.
For the FLD case study subject, 
since the reason for not detecting a landmark cannot be related to a single simulator parameter but often depends on combinations of parameters (e.g., the position of the head and the illumination angle lead to shadows on the face), 
it is impossible to determine unsafe values and therefore such a set of explanatory parameters; as a result, FLD is omitted from this experiment.

Based on the above, we address this research question by computing the percentage of clusters that are explanatory according to our definition. The higher this percentage, the more evidence we have that clustering is useful for identifying causes of DNN errors. 

\emph{Results.}  Table \ref{tab:statistical_rq13} shows the percentage of the root cause clusters that are explanatory for both \APPR and HUDD. Since we repeat the execution \APPR with ten different DNN instances for each case study subject, we report the minimum, maximum and median of the percentages obtained. 
Across all three case study subjects, \APPR shows a higher percentage of explanatory root cause clusters than HUDD. The median results with GD, OC, and HPD are 86\%, 100\%, and 90\%, respectively, compared to 86\%, 57\%, and 88\%, respectively, with HUDD. 
\MAJOR{R3.3}{Table \ref{tab:statistical_rq13} also reports the p-values resulting from performing a one-sample Wilcoxon signed rank test (see Section~\ref{sec:subj}). 
The p-values are below 0.05 for OC and HPD, which indicates 
that the percentage of \APPR's root cause clusters that present at least one
explanatory parameter is significantly larger than the one obtained with HUDD for OC and HPD. As for GD, the results are similar.}

\begin{table}[ht!]
\caption{Minimum, Maximum and Median percentage of root cause clusters that present at least one explanatory parameter with the percentage obtained by HUDD and the p-value when comparing SAFE to HUDD.}
\begin{tabular}{|c|ccc|c|c|}
\hline
    Case Study Subjects & \multicolumn{3}{c|}{SAFE}                                        & HUDD & p-value \\ \hline
    & \multicolumn{1}{c|}{Min}  & \multicolumn{1}{c|}{Max}   & Median &      &         \\ \hline
GD  & \multicolumn{1}{c|}{83\%} & \multicolumn{1}{c|}{100\%} & 86\%    & 86\% & 0.99    \\ \hline
OC  & \multicolumn{1}{c|}{93\%} & \multicolumn{1}{c|}{100\%} & 100\%    & 57\% & 0.002   \\ \hline
HPD & \multicolumn{1}{c|}{84\%} & \multicolumn{1}{c|}{96\%}  & 90\%    & 88\% & 0.02    \\ \hline
\end{tabular}
\label{tab:statistical_rq13}
\end{table}

These results show a large difference in the percentage of clusters that can be explained between \APPR or HUDD for the OC case study subject (43\% difference). Indeed, for \APPR, all the clusters have a high reduction in variance, while this is only the case for 57\% of the clusters for HUDD. As explained in the previous Section, this can be explained by the fact that OC clusters have a complex shape, more so than in other case study subjects. 



As for GD and HPD, the \APPR median is close to the result obtained with HUDD (although for HPD the difference is significant with a significance level of $0.05$). For the median, we observe a 2\% difference for HPD and no difference for GD. These results confirm the results obtained in RQ1.2. For GD and HPD, both methods show 100\% of the clusters with a parameter presenting a variance reduction above 50\%. These results are however still slightly in favor of \APPR. 
Once again, the above results are explained by the fact that the root cause clusters found by \APPR can take arbitrary shapes. Such clusters, as previously explained and in the general case, have better chances to group similar images than clusters with convex shapes. 

\subsubsection{RQ1.4} \emph{Does \APPR identify more distinct error root causes than HUDD?}
\label{sec:exp:distinctErrorCauses}

\vspace{1mm}
\emph{Design and measurements.} 
This research question investigates if \APPR identifies a larger number of possible causes of errors than HUDD. Specifically, we compare the two approaches in terms of the number of unsafe values being covered by at least one cluster. 
We say that an unsafe value $v$ is covered by a cluster $c$,
when $c$
presents a parameter $p$ with a high variance reduction and the parameter $p$ has an average value close to the unsafe value $v$. 

Since our simulators generate images having parameter values that are uniformly sampled within the input domain, every unsafe value has the same likelihood of being observed in the test set images.
Therefore, ideally, we aim for the root cause clusters to cover all such values. 


\emph{Results.} 
In Table \ref{tab:statistical_rq14}, we report the minimum, maximum and median percentage of the unsafe values covered by the root cause clusters obtained when applying \APPR to our case study subjects. 
The clusters generated by \APPR with GD, OC, and HPD cover (median) 92\%, 64\%, and 80\% of the unsafe values, respectively. The clusters generated by HUDD, instead, cover 71\%, 50\%, and 60\% of the unsafe values, respectively.
The p-values resulting from performing the one-tailed, one-sample Wilcoxon signed-rank test are always below 0.05,  which implies that the median obtained with \APPR is significantly higher than the result obtained with HUDD. 

\begin{table}[H]
\caption{Minimum, Maximum and Median percentage of the unsafe values covered by the root cause clusters with the p-value when comparing SAFE to HUDD.}
\begin{tabular}{|c|ccc|c|c|}
\hline
    & \multicolumn{3}{c|}{SAFE}                                        & HUDD & p-value \\ \hline
    & \multicolumn{1}{c|}{Min}  & \multicolumn{1}{c|}{Max}   & Median &      &         \\ \hline
GD  & \multicolumn{1}{c|}{85\%} & \multicolumn{1}{c|}{100\%} & 92\%    & 71\% & 0.002   \\ \hline
OC  & \multicolumn{1}{c|}{64\%} & \multicolumn{1}{c|}{71\%}  & 64\%    & 50\% & 0.002   \\ \hline
HPD & \multicolumn{1}{c|}{60\%} & \multicolumn{1}{c|}{80\%}  & 80\%    & 60\% & 0.004   \\ \hline
\end{tabular}
\label{tab:statistical_rq14}
\end{table} 
\begin{table}[ht!]
\centering
\caption{Coverage of the unsafe values by the root cause clusters (OC and GD case study subjects) }
\makebox[0.8\textwidth]{\begin{tabular}{|c|c|C{0.35cm}|C{0.4cm}|C{0.35cm}|C{0.4cm}|}
\hline
\multicolumn{1}{|l|}{}                   & \multicolumn{1}{l|}{}               & \multicolumn{2}{c|}{GD}                             & \multicolumn{2}{c|}{OC}                             \\ \hline
\multirow{10}{*}{Angle:}                 & \multicolumn{1}{c|}{Unsafe values}               & \multicolumn{1}{l|}{\APPR} & \multicolumn{1}{c|}{HUDD} & \multicolumn{1}{l|}{\APPR} & \multicolumn{1}{c|}{HUDD} \\ \cline{2-6} 

                                         & 337,5                               & \cmark                       & \cmark                      & \cmark                    & \xmark                         \\ \cline{2-6} 
                                         & 22,5                                & \cmark                    & \cmark                      & \xmark                       & \xmark                         \\ \cline{2-6} 
                                         & 67,5                                & \cmark                    & \cmark                      & \xmark                       & \xmark                         \\ \cline{2-6} 
                                         & 112,5                               & \cmark                    & \cmark                      & \xmark                       & \xmark                         \\ \cline{2-6} 
                                         & 157,5                               & \cmark                    & \cmark                      & \xmark                       & \xmark                         \\ \cline{2-6} 
                                         & 202,5                               & \cmark                    & \cmark                      & \cmark                    & \xmark                         \\ \cline{2-6} 
                                         & 247,5                               & \cmark                    & \cmark                      & \cmark                    & \cmark                      \\ \cline{2-6} 
                                         & 292,5                               & \cmark                    & \cmark                      & \cmark                    & \cmark                      \\ \hline
\multirow{2}{*}{H-Headpose} 
                                         & 220                                 & \cmark                    & \xmark                         & \cmark                    & \cmark                      \\ \cline{2-6} 
                                         & 160                                 & \cmark                    & \cmark                      & \cmark                    & \cmark                      \\ \hline
\multirow{2}{*}{V-Headpose} 
                                         & 20                                  & \cmark                    & \xmark                         & \cmark                    & \cmark                      \\ \cline{2-6} 
                                         & 340                                 & \cmark                    & \xmark                         & \cmark                    & \xmark                         \\ \hline
StrangeDist Top/Bot                      & -14                                 & \cmark                    & \xmark                         & \cmark                    & \cmark                      \\ \hline
Distance                                 & 25                                  & \cmark                    & \cmark                      & \cmark                    & \cmark                      \\ \hline
& \multicolumn{1}{l|}{TOTAL Coverage} & \multicolumn{1}{r|}{\textbf{14}} & \multicolumn{1}{r|}{10}   & \multicolumn{1}{r|}{\textbf{10}} & \multicolumn{1}{r|}{7}    \\ \hline
\end{tabular}}
\label{tab:coverage}
\end{table}

\begin{table}[ht!]
\caption{Coverage of the unsafe values by the root cause clusters (HPD case study subject) }
\centering
\begin{tabular}{|C{3cm}|c|C{1cm}|C{1cm}|}
\hline
\multicolumn{1}{|l|}{}      & \multicolumn{1}{l|}{}               & \multicolumn{2}{c|}{HPD}                            \\ \hline
\multicolumn{1}{|l|}{}      & \multicolumn{1}{l|}{}               & \multicolumn{1}{l|}{\APPR} & \multicolumn{1}{l|}{HUDD} \\ \hline
\multirow{5}{*}{H-Headpose} & -28                                 & \cmark                    & \cmark                      \\ \cline{2-4} 
                            & -10                                 & \cmark                    & \cmark                      \\ \cline{2-4} 
                            & 0                                   & \cmark                    & \cmark                      \\ \cline{2-4} 
                            & 10                                  & \cmark                       & \xmark                         \\ \cline{2-4} 
                            & 21                                  & \xmark                       & \xmark                         \\ \hline
\multirow{5}{*}{V-HeadPose} & -88                                 & \xmark                       & \xmark                         \\ \cline{2-4} 
                            & -10                                 & \cmark                    & \cmark                      \\ \cline{2-4} 
                            & 0                                   & \cmark                    & \cmark                      \\ \cline{2-4} 
                            & 10                                  & \cmark                    & \cmark                      \\ \cline{2-4} 
                            & 77                                  & \cmark                    & \xmark                         \\ \hline
\multicolumn{1}{|l|}{}      & \multicolumn{1}{l|}{TOTAL Coverage} & \multicolumn{1}{r|}{\textbf{8}}  & \multicolumn{1}{r|}{6}    \\ \hline
\end{tabular}
\label{tab:coverage2}
\end{table}
Below, we discuss, more in detail, the differences between \APPR and HUDD. To exemplify our discussion, we report in Table \ref{tab:coverage} (GD, OC) and Table \ref{tab:coverage2} (HPD) examples of unsafe values covered by the clusters generated in one of the runs of \APPR and HUDD.

Based on Table \ref{tab:statistical_rq14}, for GD, \APPR identifies root cause clusters covering 92\% unsafe values (median out of ten runs), compared to 71\% for HUDD. 
This is because \APPR relies on images represented with features extracted from convolutional layers, which provide a better representation than heatmaps and show aspects that are not captured by heatmaps, such as eye shape, edges, and corners. 

For OC, \APPR covers  a median of 65\% unsafe values compared to 50\% for HUDD. The uncovered unsafe values concern the parameter \textit{Angle}. However, in the OC case study subject, we mainly focus on eye openness (\textit{Distance} parameter) and the distance between the pupil and eyelid (\textit{StrangeDist Top/Bot} parameter). All of the unsafe values for these two relevant parameters were covered by \APPR and HUDD.

For the HPD case study subject, \APPR covers a median of 80\% unsafe values compared to 60\% for HUDD. None of the techniques covers values $21$ for \textit{H-Headpose} and $-88$ for \textit{V-Headpose}. However, such values can be observed if we also consider parameters with a variance below 50\% (which is an arbitrary threshold). These two values represent boundary values that we believe confuse \APPR as they correspond to situations where it is hard to see the eyes and the shape of the head. As a result, such images are sometimes clustered erroneously. Thus, these clusters show a lower variance reduction. 

\MAJOR{R2.2, R3.30}{In addition, we report that 90\% of the clusters cover a unique set of unsafe values (e.g., \emph{Angle=337.5, H-Headpose=220, V-Headpose=340, Distance=25}). Concerning the remaining 10\%, we observe that they are still unique but differ with respect to a parameter that is not unsafe. Therefore, we conclude that all the clusters generated by \APPR cover a unique set of parameter values that are useful to determine distinct failure causes.}

\subsubsection{RQ1.5} \MAJOR{R2.6}{How many images are required to identify commonalities in a cluster?}  
\label{sec:images:number}

\MAJORSTARTS{}

\emph{Design and measurements.} We aim to determine how many images from a root cause cluster an engineer should inspect to correctly identify the root cause captured by the cluster. Recall that in our previous analysis (i.e., RQ1.1), we made an assumption regarding this point to estimate cost savings that can be expected from \APPR.
To this end we conducted an online questionnaire-based experiment, following best practices~\cite{linaker2015guidelines}.
Our population of participants consists of 19 PhD students at the University of Luxembourg and University of Ottawa. 
All the selected students hold a master's degree in computer science or a corresponding engineering degree; further,  most have acquired fundamentals in machine learning and a few work on safety-related topics. Therefore, all the selected participants are competent to perform the experimental task, which does not require background knowledge, though they are less familiar with the problem domain than engineers would be in practice and therefore can be expected to make more mistakes. 

In our experiment, we asked participants to identify the commonalities in a randomly selected subset of images belonging to randomly selected root cause clusters generated by \APPR for our case study subjects. 
Such task emulates what should be done by an engineer to understand the root causes for an error.
We asked each participant to perform the task three times, on different clusters of images. Each cluster of images included respectively 5, 10, and 15 images randomly selected from a root cause cluster belonging to a different case study subject. Also, for the same subject, each participant received a different random set of images, belonging to a randomly selected cluster. In short, each participant therefore received three questions with 5 images, 10 images, and 15 images, respectively, belonging to different clusters from different case study subjects.

We provided closed-ended questions to objectively compare the obtained answers with the ground truth (i.e., likely root causes for DNN errors).
Similarly to RQ1.4, we considered only case study subjects performing classification tasks because, for these cases, the availability of the ground truth makes it possible to determine the reasons for DNN failure objectively. 

\begin{figure}[t]
\includegraphics[width=0.9\textwidth]{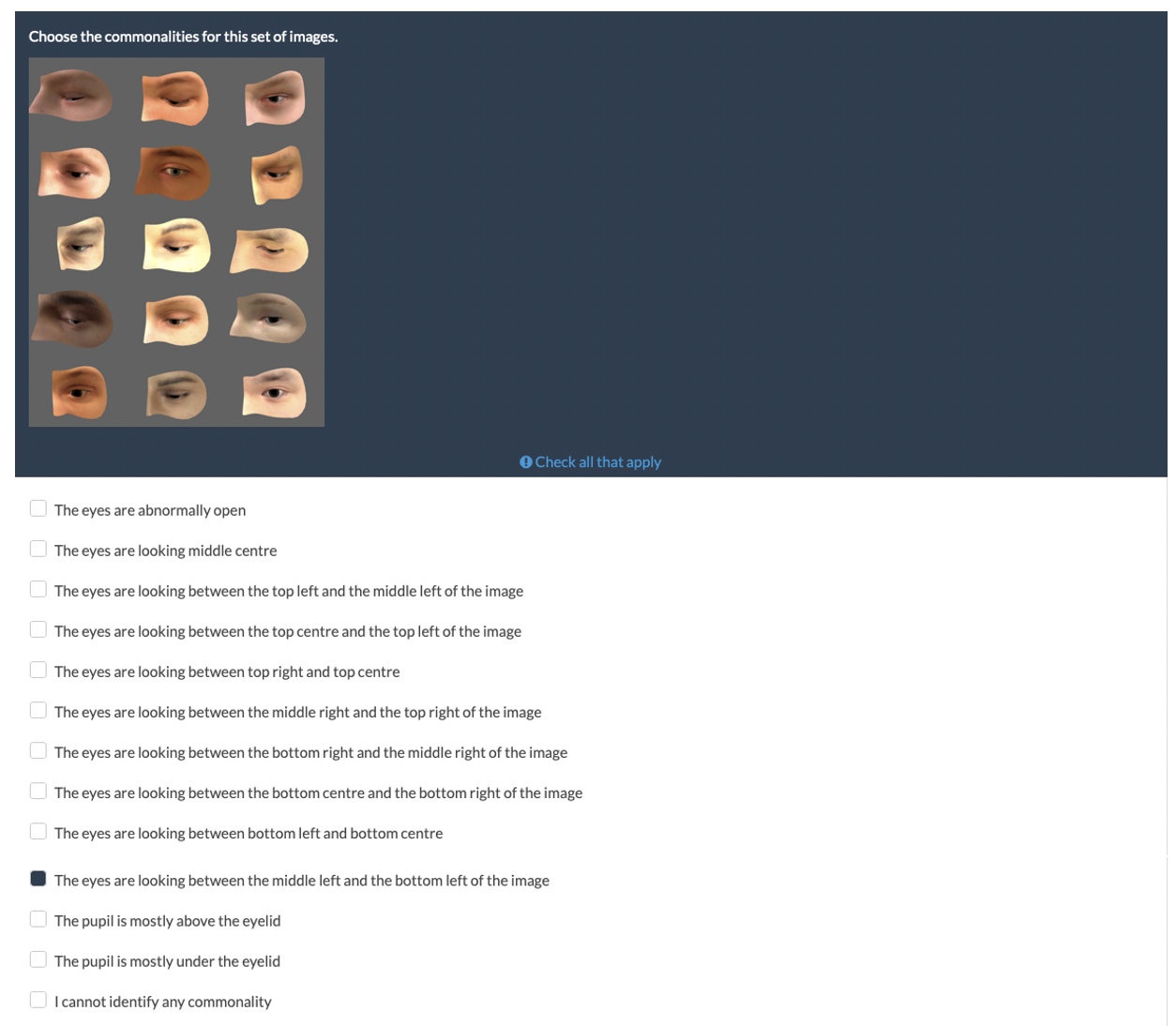}
\caption{Example question appearing in our questionnaire.}
\label{fig:survey}
\end{figure}

As commonalities to be identified by the participants, we considered all the unsafe values described in RQ1.4. For each unsafe value, we provided a descriptive sentence (see Table~\ref{unsafe_values_description}) to be selected by the participants from a checkbox list. Participants could select more than one option and we provided the option \emph{None of the above}, for participants who did not find the set of provided answers to be satisfactory.
\begin{table}[]
\centering
\caption{The descriptive sentence used in the survey for each unsafe value}
\label{unsafe_values_description}
\begin{tabular}{|C{2cm}|C{8cm}|c|}
\hline
Parameter     & Descriptive sentence                                                                  & Unsafe value                \\ \hline
Headpose-Y    & The face is straight forward                                                          & 0$^{\circ}$                 \\ \hline
Headpose-Y    & The face is partially not visible because it is inclined to the top                   & 74.17$^{\circ}$             \\ \hline
Headpose-Y    & The face is partially not visible because it is inclined to the bottom                & -88.10$^{\circ}$            \\ \hline
Headpose-X    & The face is partially not visible because it is turned to the left side of the image  & -28.88$^{\circ}$            \\ \hline
Headpose-X    & The face is partially not visible because it is turned to the right side of the image & 21.35$^{\circ}$             \\ \hline
Headpose-X    & The face is turned between the middle and the left side of the image                  & -10$^{\circ}$               \\ \hline
Headpose-X    & The face is turned between the middle and the right side of the image                 & 10$^{\circ}$                \\ \hline
Headpose-Y    & The face is turned between the centre and the top of the image                        & 10$^{\circ}$                \\ \hline
Headpose-Y    & The face is looking between the centre and the bottom of the image                    & -10$^{\circ}$               \\ \hline
Openness      & The eyes are abnormally open                                                          & \textgreater{}64$^{\circ}$  \\ \hline
DistToCenter  & The eyes are looking middle centre                                                    & \textless{}11.5$^{\circ}$   \\ \hline
Gaze Angle    & The eyes are looking between the top left and the middle left of the image            & 22.5$^{\circ}$              \\ \hline
Gaze Angle    & The eyes are looking between the top centre and the top left  of the image            & 67.5$^{\circ}$              \\ \hline
Gaze Angle    & The eyes are looking between top right and top centre                                 & 112.5$^{\circ}$             \\ \hline
Gaze Angle    & The eyes are looking between the middle right and the top right of the image          & 157.5$^{\circ}$             \\ \hline
Gaze Angle    & The eyes are looking between the bottom right and the middle right of the image       & 202.5$^{\circ}$             \\ \hline
Gaze Angle    & The eyes are looking between the bottom centre and the bottom right of the image      & 247.5$^{\circ}$             \\ \hline
Gaze Angle    & The eyes are looking between bottom left and bottom centre                            & 292.5$^{\circ}$             \\ \hline
Gaze Angle    & The eyes are looking between the middle left and the bottom left of the image         & 337.5$^{\circ}$             \\ \hline
PupilToTop    & The pupil is mostly above the eyelid                                                  & \textless{}-16$^{\circ}$    \\ \hline
PupilToBottom & The pupil is mostly under the eyelid                                                  & \textgreater{}-16$^{\circ}$ \\ \hline
\end{tabular}
\end{table}
Each questionnaire was introduced by a short description of \APPR and a detailed description of the task to perform. Further, we provided an example answer from a different case that is simple to understand (classification of animal pictures). 


Data collection was automated using Lime Survey\footnote{https://ulsurvey.uni.lu/} and its link was sent to the participants by email. Figure \ref{fig:survey} shows an example question provided to the participants (in this case, the images belong to one root cause cluster generated for the GD DNN). An example of a questionnaire sent to the participants can be found in our replicability package. 

For each set of answers (i.e., the ones provided with 5, 10, or 15 images), we counted the number of answers that matched our ground truth.
The number of images required to determine the root causes of a DNN error is the minimal number leading to a high percentage of correct answers.

To summarize, based on the experimental design above, we prevented learning effects from one question to the next. We ensured that, for the same case study, the selected cluster and images were different for each participant and selected randomly, to avoid any form of systematic bias. 

\begin{table}[H]
\caption{Percentage of correct responses by inspecting 5, 10 and 15 images for each case study subject, based on the questionnaire study.}
\begin{tabular}{|c|c|c|c|}
\hline
&\multicolumn{3}{c|}{Correct responses}\\
Case Study Subjects & 5 images inspected& 10 images inspected& 15 images inspected\\ \hline
GD                  & 80\%     & 88\%      & 67\%      \\ \hline
OC                  & 83\%     & 83\%      & 86\%      \\ \hline
HPD                 & 100\%    & 83\%      & 67\%      \\ \hline
Overall& 89\%     & 84\%      & 74\%      \\ \hline
\end{tabular}
\label{tab:survey_results}
\end{table}

\emph{Results.}
Table \ref{tab:survey_results} presents the results obtained from analyzing the questionnaire data. 
Overall, we notice that when looking at 5, 10, and 15 images, 89\% 84\%, and 74\% of the participants found the correct commonality in a cluster, respectively. Given that participants performed that type of task for the first time, with limited initial training, we can consider 89\% to be a good result and a lower bound of what experienced practitioners would achieve. 
However, surprisingly, a larger number of inspected images does not improve results. On the contrary, with a larger number of images being inspected (e.g., 15 images), the percentage of correct answers tends to drop. This result may be due to the fact that a larger set of images leads to a higher cognitive load and is also more likely to include noisy images (i.e., images that do not present the same commonalities as most of the other images in the cluster). In the presence of noisy images, a user who looks for a commonality across all the images may identify characteristics that are not a correct explanation for the DNN error.

Since we are dealing with proportions, we performed a Fisher exact test to determine if the differences observed between the pairs \emph{5 images VS 10 images} and \emph{5 images VS 15 images} are significant. The obtained p-values (i.e., \emph{1} for \emph{`5 vs 10'} and \emph{0.4} for \emph{`5 vs 15}') indicate that differences are not significant, which may be due in part to the small number of participants (19). However, our results clearly show that the inspection of five images does not lead to worse results than the inspection of a higher number of images, thus justifying our assumption in RQ1.1 (i.e., engineers inspect five images per cluster).


GD and OC lead to similar results, with 80\% (GD) and 83\% (OC) of the participants who inspected five images providing a correct answer. HPD leads to better results; indeed, 100\% of the participants found the correct commonality by inspecting five images. Such difference between HPD and the other two cases is likely due to the fact that the simulator used to generate HPD images is more realistic (i.e., generates a whole face), while the simulator used for GD and OC simply generates eye bulbs and part of the forehead (see Figure~\ref{fig:RCC_examples:two}), thus resulting in images that are more complicated to quickly understand by participants who are not familiar with the application domain. Despite such differences, which are to be expected across case studies, the generic trend is consistent: five images seem to be sufficient and not worse than larger sets of images. Therefore, we conclude that inspecting five images per cluster is an acceptable choice regardless of the case study DNN.

\MAJORENDS{}

\subsubsection{RQ2} \emph{\MAJOR{R3.26}{Does \APPR enable engineers to more effectively and efficiently retrain a DNN when compared with HUDD and baselines approaches?}}
\label{sec:rq2}

\begin{figure}[htp]
\includegraphics[width=\textwidth]{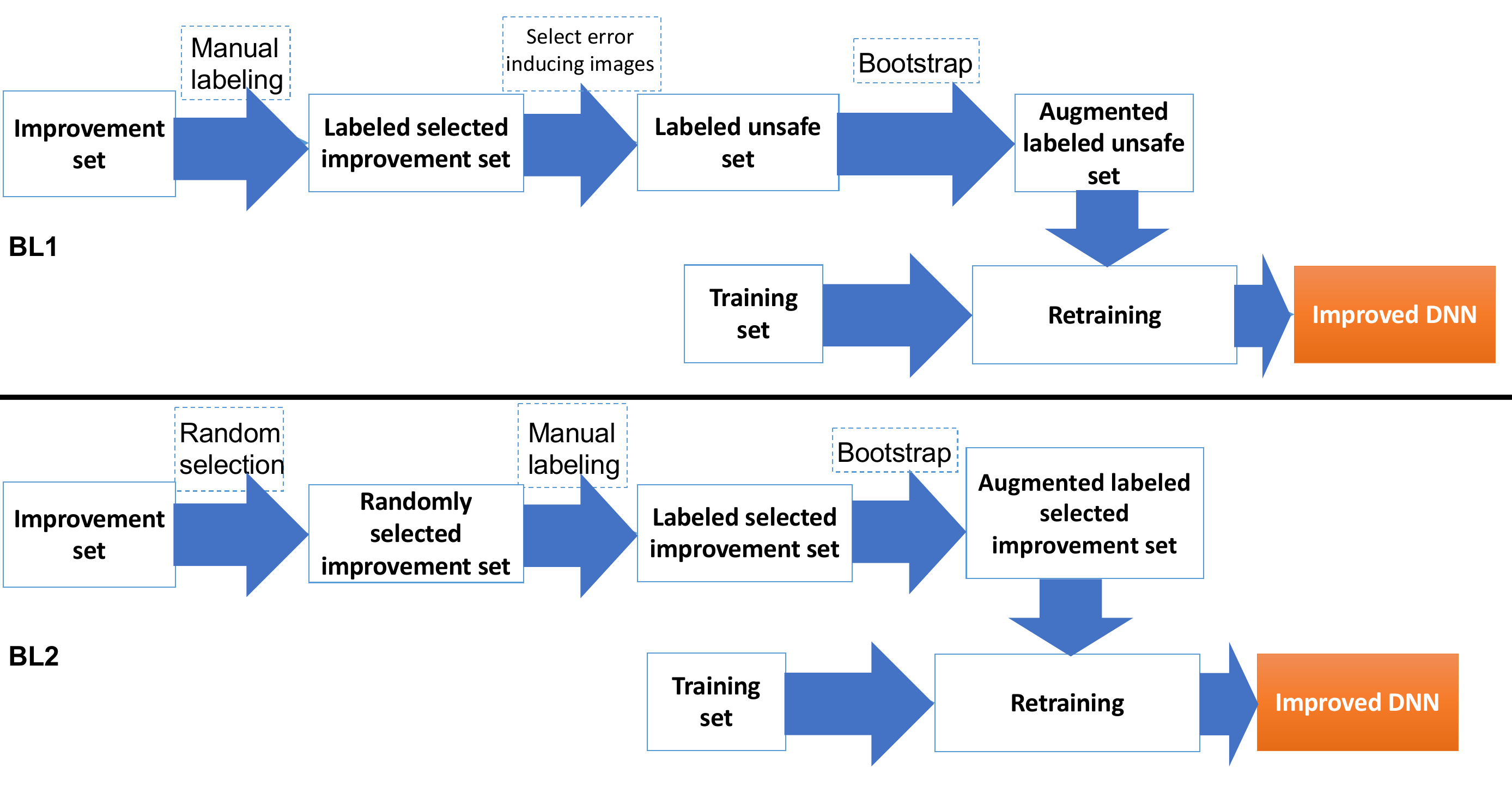}
\caption{Process of the two baselines used to compare \APPR}
\label{fig:baselines}
\end{figure}

\vspace{1mm}
\CHANGED{\emph{Design and measurements.}}
In this experiment, we investigate whether \APPR significantly improves the accuracy due to retraining the DNN, thanks to its selection method accounting for the shape of clusters. We compare these improvements to HUDD and two baselines, namely BL1 and BL2, which are depicted in Figure~\ref{fig:baselines} and explained in the following:

\textit{BL1}: In this baseline, we use the error-inducing images in the improvement set as an unsafe set.
Precisely, we label\footnote{For our experiments, labeling comes for free because we either derive images using a simulator or we rely on available datasets. However, labeling comes at a high cost in industrial practice, where new images are collected from the field.} a random subset of the improvement set and execute the DNN to identify error-inducing images.
As for HUDD, this selected unsafe set 
is augmented by applying bootstrap resampling (i.e., replicating samples in the unsafe set) in order to have a sufficiently large number of unsafe images to improve the DNN. 

\textit{BL2}: 
This baseline consists of randomly selecting a set of images from the improvement set, labeling them to obtain a \emph{labeled selected improvement set}, and using them for retraining.

We rely on the same settings and environment to run the experiments for the four different retraining strategies (i.e., SAFE, HUDD, BL1, and BL2). These experiments are repeated ten times to account for randomness.

To retrain DNNs, we rely on the approach described in Section \ref{sec:step4}. 
\ASEnn{For fair comparisons with HUDD~\cite{fahmysupporting}, BL1, and BL2, we configure bootstrap resampling to generate an \emph{augmented labeled unsafe set} and an \emph{augmented labeled selected improvement set} with the same size as the \emph{balanced labeled unsafe set} for HUDD (see Figure~\ref{fig:HUDD}).}

We compute the accuracy of the retrained models on the test sets and compare the accuracy improvement obtained by \APPR with those obtained by HUDD and the baselines. For this experiment, we consider the case study subjects presented in Table \ref{tab:dnns}.

The improvement sets for \GD and \CloseDNN have been generated through additional executions of UnityEyes. For \HPD and \FLD, they have been generated with additional executions of the IEE simulator, configured to use two new face models which were not used for generating the training and test sets. We selected images of the original datasets not used for the training and test sets for the other cases.

\emph{Results.} Table~\ref{tab:rq2_safe} provides an accuracy comparison between \APPR, HUDD, and the two baselines on the different case study DNNs. It also provides the size of the unsafe sets (number of images) selected by each method. 

\begin{table*}[tb]
\footnotesize
\centering
\caption{RQ2: Unsafe set size and the accuracy improvement of \APPR compared to HUDD}
\makebox[0.8\textwidth]{\begin{tabular}{|C{0.7cm}|C{1cm}|ccc|ccc|C{0.8cm}|C{0.8cm}|}
\hline
\textbf{DNN} & Original Model & \multicolumn{3}{c|}{SAFE}                                                                                                                        & \multicolumn{3}{c|}{HUDD}                                                                                                                        & BL1      & BL2      \\ \hline
             & Accuracy       & \multicolumn{1}{C{0.8cm}|}{Unsafe set size} & \multicolumn{1}{C{1.2cm}|}{Accuracy} & \begin{tabular}[L{1cm}]{@{}l@{}l@{}l@{}}Improvement\\ over\\ original\\ model\end{tabular} & \multicolumn{1}{C{0.8cm}|}{Unsafe set size} & \multicolumn{1}{C{1cm}|}{Accuracy} & \begin{tabular}[C{1.2cm}]{@{}l@{}l@{}l@{}}Improvement\\ over\\ original\\ model\end{tabular} & Accuracy & Accuracy \\ \hline
GD           & 95.95\%        & \multicolumn{1}{c|}{1648}            & \multicolumn{1}{c|}{96.74\%}  & \textbf{+0.79}                                                                     & \multicolumn{1}{c|}{1615}            & \multicolumn{1}{c|}{96.23\%}  & +0.28                                                                     & 95.23\%  & 95.80\%  \\ \hline
OC           & 88.03\%        & \multicolumn{1}{c|}{153}             & \multicolumn{1}{c|}{96.29\%}  & \textbf{+8.2}                                                                      & \multicolumn{1}{c|}{160}             & \multicolumn{1}{c|}{94.41\%}  & +6.38                                                                     & 91.65\%  & 92.33\%  \\ \hline
HPD          & 44.07\%        & \multicolumn{1}{c|}{438}             & \multicolumn{1}{c|}{69.58\%}  & \textbf{+25.51}                                                                      & \multicolumn{1}{c|}{481}             & \multicolumn{1}{c|}{68.13\%}  & +24.06                                                                    & 66.73\%  & 66.30\%  \\ \hline
FLD          & 44.99\%        & \multicolumn{1}{c|}{659}             & \multicolumn{1}{c|}{79.17\%}         &        \textbf{+34.18}                                                                   & \multicolumn{1}{c|}{502}             & \multicolumn{1}{c|}{75.23\%}  & +30.24                                                                    & 72.02\%  & 73.83\%  \\ \hline
TS           & 81.65\%        & \multicolumn{1}{c|}{597}             & \multicolumn{1}{c|}{93.47\%}  & \textbf{+11.82}                                                                     & \multicolumn{1}{c|}{704}             & \multicolumn{1}{c|}{93.03\%}  & +11.38                                                                    & 92.63\%  & 92.73\%  \\ \hline
OD           & 84.12\%        & \multicolumn{1}{c|}{212}             & \multicolumn{1}{c|}{97.14\%}  & \textbf{+13.02}                                                                     & \multicolumn{1}{c|}{258}             & \multicolumn{1}{c|}{97.04\%}  & +12.92                                                                    & 97.04\%  & 96.67\%  \\ \hline
\end{tabular}}
\label{tab:rq2_safe}
\end{table*}
\begin{table*}[tb]
\footnotesize
\centering

\caption{RQ2: p-values and VDA values when comparing \APPR to HUDD and the baselines}

\makebox[0.8\textwidth]{\begin{tabular}{
|@{\hspace{1pt}}>{\raggedleft\arraybackslash}p{9mm}@{\hspace{1pt}}|
@{\hspace{1pt}}>{\raggedleft\arraybackslash}p{6mm}@{\hspace{1pt}}|
@{\hspace{1pt}}>{\raggedleft\arraybackslash}p{6mm}@{\hspace{1pt}}|
@{\hspace{1pt}}>{\raggedleft\arraybackslash}p{6mm}@{\hspace{1pt}}|
@{\hspace{1pt}}>{\raggedleft\arraybackslash}p{6mm}@{\hspace{1pt}}|
@{\hspace{1pt}}>{\raggedleft\arraybackslash}p{6mm}@{\hspace{1pt}}|
@{\hspace{1pt}}>{\raggedleft\arraybackslash}p{6mm}@{\hspace{1pt}}|
@{\hspace{1pt}}>{\raggedleft\arraybackslash}p{6mm}@{\hspace{1pt}}|
@{\hspace{1pt}}>{\raggedleft\arraybackslash}p{6mm}@{\hspace{1pt}}|
@{\hspace{1pt}}>{\raggedleft\arraybackslash}p{6mm}@{\hspace{1pt}}|
@{\hspace{1pt}}>{\raggedleft\arraybackslash}p{6mm}@{\hspace{1pt}}|
@{\hspace{1pt}}>{\raggedleft\arraybackslash}p{6mm}@{\hspace{1pt}}|
@{\hspace{1pt}}>{\raggedleft\arraybackslash}p{6mm}@{\hspace{1pt}}|
@{\hspace{1pt}}>{\raggedleft\arraybackslash}p{6mm}@{\hspace{1pt}}|
@{\hspace{1pt}}>{\raggedleft\arraybackslash}p{6mm}@{\hspace{1pt}}|
@{\hspace{1pt}}>{\raggedleft\arraybackslash}p{6mm}@{\hspace{1pt}}|
@{\hspace{1pt}}>{\raggedleft\arraybackslash}p{6mm}@{\hspace{1pt}}|
@{\hspace{1pt}}>{\raggedleft\arraybackslash}p{6mm}@{\hspace{1pt}}|
@{\hspace{1pt}}>{\raggedleft\arraybackslash}p{6mm}@{\hspace{1pt}}|
}
\hline
\textbf{} &	\multicolumn{3}{c|}{GD} & \multicolumn{3}{c|}{OC} & \multicolumn{3}{c|}{TS} & \multicolumn{3}{c|}{OD} & \multicolumn{3}{c|}{HPD} & \multicolumn{3}{c|}{FLD}\\\hline
p-value & 0.003 & 0.003 & 0.001	& 0.000	& 0.000	& 0.000 & 0.007 & 0.003 & 0.001 & 0.22 & 0.002 & 0.04 & 0.003 & 0.000 & 0.000 & 0.000 & 0.000 & 0.000	\\\hline
 VDA & 0.88  & 0.88 & 0.9	& 1	&	1 & 1 & 0.85 & 0.96 & 1 & 0.66 & 0.9 & 0.77 & 0.85 & 1 & 1 & 1 & 1 & 1	\\\hline

\hline
\end{tabular}}
\label{vda_results}
\end{table*}%

For \APPR, improvements in accuracy over the original model varies from 0.79 to 34.18, compared to lower ranges for HUDD (from 0.28\% to 30.24\%), BL1 (from -0.18\% to 27.03\%) and BL2 (-0.15\% to 28.84\%). Across case study DNNs, \APPR clearly and systematically yields better accuracy results compared to HUDD and the baselines, though to varying extents. Note that unsafe set sizes are comparable across improvement strategies and that differences in accuracy are therefore due to the strategy adopted for selecting unsafe images.  

The lower improvement shown by the baselines can be explained by the fact they do not rely on root cause clusters to select images for the unsafe set, a strategy that appears to be beneficial in our context. \APPR always yields higher improvement than that of HUDD. This is because HUDD relies on the cluster's centroids to select images for the unsafe set without accounting for the cluster's shape. In contrast, \APPR relies on core points to choose the images for the unsafe set. This enables the handling of clusters with arbitrary shapes as the selected images take the shape of the cluster, as explained in Section \ref{sec:step4}.


Regarding \APPR results, we notice three types of improvement explained in the following:

For \GD, the improvement is only +0.79. This is because the accuracy of the original model for this case study subject was already very high (95.95\%), with limited room for improvement.

For \HPD and \FLD, the improvement in accuracy is +26.16 and +34.18, respectively. The original models for these two case study subjects had low accuracy and we therefore expected a large improvement. In fact, \HPD and \FLD represent two cases where retraining was very much needed.

For \CloseDNN, TS and \ODDNN, the improvement in accuracy is +8.2, +11.82 and +13.18, respectively. Though this improvement is moderate compared to the previous case study DNNs, it is still significant given the original models' relatively high accuracy. After retraining, \APPR achieved high accuracy for these case study subjects with 96.29\%, 93.47\%, and 97.30\%, respectively.

We report the significance of these results in Table \ref{vda_results}, including the values of the Vargha and Delaney's $\hat{A}_{12}$ effect size and the p-values resulting from performing a Mann-Whitney U-test between the accuracy of \APPR and other improvement strategies. Recall that we run each strategy 10 times. 


Typically, an $\hat{A}_{12}$ effect size above 0.56 is considered significant with higher thresholds for medium (0.64) and large (0.71) effects, thus suggesting the effect sizes between \APPR and other strategies are large across case study DNNs, in all but one case. 

We notice in Table \ref{vda_results} that the p-values when comparing \APPR to the baselines are always below 0.05. As for HUDD, the p-values are lower than 0.05 in 5 out of 6 case study subjects. This implies that in most cases, the null hypothesis is rejected. 


\subsubsection{RQ3} \emph{\MAJOR{R3.27}{Does \APPR provide time and memory savings compared to HUDD?}}
\label{sec:emp:executionTime}
\begin{table}[t]
\caption{Execution time of the feature extraction of the improvement sets and the memory allocations of these features for \APPR. Compared to the execution time of the generation of the heatmaps and their memory allocation.}
\label{tab:execution_time}
\footnotesize
\makebox[0.5\textwidth]{\begin{tabular}{|c|cc|cc|}
\hline
Subject& \multicolumn{2}{c|}{Execution time (s)} & \multicolumn{2}{c|}{Memory allocation (Mb)} \\ \hline
           & \multicolumn{1}{c|}{SAFE}     & HUDD    & \multicolumn{1}{c|}{SAFE}       & HUDD      \\ \hline
GD         & \multicolumn{1}{c|}{176}      & 3,920   & \multicolumn{1}{c|}{74.2}       & 78,641  \\ \hline
OC         & \multicolumn{1}{c|}{160}      & 958     & \multicolumn{1}{c|}{4.1}        & 3,551    \\ \hline
HPD        & \multicolumn{1}{c|}{161}      & 1,294   & \multicolumn{1}{c|}{1.6}        & 8,839   \\ \hline
FLD        & \multicolumn{1}{c|}{210}      & 1,883   & \multicolumn{1}{c|}{0.82}       & 11,981  \\ \hline
OD         & \multicolumn{1}{c|}{149}      & 1,059   & \multicolumn{1}{c|}{13.2}       & 5,989    \\ \hline
TS         & \multicolumn{1}{c|}{152}      & 2,335   & \multicolumn{1}{c|}{1.6}        & 16,744     \\ \hline
\end{tabular}}
\end{table}

\vspace{1mm}

\CHANGED{\emph{Design and measurements.}}
As mentioned in the previous Section, \APPR proved its effectiveness in retraining the different case study subjects as it obtained significant improvements over the original models. \APPR not only performs better than HUDD but also, and perhaps more significantly, provides very high time and memory savings. This research question investigates whether \APPR, with its black-box nature, offers significant time and memory saving.

We compare the time required by \APPR and HUDD to perform their most expensive tasks, which are the time required to extract the features by \APPR and to generate the heatmaps by HUDD; we do not report the time required to perform the other steps of the two approaches because these steps are either shared or do not have any practical impact on performance (i.e., they took few seconds in our experiments).
We also compare the memory allocation of the features and the heatmaps, which are the data types processed only by \APPR and HUDD, respectively.

\emph{Results.} Table~\ref{tab:execution_time} provide our results.  We observe a large time and memory saving for \APPR compared to HUDD. Such performance considerations have significant practical implications. For example, we can observe that SAFE requires, in the worst case, only 3.5 minutes to generate RCCs and the unsafe set to be used for retraining. HUDD, instead, requires 65 minutes to achieve the same objective. Such difference has a huge impact on the practicality of the approach as, for SAFE, the analysis of RCCs can be performed shortly after observing DNN failures. HUDD may require up to one hour to do the same. 
Such execution time savings allow engineers to conduct DNN safety analysis and improvements in a much shorter time. Memory savings also have significant implications since it prevents the need for expensive hardware to perform such analysis.  

The explanation for the above results is that \APPR is using extracted features to represent the images instead of heatmaps. Feature extraction is less costly in time and memory than the computation of heatmaps. Indeed, the distance computed on features is less computationally complex than the one calculated on heatmaps. 

Heatmaps also take a great deal of memory for storage. HUDD generates heatmaps for each layer. For instance, the heatmap for the eighth layer of AlexNet has a size of $169 \times 256$ (convolution layer), while the heatmap for the tenth layer has a size of $4096 \times C$. With this architecture, HUDD will generate eight heatmaps of size $169 \times 256$ and one heatmap of size $4096 \times C$ for every image in the dataset. In contrast, for \APPR, each image is represented by a $1 \times 256$ matrix (256 features for each image).

\subsection{Threats to validity} 
We discuss internal, conclusion, construct, and external validity according to conventional practices
~\cite{Wohlin2012}.

\subsubsection{Internal validity}

A possible internal threat is the use of the feature extraction method on which we rely, which could negatively affect our results if inadequate. Indeed, clustering relies on the similarity computed in terms of the extracted features. To mitigate this threat, we have checked that some of the features extracted by our method are consistent within clusters, by visually inspecting them. Indeed, such features contain enough information on the images if the clusters are visually consistent, thus demonstrating that the features extraction method worked.

\subsubsection{External validity}
The selection of the case study DNNs could be a threat to validity. This paper alleviates this issue by using six datasets with diverse complexity. Four subject DNNs out of six implement tasks motivated by IEE business needs that address problems that are quite common in the automotive industry. \MAJOR{R1.2}{Also, the simulators used in our experiments, though being related to IEE in-car sensing business cases, vary in terms of characteristics; indeed, they range from high-fidelity simulation of specific body parts (in our case, the human eye) to whole human body simulations (we crop the face) with lower fidelity.}
 \MAJOR{R2.3}{In our experiments we test DNNs that process cropped images (e.g., human's head, traffic signs). Cropping, which a DNN can perform, limits the number of features appearing in the images to be processed, thus potentially simplifying the task to be performed by \APPR. However, cropping was justified in our context by the expected inputs of our subject DNNs. Finally, although we focus on data sets related to in-car sensing, we believe that \APPR will perform well with other data sets since the VGG model used for the feature extraction was pre-trained on Image-Net, which means that the model can capture features related to 1000 classes, including humans, animals, and objects. In the future, we aim to extend our work to include subjects from different domains (e.g., different types of classification tasks with non-cropped images).
 }
 
 \MINOR{R1.1}{Another threat to the generalizability of our results is the dependence on ImageNet. \APPR relies on VGG16 to extract features from images. VGG16 was pre-trained on ImageNet, which is a large image database. Therefore, \APPR is expected to work better with images containing objects recognized by ImageNet. However, we believe that this characteristic does not affect the practical applicability of \APPR in the automotive context since the ImageNet VGG recognizes 1000 different objects, including  objects belonging to the automotive scenery; these objects include cars, faces, and eyes, for example \footnote{The full list of classes is available at \url{https://gist.github.com/yrevar/942d3a0ac09ec9e5eb3a}}.}

 


\subsubsection{Conclusion validity}
To avoid violating parametric assumptions in our statistical analysis, we rely on a non-parametric test and effect size measure (i.e., Mann Whitney U-test, the Vargha and Delaney’s $\hat{A}_{12}$ statistics, and the one-sample Wilcoxon signed rank test, respectively) to evaluate the statistical and practical significance of differences in results. We report both p-values and effect sizes.  

Due to the stochastic nature of \APPR (e.g., DNN retraining), the experiments conducted with the \APPR method were executed over ten runs. We reported the descriptive statistics of those runs and discussed the statistical significance and effect size of differences across methods.

\subsubsection{Construct validity}

The constructs considered in our work are effectiveness and cost. Effectiveness is measured through complementary  indicators, which include
(1) within-cluster variance reduction for at least one parameter (for RQ1.2), (2) 
 average values being close to unsafe values for parameters with high within-cluster variance reduction (for RQ1.3), (3) coverage of plausible causes of errors represented by unsafe values (for RQ1.4), (4) DNN accuracy improvement (for RQ2).
Although the effectiveness regarding the analysis of root causes (i.e., RQ1.2 to RQ1.4) might be evaluated with user studies, such evaluation might be biased by the background and experience of the selected pool of users, which shall also be sufficiently large in number. 
In our context, end-users are engineers with background in safety analysis (e.g., to determine if an input is realistic or if a DNN error may lead to a hazard) and machine learning.
However, since safety experts are generally not trained to use DNNs whereas DNN experts (e.g., recently graduated students) are typically not safety experts, it would be difficult to select a large enough set of users for the study. For this reason, we preferred to rely on reflective indicators based on the information provided by simulators, thus enabling an objective evaluation. Moreover, the quality and usefulness of our results have been confirmed by experts at our industry partner, IEE Sensing, over multiple technical and management meetings. These experts included researchers with a Ph.D. in mathematics and machine learning who develop safety-software components, safety engineers integrating DNN-based components, and chief technology officers. To measure the effectiveness of DNN retraining, we relied on improvements in accuracy, which is common practice.

Concerning cost, we discussed the feasibility of root cause analysis by reporting the number of clusters generated by the approach and the number of images that are sufficient to determine commonalities across images, based on our experience and that of our industry partners. 
We also discussed the cost of DNN retraining by reporting the time required for retraining, which affects the feasibility of the approach in practice, and memory allocation costs, which affect hardware requirements.

\MAJOR{R3.2}{Another threat is with RQ1.3 (Section \ref{sec:evaluation:RQ1.3}) where we systematically select unsafe parameters for each case study to evaluate the clusters. This can prevent us from identifying other reasons for failures if, for any reason, we miss some of these parameters. However, whether this is the case or not, this does not prevent the identification of clusters and missing parameters would then lead to clusters without clear root causes, something we have not observed in our experiments.}


\section{Related Work}
\label{sec:related}

\ASEnnn{Most of the DNN testing and analysis approaches are summarized in recent surveys~\cite{huang2020survey,zhang2019machine}.}
However, no survey on the automated debugging and retraining of DNNs has been proposed to date. 

AUTOTRAINER is a DNN monitoring and auto-repairing system~\cite{zhang2021autotrainer}. It monitors the training status of the model and automatically fixes it (by retraining) once a problem is detected. AUTOTRAINER can efficiently detect and repair five targeted training problems (i.e., vanishing gradient, exploding gradient, dying ReLU, oscillating loss, and slow convergence). Despite its effectiveness in detecting these problems, AUTOTRAINER cannot explain certain misclassifications since it cannot detect a root cause of the error that is not defined a priori.

AI-Lancet optimizes deep learning models by locating the error-inducing (EI) neurons and fixing them using either neuron-flip or neuron-fine-tuning methods \cite{zhao2021ai}. It starts by revealing the EI regions in the input sample and then extracts the EI features activated by the EI regions of the input. Finally, the EI neurons can be located with the guidance of the EI features. Unlike \APPR, AI-Lancet requires the modification and the retraining of the DNN to find the erroneous neurons. Another limitation is that it does not explain the root causes of errors. Instead, it attempts to fix them by fine-tuning or flipping neurons.

MODE \cite{Ma2018} evaluates each layer to identify buggy neurons and further generates fixed-size batches of images for retraining. However, setting such size can be difficult when dealing with datasets with complex features. Also, in the improvement set selection step, MODE requires the modification and the retraining of the DNN. In \APPR, the unsafe set is selected automatically. Another difference between MODE and \APPR is that the former cannot detect the root cause of a DNN error, which is one of the most important features of the latter. \MAJOR{R3.24}{Further, HUDD outperformed MODE based on OD (the only usable dataset to compare since MODE authors did not provide an implementation). SAFE also outperformed both HUDD and MODE based on the OD dataset (97.14\% compared to 97.04\% for HUDD and 89\% for MODE).}

Apricot \cite{zhang2019apricot} is a two-phases approach. The first phase is weight adjustment, where for each failing input $x$, it adjusts the weights of the model by running different DNN's on different subsets of the training and test sets. The second phase is retraining, where the DNN is retrained using the entire training set with the new adjusted weights. In addition to the low accuracy improvement shown by this method (less than 2\%), it also requires the manipulation of the DNN (adjusting the weights). 


Kim et al. \cite{kim2020reducing} use Surprise Adequacy (SA) to guide the selection of newly collected inputs to be added to the base training dataset for the retraining of a DNN-based semantic segmentation module for autonomous driving in the automotive industry. SA measures how \textit{Surprising} an input is to the DNN (i.e., how different this input is from the ones the network has already seen). The main limitation of this method is that it does not explain the DNN failures. 



\MAJOR{R3.8}{RobOT (Robustness-Oriented Testing) iteratively improves the robusteness of a DNN model by generating adversarial inputs that can be used to test the model and retrain it~\cite{wang2021robot}. The test generation is driven by the first-order loss, which measures the loss achieved by the input generated from a given seed. RobOT outperforms related approaches~\cite{DeepGini,Adapt}. Different from \APPR, RobOT aims to improve robustness rather than DNN accuracy; also, it does not include strategies to provide explanations.
}

\MAJOR{R3.8}{Some DNN testing approaches can provide explanations for the input regions in which DNN errors are observed.
For example, Abdessalem et al. \cite{abdessalem2018testing} rely on evolutionary search to drive the generation of test inputs using simulators; to maximize effectiveness, decision trees are used to learn, during search, the portions of the input space that are less safe and, therefore, should be targeted by testing. The decision tree leaves that characterize such unsafe portions are then presented to the end-users. Recent work further demonstrates the effectiveness of decision trees to characterize the input space, based on the results obtained during simulator-based testing~\cite{Haq:2021}.
DeepHyperion~\cite{zohdinasab2021deephyperion} relies on a metaheuristic search algorithm to configure a generative model (e.g., a simulator) to generate test inputs towards specific dimension of the input space (e.g., image orientation); then, it provides to the end-user a set of feature maps that visualize the degree of accuracy obtained for varying values in pairs of dimensions.
The main limitation of these DNN testing approaches is that, different from \APPR, they can provide explanations only for inputs generated with simulators, not for real-world inputs.}

Further, most of the previously mentioned approaches rely on white-box techniques, which means that the modification of the DNN and a specific set of its parameters is required. \APPR overcomes these limitations by proposing a black-box approach based on a pre-trained feature extraction model trained on the ImageNet dataset. The pre-trained model is used as-is to extract the features and then select an unsafe set for retraining. 

Another advantage of \APPR over the other methods is in helping detect error root causes in the DNN, using a clustering algorithm. As discussed, this is required in the context of safety analysis. Further, every cluster representing a root cause can be used to select images to improve the model by retraining it and thus making it more robust to any targeted root cause. 

Both \APPR and HUDD identify different situations in which image-processing DNNs are likely to trigger an erroneous result. However, the former fares better than HUDD with large time and memory savings due to the use of extracted features to represent images instead of heatmaps. Features require less time to extract and less memory to store. 
 
\MINOR{R1.1}{Another state-of-the-art approach for the generation of explanations is the Anchor algorithm, which derives decision rules (called \emph{anchor explanations}) that sufficiently tie a prediction locally \cite{ribeiro2018anchors}. Changes to the rest of the feature values do not matter, i.e., similar instances covered by the same anchor have the same prediction outcome. The Anchor algorithm is applied on tabular images and textual datasets. The Anchor algorithm constructs an explanation rule iteratively by interacting with the model to be explained. Each iteration alters the values associated with one input feature until it identifies a range within which the accuracy is above a given threshold. Though designed for textual datasets, the Anchor algorithm may provide decision rules that can be easily generalized to multiple inputs (e.g., the ones that match the same rule and lead to the same result); for image inputs, Anchor simply emphasizes the image chunk that is sufficient for the classifier to make the prediction.
Therefore, Anchor does not help engineers in analyzing large input datasets because it requires all the chunks belonging to each input image to be visualized. \APPR, instead, enables engineers to efficiently identify root causes from large sets of error-inducing images; further, it automatically retrains the DNN, a task not supported by Anchor.}


\section{Conclusion}
\label{sec:conclusion}

In this paper, we presented \APPR, a new black-box approach that automatically identifies the different situations in which a DNN is more likely to fail, without requiring any modification to the DNN or access to its internal information. Similar to our previous white-box approach (i.e., HUDD), \APPR characterizes such situations by generating clusters of images that likely lead to a DNN error because of the same underlying reason. We refer to these clusters as root cause clusters.
Differently from HUDD,
\APPR is based on a pre-trained model to extract features from error-inducing images. These features replace the heatmaps used by HUDD to generate the root cause clusters. Also, \APPR uses a density-based clustering algorithm to generate arbitrary-shaped clusters. 

\APPR uses a new method to select an unsafe set for retraining based on the cluster's core points. More specifically, \APPR selects images close to each cluster's core points.  These images are manually labeled and used to augment the training set to improve the DNN through retraining.

Empirical results show that \APPR derives root cause clusters that can effectively help engineers determine the root causes for DNN errors. Indeed, the number of generated clusters is low, thus making the visual inspection of a few representative images for each cluster feasible. They include images with similar characteristics that are likely related to the cause of the error. Further, for our case study subjects, the generated clusters capture all the possible causes of errors. 
Compared to HUDD, \APPR generates clusters with more similar characteristics covering a larger set of error causes.
Moreover, the DNNs retrained by \APPR achieve a higher  accuracy than that obtained with HUDD and baseline approaches.

Besides the benefits described above, \APPR also saves large amounts of execution time and memory due to its black-box nature and the reliance on the extracted features instead of heatmaps, thus making it more usable in practical contexts.

\begin{acks}
This project has received funding from IEE Luxembourg, Luxembourg’s National Research Fund (FNR) under grant BRIDGES2020/IS/14711346/FUNTASY, 
and NSERC of Canada under the Discovery and CRC programs. Authors would like to thank Thomas Stifter from IEE for his valuable support. The experiments presented in this paper were carried out using the HPC facilities of the University of Luxembourg (see http://hpc.uni.lu).
\end{acks}


\bibliographystyle{ACM-Reference-Format}
\bibliography{BlackBoxExplanation}


\begin{thebibliography}{86}


\ifx \showCODEN    \undefined \def \showCODEN     #1{\unskip}     \fi
\ifx \showDOI      \undefined \def \showDOI       #1{#1}\fi
\ifx \showISBNx    \undefined \def \showISBNx     #1{\unskip}     \fi
\ifx \showISBNxiii \undefined \def \showISBNxiii  #1{\unskip}     \fi
\ifx \showISSN     \undefined \def \showISSN      #1{\unskip}     \fi
\ifx \showLCCN     \undefined \def \showLCCN      #1{\unskip}     \fi
\ifx \shownote     \undefined \def \shownote      #1{#1}          \fi
\ifx \showarticletitle \undefined \def \showarticletitle #1{#1}   \fi
\ifx \showURL      \undefined \def \showURL       {\relax}        \fi
\providecommand\bibfield[2]{#2}
\providecommand\bibinfo[2]{#2}
\providecommand\natexlab[1]{#1}
\providecommand\showeprint[2][]{arXiv:#2}

\bibitem[\protect\citeauthoryear{Abdessalem, Nejati, Briand, and
  Stifter}{Abdessalem et~al\mbox{.}}{2018}]%
        {abdessalem2018testing}
\bibfield{author}{\bibinfo{person}{Raja~Ben Abdessalem}, \bibinfo{person}{Shiva
  Nejati}, \bibinfo{person}{Lionel~C Briand}, {and} \bibinfo{person}{Thomas
  Stifter}.} \bibinfo{year}{2018}\natexlab{}.
\newblock \showarticletitle{Testing vision-based control systems using
  learnable evolutionary algorithms}. In \bibinfo{booktitle}{\emph{2018
  IEEE/ACM 40th International Conference on Software Engineering (ICSE)}}.
  IEEE, \bibinfo{pages}{1016--1026}.
\newblock


\bibitem[\protect\citeauthoryear{Alaeddini and Dogan}{Alaeddini and
  Dogan}{2011}]%
        {alaeddini2011using}
\bibfield{author}{\bibinfo{person}{Adel Alaeddini} {and}
  \bibinfo{person}{Ibrahim Dogan}.} \bibinfo{year}{2011}\natexlab{}.
\newblock \showarticletitle{Using Bayesian networks for root cause analysis in
  statistical process control}.
\newblock \bibinfo{journal}{\emph{Expert Systems with Applications}}
  \bibinfo{volume}{38}, \bibinfo{number}{9} (\bibinfo{year}{2011}),
  \bibinfo{pages}{11230--11243}.
\newblock


\bibitem[\protect\citeauthoryear{Albawi, Mohammed, and Al-Zawi}{Albawi
  et~al\mbox{.}}{2017}]%
        {albawi2017understanding}
\bibfield{author}{\bibinfo{person}{Saad Albawi}, \bibinfo{person}{Tareq~Abed
  Mohammed}, {and} \bibinfo{person}{Saad Al-Zawi}.}
  \bibinfo{year}{2017}\natexlab{}.
\newblock \showarticletitle{Understanding of a convolutional neural network}.
  In \bibinfo{booktitle}{\emph{2017 international conference on engineering and
  technology (ICET)}}. Ieee, \bibinfo{pages}{1--6}.
\newblock


\bibitem[\protect\citeauthoryear{Alber, Lapuschkin, Seegerer, H{{\"a}}gele,
  Sch{{\"u}}tt, Montavon, Samek, M{{\"u}}ller, D{{\"a}}hne, and
  Kindermans}{Alber et~al\mbox{.}}{2019}]%
        {iNNvestigate}
\bibfield{author}{\bibinfo{person}{Maximilian Alber},
  \bibinfo{person}{Sebastian Lapuschkin}, \bibinfo{person}{Philipp Seegerer},
  \bibinfo{person}{Miriam H{{\"a}}gele}, \bibinfo{person}{Kristof~T.
  Sch{{\"u}}tt}, \bibinfo{person}{Gr{{\'e}}goire Montavon},
  \bibinfo{person}{Wojciech Samek}, \bibinfo{person}{Klaus-Robert
  M{{\"u}}ller}, \bibinfo{person}{Sven D{{\"a}}hne}, {and}
  \bibinfo{person}{Pieter-Jan Kindermans}.} \bibinfo{year}{2019}\natexlab{}.
\newblock \showarticletitle{iNNvestigate Neural Networks!}
\newblock \bibinfo{journal}{\emph{Journal of Machine Learning Research}}
  \bibinfo{volume}{20}, \bibinfo{number}{93} (\bibinfo{year}{2019}),
  \bibinfo{pages}{1--8}.
\newblock
\urldef\tempurl%
\url{http://jmlr.org/papers/v20/18-540.html}
\showURL{%
\tempurl}


\bibitem[\protect\citeauthoryear{{Authors of this paper}}{{Authors of this
  paper}}{2022}]%
        {REPLICABILITY}
\bibfield{author}{\bibinfo{person}{{Authors of this paper}}.}
  \bibinfo{year}{2022}\natexlab{}.
\newblock \bibinfo{title}{{SAFE: toolset and replicability package}}.
\newblock
\newblock
\urldef\tempurl%
\url{https://zenodo.org/record/6619279}
\showURL{%
\tempurl}


\bibitem[\protect\citeauthoryear{Bholowalia and Kumar}{Bholowalia and
  Kumar}{2014}]%
        {bholowalia2014ebk}
\bibfield{author}{\bibinfo{person}{Purnima Bholowalia} {and}
  \bibinfo{person}{Arvind Kumar}.} \bibinfo{year}{2014}\natexlab{}.
\newblock \showarticletitle{EBK-means: A clustering technique based on elbow
  method and k-means in WSN}.
\newblock \bibinfo{journal}{\emph{International Journal of Computer
  Applications}} \bibinfo{volume}{105}, \bibinfo{number}{9}
  (\bibinfo{year}{2014}).
\newblock


\bibitem[\protect\citeauthoryear{Dabkowski and Gal}{Dabkowski and Gal}{2017}]%
        {Dabkowski17}
\bibfield{author}{\bibinfo{person}{Piotr Dabkowski} {and}
  \bibinfo{person}{Yarin Gal}.} \bibinfo{year}{2017}\natexlab{}.
\newblock \showarticletitle{Real Time Image Saliency for Black Box
  Classifiers}. In \bibinfo{booktitle}{\emph{Proceedings of the 31st
  International Conference on Neural Information Processing Systems}}
  \emph{(\bibinfo{series}{NIPS?17})}. \bibinfo{publisher}{Curran Associates
  Inc.}, \bibinfo{address}{Red Hook, NY, USA}, \bibinfo{pages}{6970--6979}.
\newblock
\showISBNx{9781510860964}


\bibitem[\protect\citeauthoryear{Davies and Bouldin}{Davies and
  Bouldin}{1979}]%
        {davies1979cluster}
\bibfield{author}{\bibinfo{person}{David~L Davies} {and}
  \bibinfo{person}{Donald~W Bouldin}.} \bibinfo{year}{1979}\natexlab{}.
\newblock \showarticletitle{A cluster separation measure}.
\newblock \bibinfo{journal}{\emph{IEEE transactions on pattern analysis and
  machine intelligence}} \bibinfo{number}{2} (\bibinfo{year}{1979}),
  \bibinfo{pages}{224--227}.
\newblock


\bibitem[\protect\citeauthoryear{Devlin, Chang, Lee, and Toutanova}{Devlin
  et~al\mbox{.}}{2019}]%
        {devlin2018bert}
\bibfield{author}{\bibinfo{person}{Jacob Devlin}, \bibinfo{person}{Ming-Wei
  Chang}, \bibinfo{person}{Kenton Lee}, {and} \bibinfo{person}{Kristina
  Toutanova}.} \bibinfo{year}{2019}\natexlab{}.
\newblock \showarticletitle{{BERT}: Pre-training of Deep Bidirectional
  Transformers for Language Understanding}.
\newblock  (\bibinfo{date}{June} \bibinfo{year}{2019}),
  \bibinfo{pages}{4171--4186}.
\newblock
\urldef\tempurl%
\url{https://doi.org/10.18653/v1/N19-1423}
\showDOI{\tempurl}


\bibitem[\protect\citeauthoryear{Dif, Attaoui, Elberrichi, Lebbah, and
  Azzag}{Dif et~al\mbox{.}}{2021}]%
        {dif2021transfer}
\bibfield{author}{\bibinfo{person}{Nassima Dif},
  \bibinfo{person}{Mohammed~Oualid Attaoui}, \bibinfo{person}{Zakaria
  Elberrichi}, \bibinfo{person}{Mustapha Lebbah}, {and} \bibinfo{person}{Hanene
  Azzag}.} \bibinfo{year}{2021}\natexlab{}.
\newblock \showarticletitle{Transfer learning from synthetic labels for
  histopathological images classification}.
\newblock \bibinfo{journal}{\emph{Applied Intelligence}}
  (\bibinfo{year}{2021}), \bibinfo{pages}{1--20}.
\newblock


\bibitem[\protect\citeauthoryear{Ester, Kriegel, Sander, and Xu}{Ester
  et~al\mbox{.}}{1996}]%
        {ester1996density}
\bibfield{author}{\bibinfo{person}{Martin Ester}, \bibinfo{person}{Hans-Peter
  Kriegel}, \bibinfo{person}{J\"{o}rg Sander}, {and} \bibinfo{person}{Xiaowei
  Xu}.} \bibinfo{year}{1996}\natexlab{}.
\newblock \showarticletitle{A Density-Based Algorithm for Discovering Clusters
  in Large Spatial Databases with Noise}. In
  \bibinfo{booktitle}{\emph{Proceedings of the Second International Conference
  on Knowledge Discovery and Data Mining}} \emph{(\bibinfo{series}{KDD'96})}.
  \bibinfo{publisher}{AAAI Press}, \bibinfo{pages}{226–231}.
\newblock


\bibitem[\protect\citeauthoryear{Fahmy, Pastore, Bagherzadeh, and Briand}{Fahmy
  et~al\mbox{.}}{2021}]%
        {fahmysupporting}
\bibfield{author}{\bibinfo{person}{Hazem Fahmy}, \bibinfo{person}{Fabrizio
  Pastore}, \bibinfo{person}{Mojtaba Bagherzadeh}, {and}
  \bibinfo{person}{Lionel Briand}.} \bibinfo{year}{2021}\natexlab{}.
\newblock \showarticletitle{Supporting Deep Neural Network Safety Analysis and
  Retraining Through Heatmap-Based Unsupervised Learning}.
\newblock \bibinfo{journal}{\emph{IEEE Transactions on Reliability}}
  (\bibinfo{year}{2021}), \bibinfo{pages}{1--17}.
\newblock
\urldef\tempurl%
\url{https://doi.org/10.1109/TR.2021.3074750}
\showDOI{\tempurl}


\bibitem[\protect\citeauthoryear{Feng, Shi, Gao, Wan, Fang, and Chen}{Feng
  et~al\mbox{.}}{2020}]%
        {DeepGini}
\bibfield{author}{\bibinfo{person}{Yang Feng}, \bibinfo{person}{Qingkai Shi},
  \bibinfo{person}{Xinyu Gao}, \bibinfo{person}{Jun Wan},
  \bibinfo{person}{Chunrong Fang}, {and} \bibinfo{person}{Zhenyu Chen}.}
  \bibinfo{year}{2020}\natexlab{}.
\newblock \showarticletitle{DeepGini: Prioritizing Massive Tests to Enhance the
  Robustness of Deep Neural Networks}. In \bibinfo{booktitle}{\emph{Proceedings
  of the 29th ACM SIGSOFT International Symposium on Software Testing and
  Analysis}} \emph{(\bibinfo{series}{ISSTA 2020})}.
  \bibinfo{publisher}{Association for Computing Machinery},
  \bibinfo{address}{New York, NY, USA}, \bibinfo{pages}{177–188}.
\newblock
\showISBNx{9781450380089}
\urldef\tempurl%
\url{https://doi.org/10.1145/3395363.3397357}
\showDOI{\tempurl}


\bibitem[\protect\citeauthoryear{Fu, Tang, Yu, Li, Sun, and Liu}{Fu
  et~al\mbox{.}}{2021}]%
        {fu2021dvqshare}
\bibfield{author}{\bibinfo{person}{Hao Fu}, \bibinfo{person}{Shanjiang Tang},
  \bibinfo{person}{Ce Yu}, \bibinfo{person}{Yusen Li}, \bibinfo{person}{Jizhou
  Sun}, {and} \bibinfo{person}{Yanjie Liu}.} \bibinfo{year}{2021}\natexlab{}.
\newblock \showarticletitle{DVQShare: An Analytics System for DNN-based Video
  Queries}. In \bibinfo{booktitle}{\emph{2021 IEEE/ACM 21st International
  Symposium on Cluster, Cloud and Internet Computing (CCGrid)}}. IEEE,
  \bibinfo{pages}{166--175}.
\newblock


\bibitem[\protect\citeauthoryear{Garcia, Telea, da~Silva, Torresen, and
  Comba}{Garcia et~al\mbox{.}}{2018}]%
        {GARCIA2018}
\bibfield{author}{\bibinfo{person}{Rafael Garcia},
  \bibinfo{person}{Alexandru~C. Telea}, \bibinfo{person}{Bruno~Castro da
  Silva}, \bibinfo{person}{Jim Torresen}, {and} \bibinfo{person}{Joao Luiz~Dihl
  Comba}.} \bibinfo{year}{2018}\natexlab{}.
\newblock \showarticletitle{A task-and-technique centered survey on visual
  analytics for deep learning model engineering}.
\newblock \bibinfo{journal}{\emph{Computers and Graphics}}
  \bibinfo{volume}{77} (\bibinfo{year}{2018}), \bibinfo{pages}{30 -- 49}.
\newblock
\showISSN{0097-8493}
\urldef\tempurl%
\url{https://doi.org/10.1016/j.cag.2018.09.018}
\showDOI{\tempurl}


\bibitem[\protect\citeauthoryear{Gilpin, Bau, Yuan, Bajwa, Specter, and
  Kagal}{Gilpin et~al\mbox{.}}{2018}]%
        {gilpin2018explaining}
\bibfield{author}{\bibinfo{person}{Leilani~H Gilpin}, \bibinfo{person}{David
  Bau}, \bibinfo{person}{Ben~Z Yuan}, \bibinfo{person}{Ayesha Bajwa},
  \bibinfo{person}{Michael Specter}, {and} \bibinfo{person}{Lalana Kagal}.}
  \bibinfo{year}{2018}\natexlab{}.
\newblock \showarticletitle{Explaining explanations: An overview of
  interpretability of machine learning}. In \bibinfo{booktitle}{\emph{2018 IEEE
  5th International Conference on data science and advanced analytics (DSAA)}}.
  IEEE, \bibinfo{pages}{80--89}.
\newblock


\bibitem[\protect\citeauthoryear{G{\'o}mez-Andrades, Munoz, Serrano, and
  Barco}{G{\'o}mez-Andrades et~al\mbox{.}}{2015}]%
        {gomez2015automatic}
\bibfield{author}{\bibinfo{person}{Ana G{\'o}mez-Andrades},
  \bibinfo{person}{Pablo Munoz}, \bibinfo{person}{Inmaculada Serrano}, {and}
  \bibinfo{person}{Raquel Barco}.} \bibinfo{year}{2015}\natexlab{}.
\newblock \showarticletitle{Automatic root cause analysis for LTE networks
  based on unsupervised techniques}.
\newblock \bibinfo{journal}{\emph{IEEE Transactions on Vehicular Technology}}
  \bibinfo{volume}{65}, \bibinfo{number}{4} (\bibinfo{year}{2015}),
  \bibinfo{pages}{2369--2386}.
\newblock


\bibitem[\protect\citeauthoryear{Gorban and Zinovyev}{Gorban and
  Zinovyev}{2010}]%
        {gorban2010principal}
\bibfield{author}{\bibinfo{person}{Alexander~N Gorban} {and}
  \bibinfo{person}{Andrei~Y Zinovyev}.} \bibinfo{year}{2010}\natexlab{}.
\newblock \showarticletitle{Principal graphs and manifolds}.
\newblock In \bibinfo{booktitle}{\emph{Handbook of research on machine learning
  applications and trends: algorithms, methods, and techniques}}.
  \bibinfo{publisher}{IGI Global}, \bibinfo{pages}{28--59}.
\newblock


\bibitem[\protect\citeauthoryear{Gosztolya, Busa-Fekete, Gr{\'o}sz, and
  T{\'o}th}{Gosztolya et~al\mbox{.}}{2017}]%
        {gosztolya2017dnn}
\bibfield{author}{\bibinfo{person}{G{\'a}bor Gosztolya},
  \bibinfo{person}{R{\'o}bert Busa-Fekete}, \bibinfo{person}{Tam{\'a}s
  Gr{\'o}sz}, {and} \bibinfo{person}{L{\'a}szl{\'o} T{\'o}th}.}
  \bibinfo{year}{2017}\natexlab{}.
\newblock \showarticletitle{{DNN-Based Feature Extraction and Classifier
  Combination for Child-Directed Speech, Cold and Snoring Identification}}. In
  \bibinfo{booktitle}{\emph{Proceeding of Interspeech 2017}}.
  \bibinfo{publisher}{International Speech Communication Association (ISCA)},
  \bibinfo{pages}{3522--3526}.
\newblock
\urldef\tempurl%
\url{https://doi.org/10.21437/Interspeech.2017-905}
\showDOI{\tempurl}


\bibitem[\protect\citeauthoryear{Haq, Shin, Briand, Stifter, and Wang}{Haq
  et~al\mbox{.}}{2021}]%
        {Haq:2021}
\bibfield{author}{\bibinfo{person}{Fitash~Ul Haq}, \bibinfo{person}{Donghwan
  Shin}, \bibinfo{person}{Lionel~C. Briand}, \bibinfo{person}{Thomas Stifter},
  {and} \bibinfo{person}{Jun Wang}.} \bibinfo{year}{2021}\natexlab{}.
\newblock \showarticletitle{Automatic Test Suite Generation for Key-Points
  Detection DNNs Using Many-Objective Search (Experience Paper)}. In
  \bibinfo{booktitle}{\emph{Proceedings of the 30th ACM SIGSOFT International
  Symposium on Software Testing and Analysis}} \emph{(\bibinfo{series}{ISSTA
  2021})}. \bibinfo{publisher}{Association for Computing Machinery},
  \bibinfo{address}{New York, NY, USA}, \bibinfo{pages}{91–102}.
\newblock
\showISBNx{9781450384599}
\urldef\tempurl%
\url{https://doi.org/10.1145/3460319.3464802}
\showDOI{\tempurl}


\bibitem[\protect\citeauthoryear{He, Zhang, Ren, and Sun}{He
  et~al\mbox{.}}{2016b}]%
        {he2016deep}
\bibfield{author}{\bibinfo{person}{Kaiming He}, \bibinfo{person}{Xiangyu
  Zhang}, \bibinfo{person}{Shaoqing Ren}, {and} \bibinfo{person}{Jian Sun}.}
  \bibinfo{year}{2016}\natexlab{b}.
\newblock \showarticletitle{Deep residual learning for image recognition}. In
  \bibinfo{booktitle}{\emph{Proceedings of the IEEE conference on computer
  vision and pattern recognition}}. \bibinfo{pages}{770--778}.
\newblock


\bibitem[\protect\citeauthoryear{He, He, and Wei}{He et~al\mbox{.}}{2016a}]%
        {he2016big}
\bibfield{author}{\bibinfo{person}{Zhenzhen He}, \bibinfo{person}{Yihai He},
  {and} \bibinfo{person}{Yi Wei}.} \bibinfo{year}{2016}\natexlab{a}.
\newblock \showarticletitle{Big data oriented root cause identification
  approach based on PCA and SVM for product infant failure}. In
  \bibinfo{booktitle}{\emph{2016 Prognostics and System Health Management
  Conference (PHM-Chengdu)}}. IEEE, \bibinfo{pages}{1--5}.
\newblock


\bibitem[\protect\citeauthoryear{Howard, Zhu, Chen, Kalenichenko, Wang, Weyand,
  Andreetto, and Adam}{Howard et~al\mbox{.}}{2017}]%
        {howard2017mobilenets}
\bibfield{author}{\bibinfo{person}{Andrew~G Howard}, \bibinfo{person}{Menglong
  Zhu}, \bibinfo{person}{Bo Chen}, \bibinfo{person}{Dmitry Kalenichenko},
  \bibinfo{person}{Weijun Wang}, \bibinfo{person}{Tobias Weyand},
  \bibinfo{person}{Marco Andreetto}, {and} \bibinfo{person}{Hartwig Adam}.}
  \bibinfo{year}{2017}\natexlab{}.
\newblock \showarticletitle{MobileNets: Efficient Convolutional Neural Networks
  for Mobile Vision Applications}.
\newblock \bibinfo{journal}{\emph{CoRR}}  \bibinfo{volume}{abs/1704.04861}
  (\bibinfo{year}{2017}).
\newblock
\showeprint{1704.04861}


\bibitem[\protect\citeauthoryear{Huang, Kroening, Ruan, Sharp, Sun, Thamo, Wu,
  and Yi}{Huang et~al\mbox{.}}{2020}]%
        {huang2020survey}
\bibfield{author}{\bibinfo{person}{Xiaowei Huang}, \bibinfo{person}{Daniel
  Kroening}, \bibinfo{person}{Wenjie Ruan}, \bibinfo{person}{James Sharp},
  \bibinfo{person}{Youcheng Sun}, \bibinfo{person}{Emese Thamo},
  \bibinfo{person}{Min Wu}, {and} \bibinfo{person}{Xinping Yi}.}
  \bibinfo{year}{2020}\natexlab{}.
\newblock \showarticletitle{A survey of safety and trustworthiness of deep
  neural networks: Verification, testing, adversarial attack and defence, and
  interpretability}.
\newblock \bibinfo{journal}{\emph{Computer Science Review}}
  \bibinfo{volume}{37} (\bibinfo{year}{2020}), \bibinfo{pages}{100270}.
\newblock


\bibitem[\protect\citeauthoryear{Hubert and Arabie}{Hubert and Arabie}{1985}]%
        {hubert1985comparing}
\bibfield{author}{\bibinfo{person}{Lawrence Hubert} {and}
  \bibinfo{person}{Phipps Arabie}.} \bibinfo{year}{1985}\natexlab{}.
\newblock \showarticletitle{Comparing partitions}.
\newblock \bibinfo{journal}{\emph{Journal of classification}}
  \bibinfo{volume}{2}, \bibinfo{number}{1} (\bibinfo{year}{1985}),
  \bibinfo{pages}{193--218}.
\newblock


\bibitem[\protect\citeauthoryear{Humbatova, Jahangirova, Bavota, Riccio,
  Stocco, and Tonella}{Humbatova et~al\mbox{.}}{2020}]%
        {humbatova:2019}
\bibfield{author}{\bibinfo{person}{Nargiz Humbatova}, \bibinfo{person}{Gunel
  Jahangirova}, \bibinfo{person}{Gabriele Bavota}, \bibinfo{person}{Vincenzo
  Riccio}, \bibinfo{person}{Andrea Stocco}, {and} \bibinfo{person}{Paolo
  Tonella}.} \bibinfo{year}{2020}\natexlab{}.
\newblock \showarticletitle{Taxonomy of Real Faults in Deep Learning Systems}.
  In \bibinfo{booktitle}{\emph{Proceedings of the 42nd International Conference
  on Software Engineering}}. \bibinfo{publisher}{Association for Computing
  Machinery}, \bibinfo{address}{New York, NY, USA}, 10.
\newblock


\bibitem[\protect\citeauthoryear{{IEE}}{{IEE}}{2020}]%
        {IEE}
\bibfield{author}{\bibinfo{person}{{IEE}}.} \bibinfo{year}{2020}\natexlab{}.
\newblock \bibinfo{title}{{IEE} Sensing solutions. www.iee.lu}.
\newblock
\newblock


\bibitem[\protect\citeauthoryear{{INI}}{{INI}}{2020}]%
        {TRAFFICdataset}
\bibfield{author}{\bibinfo{person}{{INI}}.} \bibinfo{year}{2020}\natexlab{}.
\newblock \bibinfo{title}{{TRaffic Sign Dataset}}.
\newblock
\newblock
\urldef\tempurl%
\url{http://benchmark.ini.rub.de/?section=gtsrb&subsection=dataset}
\showURL{%
\tempurl}


\bibitem[\protect\citeauthoryear{{International Organization for
  Standardization}}{{International Organization for Standardization}}{2020a}]%
        {ISO24765}
\bibfield{author}{\bibinfo{person}{{International Organization for
  Standardization}}.} \bibinfo{year}{2020}\natexlab{a}.
\newblock \bibinfo{title}{{ISO, ISO-24765-2017, Systems and software
  engineering - Vocabulary}}.
\newblock
\newblock


\bibitem[\protect\citeauthoryear{{International Organization for
  Standardization}}{{International Organization for Standardization}}{2020b}]%
        {ISO26262}
\bibfield{author}{\bibinfo{person}{{International Organization for
  Standardization}}.} \bibinfo{year}{2020}\natexlab{b}.
\newblock \bibinfo{title}{{ISO, ISO26262-1:2018, Road vehicles: Functional
  safety}}.
\newblock
\newblock


\bibitem[\protect\citeauthoryear{Jeyakumar, Noor, Cheng, Garcia, and
  Srivastava}{Jeyakumar et~al\mbox{.}}{2020}]%
        {jeyakumar2020can}
\bibfield{author}{\bibinfo{person}{Jeya~Vikranth Jeyakumar},
  \bibinfo{person}{Joseph Noor}, \bibinfo{person}{Yu-Hsi Cheng},
  \bibinfo{person}{Luis Garcia}, {and} \bibinfo{person}{Mani Srivastava}.}
  \bibinfo{year}{2020}\natexlab{}.
\newblock \showarticletitle{{How Can I Explain This to You? An Empirical Study
  of Deep Neural Network Explanation Methods}}.
\newblock   \bibinfo{volume}{33} (\bibinfo{year}{2020}),
  \bibinfo{pages}{4211--4222}.
\newblock


\bibitem[\protect\citeauthoryear{Kabir, Pamir, Ullah, Munawar, Asif, and
  Javaid}{Kabir et~al\mbox{.}}{2021}]%
        {kabir2021detection}
\bibfield{author}{\bibinfo{person}{Benish Kabir}, \bibinfo{person}{Pamir},
  \bibinfo{person}{Ashraf Ullah}, \bibinfo{person}{Shoaib Munawar},
  \bibinfo{person}{Muhammad Asif}, {and} \bibinfo{person}{Nadeem Javaid}.}
  \bibinfo{year}{2021}\natexlab{}.
\newblock \showarticletitle{Detection of Non-Technical Losses Using MLP-GRU
  Based Neural Network to Secure Smart Grids}. In
  \bibinfo{booktitle}{\emph{Complex, Intelligent and Software Intensive
  Systems}}, \bibfield{editor}{\bibinfo{person}{Leonard Barolli},
  \bibinfo{person}{Kangbin Yim}, {and} \bibinfo{person}{Tomoya Enokido}}
  (Eds.). \bibinfo{publisher}{Springer International Publishing},
  \bibinfo{address}{Cham}, \bibinfo{pages}{383--394}.
\newblock
\showISBNx{978-3-030-79725-6}


\bibitem[\protect\citeauthoryear{Kim, Ju, Feldt, and Yoo}{Kim
  et~al\mbox{.}}{2020}]%
        {kim2020reducing}
\bibfield{author}{\bibinfo{person}{Jinhan Kim}, \bibinfo{person}{Jeongil Ju},
  \bibinfo{person}{Robert Feldt}, {and} \bibinfo{person}{Shin Yoo}.}
  \bibinfo{year}{2020}\natexlab{}.
\newblock \showarticletitle{Reducing dnn labelling cost using surprise
  adequacy: An industrial case study for autonomous driving}. In
  \bibinfo{booktitle}{\emph{Proceedings of the 28th ACM Joint Meeting on
  European Software Engineering Conference and Symposium on the Foundations of
  Software Engineering}}. \bibinfo{pages}{1466--1476}.
\newblock


\bibitem[\protect\citeauthoryear{Kriegel, Kr{\"o}ger, Sander, and
  Zimek}{Kriegel et~al\mbox{.}}{2011}]%
        {kriegel2011density}
\bibfield{author}{\bibinfo{person}{Hans-Peter Kriegel}, \bibinfo{person}{Peer
  Kr{\"o}ger}, \bibinfo{person}{J{\"o}rg Sander}, {and} \bibinfo{person}{Arthur
  Zimek}.} \bibinfo{year}{2011}\natexlab{}.
\newblock \showarticletitle{Density-based clustering}.
\newblock \bibinfo{journal}{\emph{Wiley Interdisciplinary Reviews: Data Mining
  and Knowledge Discovery}} \bibinfo{volume}{1}, \bibinfo{number}{3}
  (\bibinfo{year}{2011}), \bibinfo{pages}{231--240}.
\newblock


\bibitem[\protect\citeauthoryear{Krizhevsky and Hinton}{Krizhevsky and
  Hinton}{2009}]%
        {cifar}
\bibfield{author}{\bibinfo{person}{A. Krizhevsky} {and} \bibinfo{person}{G.
  Hinton}.} \bibinfo{year}{2009}\natexlab{}.
\newblock \bibinfo{booktitle}{\emph{Learning multiple layers of features from
  tiny images}}.
\newblock \bibinfo{type}{{T}echnical {R}eport}.
  \bibinfo{institution}{Department of Computer Science, University of Toronto}.
\newblock


\bibitem[\protect\citeauthoryear{Krizhevsky, Sutskever, and Hinton}{Krizhevsky
  et~al\mbox{.}}{2017}]%
        {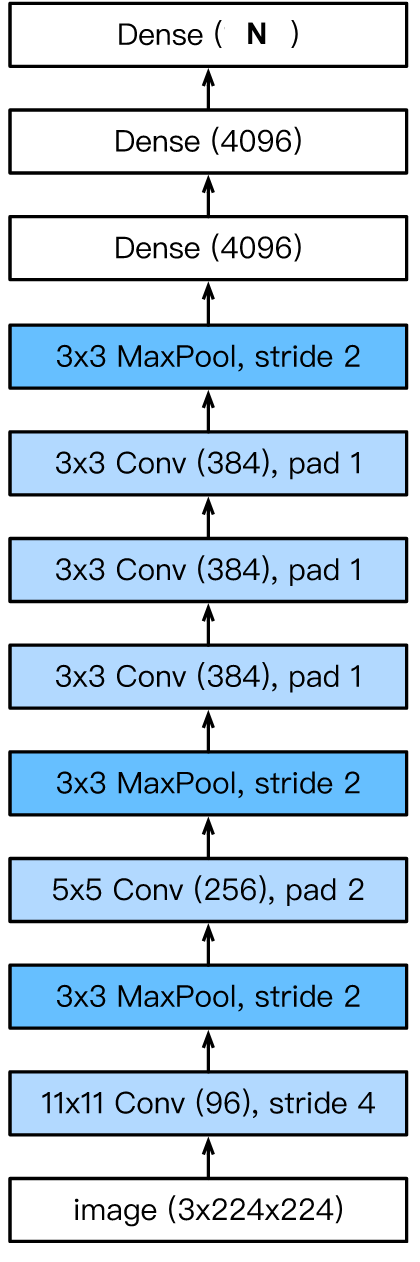}
\bibfield{author}{\bibinfo{person}{Alex Krizhevsky}, \bibinfo{person}{Ilya
  Sutskever}, {and} \bibinfo{person}{Geoffrey~E. Hinton}.}
  \bibinfo{year}{2017}\natexlab{}.
\newblock \showarticletitle{ImageNet Classification with Deep Convolutional
  Neural Networks}.
\newblock \bibinfo{journal}{\emph{Commun. ACM}} \bibinfo{volume}{60},
  \bibinfo{number}{6} (\bibinfo{date}{May} \bibinfo{year}{2017}),
  \bibinfo{pages}{84--90}.
\newblock
\showISSN{0001-0782}
\urldef\tempurl%
\url{https://doi.org/10.1145/3065386}
\showDOI{\tempurl}


\bibitem[\protect\citeauthoryear{Lee, Cha, Lee, and Oh}{Lee
  et~al\mbox{.}}{2020}]%
        {Adapt}
\bibfield{author}{\bibinfo{person}{Seokhyun Lee}, \bibinfo{person}{Sooyoung
  Cha}, \bibinfo{person}{Dain Lee}, {and} \bibinfo{person}{Hakjoo Oh}.}
  \bibinfo{year}{2020}\natexlab{}.
\newblock \showarticletitle{Effective White-Box Testing of Deep Neural Networks
  with Adaptive Neuron-Selection Strategy}. In
  \bibinfo{booktitle}{\emph{Proceedings of the 29th ACM SIGSOFT International
  Symposium on Software Testing and Analysis}} \emph{(\bibinfo{series}{ISSTA
  2020})}. \bibinfo{publisher}{Association for Computing Machinery},
  \bibinfo{address}{New York, NY, USA}, \bibinfo{pages}{165–176}.
\newblock
\showISBNx{9781450380089}
\urldef\tempurl%
\url{https://doi.org/10.1145/3395363.3397346}
\showDOI{\tempurl}


\bibitem[\protect\citeauthoryear{Lei, Jia, Lin, Xing, and Ding}{Lei
  et~al\mbox{.}}{2016}]%
        {lei2016intelligent}
\bibfield{author}{\bibinfo{person}{Yaguo Lei}, \bibinfo{person}{Feng Jia},
  \bibinfo{person}{Jing Lin}, \bibinfo{person}{Saibo Xing}, {and}
  \bibinfo{person}{Steven~X Ding}.} \bibinfo{year}{2016}\natexlab{}.
\newblock \showarticletitle{An intelligent fault diagnosis method using
  unsupervised feature learning towards mechanical big data}.
\newblock \bibinfo{journal}{\emph{IEEE Transactions on Industrial Electronics}}
  \bibinfo{volume}{63}, \bibinfo{number}{5} (\bibinfo{year}{2016}),
  \bibinfo{pages}{3137--3147}.
\newblock


\bibitem[\protect\citeauthoryear{Li, Pan, Zhang, and Li}{Li
  et~al\mbox{.}}{2021}]%
        {li2021testing}
\bibfield{author}{\bibinfo{person}{Zhong Li}, \bibinfo{person}{Minxue Pan},
  \bibinfo{person}{Tian Zhang}, {and} \bibinfo{person}{Xuandong Li}.}
  \bibinfo{year}{2021}\natexlab{}.
\newblock \showarticletitle{Testing DNN-based Autonomous Driving Systems under
  Critical Environmental Conditions}. In
  \bibinfo{booktitle}{\emph{International Conference on Machine Learning}}.
  PMLR, \bibinfo{pages}{6471--6482}.
\newblock


\bibitem[\protect\citeauthoryear{Lin, Maire, Belongie, Bourdev, Girshick, Hays,
  Perona, Ramanan, Doll{\'{a}}r, and Zitnick}{Lin et~al\mbox{.}}{2014}]%
        {LinMBHPRDZ14}
\bibfield{author}{\bibinfo{person}{Tsung{-}Yi Lin}, \bibinfo{person}{Michael
  Maire}, \bibinfo{person}{Serge~J. Belongie}, \bibinfo{person}{Lubomir~D.
  Bourdev}, \bibinfo{person}{Ross~B. Girshick}, \bibinfo{person}{James Hays},
  \bibinfo{person}{Pietro Perona}, \bibinfo{person}{Deva Ramanan},
  \bibinfo{person}{Piotr Doll{\'{a}}r}, {and} \bibinfo{person}{C.~Lawrence
  Zitnick}.} \bibinfo{year}{2014}\natexlab{}.
\newblock \showarticletitle{Microsoft {COCO:} Common Objects in Context}. In
  \bibinfo{booktitle}{\emph{Computer Vision -- ECCV 2014}}.
  \bibinfo{publisher}{Springer International Publishing},
  \bibinfo{address}{Cham}, \bibinfo{pages}{740--755}.
\newblock
\showISBNx{978-3-319-10602-1}


\bibitem[\protect\citeauthoryear{Linaker, Sulaman, H{\"o}st, and
  de~Mello}{Linaker et~al\mbox{.}}{2015}]%
        {linaker2015guidelines}
\bibfield{author}{\bibinfo{person}{Johan Linaker},
  \bibinfo{person}{Sardar~Muhammad Sulaman}, \bibinfo{person}{Martin H{\"o}st},
  {and} \bibinfo{person}{Rafael~Maiani de Mello}.}
  \bibinfo{year}{2015}\natexlab{}.
\newblock \showarticletitle{Guidelines for conducting surveys in software
  engineering v. 1.1}.
\newblock \bibinfo{journal}{\emph{Lund University}} (\bibinfo{year}{2015}).
\newblock


\bibitem[\protect\citeauthoryear{{Liu}, {Luo}, {Wang}, and {Tang}}{{Liu}
  et~al\mbox{.}}{2015}]%
        {Liu:15}
\bibfield{author}{\bibinfo{person}{Z. {Liu}}, \bibinfo{person}{P. {Luo}},
  \bibinfo{person}{X. {Wang}}, {and} \bibinfo{person}{X. {Tang}}.}
  \bibinfo{year}{2015}\natexlab{}.
\newblock \showarticletitle{Deep Learning Face Attributes in the Wild}. In
  \bibinfo{booktitle}{\emph{2015 IEEE International Conference on Computer
  Vision (ICCV)}}. \bibinfo{pages}{3730--3738}.
\newblock


\bibitem[\protect\citeauthoryear{Loey, Manogaran, Taha, and Khalifa}{Loey
  et~al\mbox{.}}{2021}]%
        {loey2021hybrid}
\bibfield{author}{\bibinfo{person}{Mohamed Loey}, \bibinfo{person}{Gunasekaran
  Manogaran}, \bibinfo{person}{Mohamed Hamed~N Taha}, {and}
  \bibinfo{person}{Nour Eldeen~M Khalifa}.} \bibinfo{year}{2021}\natexlab{}.
\newblock \showarticletitle{A hybrid deep transfer learning model with machine
  learning methods for face mask detection in the era of the COVID-19
  pandemic}.
\newblock \bibinfo{journal}{\emph{Measurement}}  \bibinfo{volume}{167}
  (\bibinfo{year}{2021}), \bibinfo{pages}{108288}.
\newblock


\bibitem[\protect\citeauthoryear{Ma, Liu, Lee, Zhang, and Grama}{Ma
  et~al\mbox{.}}{2018}]%
        {Ma2018}
\bibfield{author}{\bibinfo{person}{Shiqing Ma}, \bibinfo{person}{Yingqi Liu},
  \bibinfo{person}{Wen-Chuan Lee}, \bibinfo{person}{Xiangyu Zhang}, {and}
  \bibinfo{person}{Ananth Grama}.} \bibinfo{year}{2018}\natexlab{}.
\newblock \showarticletitle{MODE: Automated Neural Network Model Debugging via
  State Differential Analysis and Input Selection}. In
  \bibinfo{booktitle}{\emph{Proceedings of the 2018 26th ACM Joint Meeting on
  European Software Engineering Conference and Symposium on the Foundations of
  Software Engineering}} \emph{(\bibinfo{series}{ESEC/FSE 2018})}.
  \bibinfo{publisher}{ACM}, \bibinfo{address}{New York, NY, USA},
  \bibinfo{pages}{175--186}.
\newblock
\showISBNx{978-1-4503-5573-5}
\urldef\tempurl%
\url{https://doi.org/10.1145/3236024.3236082}
\showDOI{\tempurl}


\bibitem[\protect\citeauthoryear{Marc, Levoy, Szymon, Tim, Hanspeter, Nina,
  Jianhua, Lo{\"\i}c, Matthias, Leif, et~al\mbox{.}}{Marc
  et~al\mbox{.}}{2007}]%
        {marc20072}
\bibfield{author}{\bibinfo{person}{Levoy Marc}, \bibinfo{person}{Marc Levoy},
  \bibinfo{person}{Rusinkiewicz Szymon}, \bibinfo{person}{Weyrich Tim},
  \bibinfo{person}{Pfister Hanspeter}, \bibinfo{person}{Amenta Nina},
  \bibinfo{person}{Wu Jianhua}, \bibinfo{person}{Barthe Lo{\"\i}c},
  \bibinfo{person}{Zwicker Matthias}, \bibinfo{person}{Kobbelt Leif},
  {et~al\mbox{.}}} \bibinfo{year}{2007}\natexlab{}.
\newblock \showarticletitle{2-THE EARLY HISTORY OF POINT-BASED GRAPHICS}.
\newblock In \bibinfo{booktitle}{\emph{Point-Based Graphics}}.
  \bibinfo{publisher}{Elsevier}, \bibinfo{pages}{8--16}.
\newblock


\bibitem[\protect\citeauthoryear{McInnes, Healy, and Astels}{McInnes
  et~al\mbox{.}}{2017}]%
        {mcinnes2017hdbscan}
\bibfield{author}{\bibinfo{person}{Leland McInnes}, \bibinfo{person}{John
  Healy}, {and} \bibinfo{person}{Steve Astels}.}
  \bibinfo{year}{2017}\natexlab{}.
\newblock \showarticletitle{hdbscan: Hierarchical density based clustering}.
\newblock \bibinfo{journal}{\emph{Journal of Open Source Software}}
  \bibinfo{volume}{2}, \bibinfo{number}{11} (\bibinfo{year}{2017}),
  \bibinfo{pages}{205}.
\newblock


\bibitem[\protect\citeauthoryear{McInnes, Healy, and Melville}{McInnes
  et~al\mbox{.}}{2020}]%
        {mcinnes2020umap}
\bibfield{author}{\bibinfo{person}{Leland McInnes}, \bibinfo{person}{John
  Healy}, {and} \bibinfo{person}{James Melville}.}
  \bibinfo{year}{2020}\natexlab{}.
\newblock \showarticletitle{UMAP: uniform manifold approximation and projection
  for dimension reduction}.
\newblock  (\bibinfo{year}{2020}).
\newblock


\bibitem[\protect\citeauthoryear{Montavon, Binder, Lapuschkin, Samek, and
  M{\"u}ller}{Montavon et~al\mbox{.}}{2019}]%
        {Montavon2019}
\bibfield{author}{\bibinfo{person}{Gr{\'e}goire Montavon},
  \bibinfo{person}{Alexander Binder}, \bibinfo{person}{Sebastian Lapuschkin},
  \bibinfo{person}{Wojciech Samek}, {and} \bibinfo{person}{Klaus~Robert
  M{\"u}ller}.} \bibinfo{year}{2019}\natexlab{}.
\newblock \bibinfo{booktitle}{\emph{Layer-Wise Relevance Propagation: An
  Overview}}.
\newblock \bibinfo{publisher}{Springer International Publishing},
  \bibinfo{address}{Cham}, \bibinfo{pages}{193--209}.
\newblock
\showISBNx{978-3-030-28954-6}
\urldef\tempurl%
\url{https://doi.org/10.1007/978-3-030-28954-6_10}
\showDOI{\tempurl}


\bibitem[\protect\citeauthoryear{Mukherjee, Li, Chen, Chu, and Wang}{Mukherjee
  et~al\mbox{.}}{2018}]%
        {mukherjee2018neuraldrop}
\bibfield{author}{\bibinfo{person}{Rajaditya Mukherjee},
  \bibinfo{person}{Qingyang Li}, \bibinfo{person}{Zhili Chen},
  \bibinfo{person}{Shicheng Chu}, {and} \bibinfo{person}{Huamin Wang}.}
  \bibinfo{year}{2018}\natexlab{}.
\newblock \showarticletitle{Neuraldrop: Dnn-based simulation of small-scale
  liquid flows on solids}.
\newblock \bibinfo{journal}{\emph{arXiv preprint arXiv:1811.02517}}
  (\bibinfo{year}{2018}).
\newblock


\bibitem[\protect\citeauthoryear{Newell, Yang, and Deng}{Newell
  et~al\mbox{.}}{2016}]%
        {Hourglass}
\bibfield{author}{\bibinfo{person}{Alejandro Newell}, \bibinfo{person}{Kaiyu
  Yang}, {and} \bibinfo{person}{Jia Deng}.} \bibinfo{year}{2016}\natexlab{}.
\newblock \showarticletitle{Stacked Hourglass Networks for Human Pose
  Estimation}. In \bibinfo{booktitle}{\emph{Computer Vision -- ECCV 2016}},
  \bibfield{editor}{\bibinfo{person}{Bastian Leibe}, \bibinfo{person}{Jiri
  Matas}, \bibinfo{person}{Nicu Sebe}, {and} \bibinfo{person}{Max Welling}}
  (Eds.). \bibinfo{publisher}{Springer International Publishing},
  \bibinfo{address}{Cham}, \bibinfo{pages}{483--499}.
\newblock


\bibitem[\protect\citeauthoryear{Pan, Zhang, Li, Chakrabarty, and Gu}{Pan
  et~al\mbox{.}}{2021}]%
        {pan2021unsupervised}
\bibfield{author}{\bibinfo{person}{Renjian Pan}, \bibinfo{person}{Zhaobo
  Zhang}, \bibinfo{person}{Xin Li}, \bibinfo{person}{Krishnendu Chakrabarty},
  {and} \bibinfo{person}{Xinli Gu}.} \bibinfo{year}{2021}\natexlab{}.
\newblock \showarticletitle{Unsupervised Two-Stage Root-Cause Analysis for
  Integrated Systems}.
\newblock \bibinfo{journal}{\emph{IEEE Transactions on Computer-Aided Design of
  Integrated Circuits and Systems}} (\bibinfo{year}{2021}).
\newblock


\bibitem[\protect\citeauthoryear{Pearson}{Pearson}{1901}]%
        {pearson1901liii}
\bibfield{author}{\bibinfo{person}{Karl Pearson}.}
  \bibinfo{year}{1901}\natexlab{}.
\newblock \showarticletitle{LIII. On lines and planes of closest fit to systems
  of points in space}.
\newblock \bibinfo{journal}{\emph{The London, Edinburgh, and Dublin
  philosophical magazine and journal of science}} \bibinfo{volume}{2},
  \bibinfo{number}{11} (\bibinfo{year}{1901}), \bibinfo{pages}{559--572}.
\newblock


\bibitem[\protect\citeauthoryear{Petsiuk, Das, and Saenko}{Petsiuk
  et~al\mbox{.}}{2018}]%
        {Petsiuk2018rise}
\bibfield{author}{\bibinfo{person}{Vitali Petsiuk}, \bibinfo{person}{Abir Das},
  {and} \bibinfo{person}{Kate Saenko}.} \bibinfo{year}{2018}\natexlab{}.
\newblock \showarticletitle{RISE: Randomized Input Sampling for Explanation of
  Black-box Models}. In \bibinfo{booktitle}{\emph{Proceedings of the British
  Machine Vision Conference (BMVC)}}.
\newblock


\bibitem[\protect\citeauthoryear{{PyTorch}}{{PyTorch}}{2020}]%
        {PyTorch}
\bibfield{author}{\bibinfo{person}{{PyTorch}}.}
  \bibinfo{year}{2020}\natexlab{}.
\newblock \bibinfo{title}{{PyTorch DNN framework}}.
\newblock
\newblock
\urldef\tempurl%
\url{https://pytorch.org}
\showURL{%
\tempurl}


\bibitem[\protect\citeauthoryear{Rahmah and Sitanggang}{Rahmah and
  Sitanggang}{2016}]%
        {rahmah2016determination}
\bibfield{author}{\bibinfo{person}{Nadia Rahmah} {and}
  \bibinfo{person}{Imas~Sukaesih Sitanggang}.} \bibinfo{year}{2016}\natexlab{}.
\newblock \showarticletitle{Determination of optimal epsilon (eps) value on
  dbscan algorithm to clustering data on peatland hotspots in sumatra}. In
  \bibinfo{booktitle}{\emph{IOP conference series: earth and environmental
  science}}, Vol.~\bibinfo{volume}{31}. IOP Publishing,
  \bibinfo{pages}{012012}.
\newblock


\bibitem[\protect\citeauthoryear{Ribeiro, Singh, and Guestrin}{Ribeiro
  et~al\mbox{.}}{2018}]%
        {ribeiro2018anchors}
\bibfield{author}{\bibinfo{person}{Marco~Tulio Ribeiro},
  \bibinfo{person}{Sameer Singh}, {and} \bibinfo{person}{Carlos Guestrin}.}
  \bibinfo{year}{2018}\natexlab{}.
\newblock \showarticletitle{Anchors: High-precision model-agnostic
  explanations}. In \bibinfo{booktitle}{\emph{Proceedings of the AAAI
  conference on artificial intelligence}}, Vol.~\bibinfo{volume}{32}.
\newblock


\bibitem[\protect\citeauthoryear{Rousseeuw}{Rousseeuw}{1987}]%
        {rousseeuw1987silhouettes}
\bibfield{author}{\bibinfo{person}{Peter~J Rousseeuw}.}
  \bibinfo{year}{1987}\natexlab{}.
\newblock \showarticletitle{Silhouettes: a graphical aid to the interpretation
  and validation of cluster analysis}.
\newblock \bibinfo{journal}{\emph{Journal of computational and applied
  mathematics}}  \bibinfo{volume}{20} (\bibinfo{year}{1987}),
  \bibinfo{pages}{53--65}.
\newblock


\bibitem[\protect\citeauthoryear{Sahaai et~al\mbox{.}}{Sahaai
  et~al\mbox{.}}{2021}]%
        {sahaai2021brain}
\bibfield{author}{\bibinfo{person}{Madona~B Sahaai} {et~al\mbox{.}}}
  \bibinfo{year}{2021}\natexlab{}.
\newblock \showarticletitle{Brain Tumor Detection using DNN Algorithm}.
\newblock \bibinfo{journal}{\emph{Turkish Journal of Computer and Mathematics
  Education (TURCOMAT)}} \bibinfo{volume}{12}, \bibinfo{number}{11}
  (\bibinfo{year}{2021}), \bibinfo{pages}{3338--3345}.
\newblock


\bibitem[\protect\citeauthoryear{Schubert, Sander, Ester, Kriegel, and
  Xu}{Schubert et~al\mbox{.}}{2017}]%
        {DBSCAN:revisited}
\bibfield{author}{\bibinfo{person}{Erich Schubert}, \bibinfo{person}{J\"{o}rg
  Sander}, \bibinfo{person}{Martin Ester}, \bibinfo{person}{Hans~Peter
  Kriegel}, {and} \bibinfo{person}{Xiaowei Xu}.}
  \bibinfo{year}{2017}\natexlab{}.
\newblock \showarticletitle{DBSCAN Revisited, Revisited: Why and How You Should
  (Still) Use DBSCAN}.
\newblock \bibinfo{journal}{\emph{ACM Trans. Database Syst.}}
  \bibinfo{volume}{42}, \bibinfo{number}{3}, Article \bibinfo{articleno}{19}
  (\bibinfo{date}{jul} \bibinfo{year}{2017}), \bibinfo{numpages}{21}~pages.
\newblock
\showISSN{0362-5915}
\urldef\tempurl%
\url{https://doi.org/10.1145/3068335}
\showDOI{\tempurl}


\bibitem[\protect\citeauthoryear{{SciPy}}{{SciPy}}{2020}]%
        {SciPy}
\bibfield{author}{\bibinfo{person}{{SciPy}}.} \bibinfo{year}{2020}\natexlab{}.
\newblock \bibinfo{title}{{Pyton framework for mathematics, science, and
  engineering.}}
\newblock
\newblock
\urldef\tempurl%
\url{https://scipy.org/}
\showURL{%
\tempurl}


\bibitem[\protect\citeauthoryear{{Selvaraju}, {Cogswell}, {Das}, {Vedantam},
  {Parikh}, and {Batra}}{{Selvaraju} et~al\mbox{.}}{2017}]%
        {Selvaraju17}
\bibfield{author}{\bibinfo{person}{R.~R. {Selvaraju}}, \bibinfo{person}{M.
  {Cogswell}}, \bibinfo{person}{A. {Das}}, \bibinfo{person}{R. {Vedantam}},
  \bibinfo{person}{D. {Parikh}}, {and} \bibinfo{person}{D. {Batra}}.}
  \bibinfo{year}{2017}\natexlab{}.
\newblock \showarticletitle{Grad-CAM: Visual Explanations from Deep Networks
  via Gradient-Based Localization}. In \bibinfo{booktitle}{\emph{2017 IEEE
  International Conference on Computer Vision (ICCV)}}.
  \bibinfo{pages}{618--626}.
\newblock
\urldef\tempurl%
\url{https://doi.org/10.1109/ICCV.2017.74}
\showDOI{\tempurl}


\bibitem[\protect\citeauthoryear{Shlens}{Shlens}{2014}]%
        {shlens2014tutorial}
\bibfield{author}{\bibinfo{person}{Jonathon Shlens}.}
  \bibinfo{year}{2014}\natexlab{}.
\newblock \showarticletitle{A tutorial on principal component analysis}.
\newblock \bibinfo{journal}{\emph{arXiv preprint arXiv:1404.1100}}
  (\bibinfo{year}{2014}).
\newblock


\bibitem[\protect\citeauthoryear{Simonyan and Zisserman}{Simonyan and
  Zisserman}{2014}]%
        {simonyan2014very}
\bibfield{author}{\bibinfo{person}{Karen Simonyan} {and}
  \bibinfo{person}{Andrew Zisserman}.} \bibinfo{year}{2014}\natexlab{}.
\newblock \showarticletitle{Very deep convolutional networks for large-scale
  image recognition}.
\newblock \bibinfo{journal}{\emph{arXiv preprint arXiv:1409.1556}}
  (\bibinfo{year}{2014}).
\newblock


\bibitem[\protect\citeauthoryear{Sony, Dunphy, Sadhu, and Capretz}{Sony
  et~al\mbox{.}}{2021}]%
        {sony2021systematic}
\bibfield{author}{\bibinfo{person}{Sandeep Sony}, \bibinfo{person}{Kyle
  Dunphy}, \bibinfo{person}{Ayan Sadhu}, {and} \bibinfo{person}{Miriam
  Capretz}.} \bibinfo{year}{2021}\natexlab{}.
\newblock \showarticletitle{A systematic review of convolutional neural
  network-based structural condition assessment techniques}.
\newblock \bibinfo{journal}{\emph{Engineering Structures}}
  \bibinfo{volume}{226} (\bibinfo{year}{2021}), \bibinfo{pages}{111347}.
\newblock


\bibitem[\protect\citeauthoryear{Springenberg, Dosovitskiy, Brox, and
  Riedmiller}{Springenberg et~al\mbox{.}}{2015}]%
        {DB15a}
\bibfield{author}{\bibinfo{person}{J.T. Springenberg}, \bibinfo{person}{A.
  Dosovitskiy}, \bibinfo{person}{T. Brox}, {and} \bibinfo{person}{M.
  Riedmiller}.} \bibinfo{year}{2015}\natexlab{}.
\newblock \showarticletitle{Striving for Simplicity: The All Convolutional
  Net}. In \bibinfo{booktitle}{\emph{ICLR (workshop track)}}.
\newblock


\bibitem[\protect\citeauthoryear{{Stanford Vision Lab}}{{Stanford Vision
  Lab}}{2022}]%
        {IMAGENET}
\bibfield{author}{\bibinfo{person}{{Stanford Vision Lab}}.}
  \bibinfo{year}{2022}\natexlab{}.
\newblock \bibinfo{title}{{ImageNet}, image database organized according to the
  WordNet hierarchy}.
\newblock \bibinfo{howpublished}{\url{https://www.image-net.org}}.
\newblock


\bibitem[\protect\citeauthoryear{Strehl and Ghosh}{Strehl and Ghosh}{2002}]%
        {strehl2002cluster}
\bibfield{author}{\bibinfo{person}{Alexander Strehl} {and}
  \bibinfo{person}{Joydeep Ghosh}.} \bibinfo{year}{2002}\natexlab{}.
\newblock \showarticletitle{Cluster ensembles---a knowledge reuse framework for
  combining multiple partitions}.
\newblock \bibinfo{journal}{\emph{Journal of machine learning research}}
  \bibinfo{volume}{3}, \bibinfo{number}{Dec} (\bibinfo{year}{2002}),
  \bibinfo{pages}{583--617}.
\newblock


\bibitem[\protect\citeauthoryear{Szegedy, Liu, Jia, Sermanet, Reed, Anguelov,
  Erhan, Vanhoucke, and Rabinovich}{Szegedy et~al\mbox{.}}{2015}]%
        {szegedy2015going}
\bibfield{author}{\bibinfo{person}{Christian Szegedy}, \bibinfo{person}{Wei
  Liu}, \bibinfo{person}{Yangqing Jia}, \bibinfo{person}{Pierre Sermanet},
  \bibinfo{person}{Scott Reed}, \bibinfo{person}{Dragomir Anguelov},
  \bibinfo{person}{Dumitru Erhan}, \bibinfo{person}{Vincent Vanhoucke}, {and}
  \bibinfo{person}{Andrew Rabinovich}.} \bibinfo{year}{2015}\natexlab{}.
\newblock \showarticletitle{Going deeper with convolutions}. In
  \bibinfo{booktitle}{\emph{Proceedings of the IEEE conference on computer
  vision and pattern recognition}}. \bibinfo{pages}{1--9}.
\newblock


\bibitem[\protect\citeauthoryear{Talo}{Talo}{2019}]%
        {talo2019automated}
\bibfield{author}{\bibinfo{person}{Muhammed Talo}.}
  \bibinfo{year}{2019}\natexlab{}.
\newblock \showarticletitle{Automated classification of histopathology images
  using transfer learning}.
\newblock \bibinfo{journal}{\emph{Artificial Intelligence in Medicine}}
  \bibinfo{volume}{101} (\bibinfo{year}{2019}), \bibinfo{pages}{101743}.
\newblock


\bibitem[\protect\citeauthoryear{Tian}{Tian}{2021}]%
        {tian2021detect}
\bibfield{author}{\bibinfo{person}{Yuchi Tian}.}
  \bibinfo{year}{2021}\natexlab{}.
\newblock \emph{\bibinfo{title}{Detect and Repair Errors for DNN-based
  Software}}.
\newblock \bibinfo{thesistype}{Ph.D. Dissertation}. \bibinfo{school}{Columbia
  University}.
\newblock


\bibitem[\protect\citeauthoryear{Tian, Pei, Jana, and Ray}{Tian
  et~al\mbox{.}}{2018}]%
        {tian2018deeptest}
\bibfield{author}{\bibinfo{person}{Yuchi Tian}, \bibinfo{person}{Kexin Pei},
  \bibinfo{person}{Suman Jana}, {and} \bibinfo{person}{Baishakhi Ray}.}
  \bibinfo{year}{2018}\natexlab{}.
\newblock \showarticletitle{Deeptest: Automated testing of
  deep-neural-network-driven autonomous cars}. In
  \bibinfo{booktitle}{\emph{Proceedings of the 40th international conference on
  software engineering}}. \bibinfo{pages}{303--314}.
\newblock


\bibitem[\protect\citeauthoryear{Vallathan, John, Thirumalai, Mohan,
  Srivastava, and Lin}{Vallathan et~al\mbox{.}}{2021}]%
        {vallathan2021suspicious}
\bibfield{author}{\bibinfo{person}{G Vallathan}, \bibinfo{person}{A John},
  \bibinfo{person}{Chandrasegar Thirumalai}, \bibinfo{person}{SenthilKumar
  Mohan}, \bibinfo{person}{Gautam Srivastava}, {and} \bibinfo{person}{Jerry
  Chun-Wei Lin}.} \bibinfo{year}{2021}\natexlab{}.
\newblock \showarticletitle{Suspicious activity detection using deep learning
  in secure assisted living IoT environments}.
\newblock \bibinfo{journal}{\emph{The Journal of Supercomputing}}
  \bibinfo{volume}{77}, \bibinfo{number}{4} (\bibinfo{year}{2021}),
  \bibinfo{pages}{3242--3260}.
\newblock


\bibitem[\protect\citeauthoryear{Wan, Yang, Huang, Zeng, and Liu}{Wan
  et~al\mbox{.}}{2021}]%
        {wan2021review}
\bibfield{author}{\bibinfo{person}{Zitong Wan}, \bibinfo{person}{Rui Yang},
  \bibinfo{person}{Mengjie Huang}, \bibinfo{person}{Nianyin Zeng}, {and}
  \bibinfo{person}{Xiaohui Liu}.} \bibinfo{year}{2021}\natexlab{}.
\newblock \showarticletitle{A review on transfer learning in EEG signal
  analysis}.
\newblock \bibinfo{journal}{\emph{Neurocomputing}}  \bibinfo{volume}{421}
  (\bibinfo{year}{2021}), \bibinfo{pages}{1--14}.
\newblock


\bibitem[\protect\citeauthoryear{Wang, Chen, Sun, Ma, Wang, Sun, and
  Cheng}{Wang et~al\mbox{.}}{2021}]%
        {wang2021robot}
\bibfield{author}{\bibinfo{person}{Jingyi Wang}, \bibinfo{person}{Jialuo Chen},
  \bibinfo{person}{Youcheng Sun}, \bibinfo{person}{Xingjun Ma},
  \bibinfo{person}{Dongxia Wang}, \bibinfo{person}{Jun Sun}, {and}
  \bibinfo{person}{Peng Cheng}.} \bibinfo{year}{2021}\natexlab{}.
\newblock \showarticletitle{Robot: robustness-oriented testing for deep
  learning systems}. In \bibinfo{booktitle}{\emph{2021 IEEE/ACM 43rd
  International Conference on Software Engineering (ICSE)}}. IEEE,
  \bibinfo{pages}{300--311}.
\newblock


\bibitem[\protect\citeauthoryear{Wattenberg, Vi{\'e}gas, and
  Johnson}{Wattenberg et~al\mbox{.}}{2016}]%
        {wattenberg2016use}
\bibfield{author}{\bibinfo{person}{Martin Wattenberg},
  \bibinfo{person}{Fernanda Vi{\'e}gas}, {and} \bibinfo{person}{Ian Johnson}.}
  \bibinfo{year}{2016}\natexlab{}.
\newblock \showarticletitle{How to use t-SNE effectively}.
\newblock \bibinfo{journal}{\emph{Distill}} \bibinfo{volume}{1},
  \bibinfo{number}{10} (\bibinfo{year}{2016}), \bibinfo{pages}{e2}.
\newblock


\bibitem[\protect\citeauthoryear{Wohlin, Runeson, H{\"{o}}st, Ohlsson, Regnell,
  and Wessl{\'{e}}n}{Wohlin et~al\mbox{.}}{2012}]%
        {Wohlin2012}
\bibfield{author}{\bibinfo{person}{Claes Wohlin}, \bibinfo{person}{Per
  Runeson}, \bibinfo{person}{Martin H{\"{o}}st}, \bibinfo{person}{Magnus~C.
  Ohlsson}, \bibinfo{person}{Bj{\"{o}}rn Regnell}, {and}
  \bibinfo{person}{Anders Wessl{\'{e}}n}.} \bibinfo{year}{2012}\natexlab{}.
\newblock \bibinfo{booktitle}{\emph{{Experimentation in software
  engineering}}}. Vol.~\bibinfo{volume}{9783642290}.
\newblock 1--236 pages.
\newblock
\showISBNx{9783642290442}
\urldef\tempurl%
\url{https://doi.org/10.1007/978-3-642-29044-2}
\showDOI{\tempurl}


\bibitem[\protect\citeauthoryear{Xu, Guo, Wen, and Zhang}{Xu
  et~al\mbox{.}}{2021}]%
        {xu2021stealthy}
\bibfield{author}{\bibinfo{person}{Bowen Xu}, \bibinfo{person}{Fanghong Guo},
  \bibinfo{person}{Changyun Wen}, {and} \bibinfo{person}{Wen-An Zhang}.}
  \bibinfo{year}{2021}\natexlab{}.
\newblock \showarticletitle{Stealthy False Data Injection Attack Detection in
  Smart Grids with Uncertainties: A Deep Transfer Learning Based Approach}.
\newblock \bibinfo{journal}{\emph{arXiv preprint arXiv:2104.06307}}
  (\bibinfo{year}{2021}).
\newblock


\bibitem[\protect\citeauthoryear{Yang and Yang}{Yang and Yang}{2003}]%
        {yang2003can}
\bibfield{author}{\bibinfo{person}{Jian Yang} {and} \bibinfo{person}{Jing-yu
  Yang}.} \bibinfo{year}{2003}\natexlab{}.
\newblock \showarticletitle{Why can LDA be performed in PCA transformed space?}
\newblock \bibinfo{journal}{\emph{Pattern recognition}} \bibinfo{volume}{36},
  \bibinfo{number}{2} (\bibinfo{year}{2003}), \bibinfo{pages}{563--566}.
\newblock


\bibitem[\protect\citeauthoryear{Zeiler and Fergus}{Zeiler and Fergus}{2014}]%
        {Zeiler14}
\bibfield{author}{\bibinfo{person}{Matthew~D. Zeiler} {and}
  \bibinfo{person}{Rob Fergus}.} \bibinfo{year}{2014}\natexlab{}.
\newblock \showarticletitle{Visualizing and Understanding Convolutional
  Networks}. In \bibinfo{booktitle}{\emph{Computer Vision -- ECCV 2014}},
  \bibfield{editor}{\bibinfo{person}{David Fleet}, \bibinfo{person}{Tomas
  Pajdla}, \bibinfo{person}{Bernt Schiele}, {and} \bibinfo{person}{Tinne
  Tuytelaars}} (Eds.). \bibinfo{publisher}{Springer International Publishing},
  \bibinfo{address}{Cham}, \bibinfo{pages}{818--833}.
\newblock
\showISBNx{978-3-319-10590-1}


\bibitem[\protect\citeauthoryear{Zhang and Chan}{Zhang and Chan}{2019}]%
        {zhang2019apricot}
\bibfield{author}{\bibinfo{person}{Hao Zhang} {and} \bibinfo{person}{WK Chan}.}
  \bibinfo{year}{2019}\natexlab{}.
\newblock \showarticletitle{Apricot: a weight-adaptation approach to fixing
  deep learning models}. In \bibinfo{booktitle}{\emph{2019 34th IEEE/ACM
  International Conference on Automated Software Engineering (ASE)}}. IEEE,
  \bibinfo{pages}{376--387}.
\newblock


\bibitem[\protect\citeauthoryear{Zhang, Harman, Ma, and Liu}{Zhang
  et~al\mbox{.}}{2020}]%
        {zhang2019machine}
\bibfield{author}{\bibinfo{person}{Jie~M. Zhang}, \bibinfo{person}{Mark
  Harman}, \bibinfo{person}{Lei Ma}, {and} \bibinfo{person}{Yang Liu}.}
  \bibinfo{year}{2020}\natexlab{}.
\newblock \showarticletitle{Machine Learning Testing: Survey, Landscapes and
  Horizons}.
\newblock \bibinfo{journal}{\emph{IEEE Transactions on Software Engineering}}
  (\bibinfo{year}{2020}), \bibinfo{pages}{1--1}.
\newblock
\urldef\tempurl%
\url{https://doi.org/10.1109/TSE.2019.2962027}
\showDOI{\tempurl}


\bibitem[\protect\citeauthoryear{Zhang, Zhai, Ma, and Shen}{Zhang
  et~al\mbox{.}}{2021b}]%
        {zhang2021autotrainer}
\bibfield{author}{\bibinfo{person}{Xiaoyu Zhang}, \bibinfo{person}{Juan Zhai},
  \bibinfo{person}{Shiqing Ma}, {and} \bibinfo{person}{Chao Shen}.}
  \bibinfo{year}{2021}\natexlab{b}.
\newblock \showarticletitle{AUTOTRAINER: An Automatic DNN Training Problem
  Detection and Repair System}. In \bibinfo{booktitle}{\emph{2021 IEEE/ACM 43rd
  International Conference on Software Engineering (ICSE)}}. IEEE,
  \bibinfo{pages}{359--371}.
\newblock


\bibitem[\protect\citeauthoryear{Zhang, Satapathy, Guttery, G{\'o}rriz, and
  Wang}{Zhang et~al\mbox{.}}{2021a}]%
        {zhang2021improved}
\bibfield{author}{\bibinfo{person}{Yu-Dong Zhang},
  \bibinfo{person}{Suresh~Chandra Satapathy}, \bibinfo{person}{David~S
  Guttery}, \bibinfo{person}{Juan~Manuel G{\'o}rriz}, {and}
  \bibinfo{person}{Shui-Hua Wang}.} \bibinfo{year}{2021}\natexlab{a}.
\newblock \showarticletitle{Improved breast cancer classification through
  combining graph convolutional network and convolutional neural network}.
\newblock \bibinfo{journal}{\emph{Information Processing \& Management}}
  \bibinfo{volume}{58}, \bibinfo{number}{2} (\bibinfo{year}{2021}),
  \bibinfo{pages}{102439}.
\newblock


\bibitem[\protect\citeauthoryear{Zhao, Zhu, Chen, and Zhang}{Zhao
  et~al\mbox{.}}{2021}]%
        {zhao2021ai}
\bibfield{author}{\bibinfo{person}{Yue Zhao}, \bibinfo{person}{Hong Zhu},
  \bibinfo{person}{Kai Chen}, {and} \bibinfo{person}{Shengzhi Zhang}.}
  \bibinfo{year}{2021}\natexlab{}.
\newblock \showarticletitle{AI-Lancet: Locating Error-inducing Neurons to
  Optimize Neural Networks}. In \bibinfo{booktitle}{\emph{Proceedings of the
  2021 ACM SIGSAC Conference on Computer and Communications Security}}.
  \bibinfo{pages}{141--158}.
\newblock


\bibitem[\protect\citeauthoryear{{Zhou}, {Khosla}, {Lapedriza}, {Oliva}, and
  {Torralba}}{{Zhou} et~al\mbox{.}}{2016}]%
        {Zhou16}
\bibfield{author}{\bibinfo{person}{B. {Zhou}}, \bibinfo{person}{A. {Khosla}},
  \bibinfo{person}{A. {Lapedriza}}, \bibinfo{person}{A. {Oliva}}, {and}
  \bibinfo{person}{A. {Torralba}}.} \bibinfo{year}{2016}\natexlab{}.
\newblock \showarticletitle{Learning Deep Features for Discriminative
  Localization}. In \bibinfo{booktitle}{\emph{2016 IEEE Conference on Computer
  Vision and Pattern Recognition (CVPR)}}. \bibinfo{pages}{2921--2929}.
\newblock
\showISSN{1063-6919}
\urldef\tempurl%
\url{https://doi.org/10.1109/CVPR.2016.319}
\showDOI{\tempurl}


\bibitem[\protect\citeauthoryear{Zohdinasab, Riccio, Gambi, and
  Tonella}{Zohdinasab et~al\mbox{.}}{2021}]%
        {zohdinasab2021deephyperion}
\bibfield{author}{\bibinfo{person}{Tahereh Zohdinasab},
  \bibinfo{person}{Vincenzo Riccio}, \bibinfo{person}{Alessio Gambi}, {and}
  \bibinfo{person}{Paolo Tonella}.} \bibinfo{year}{2021}\natexlab{}.
\newblock \showarticletitle{Deephyperion: exploring the feature space of deep
  learning-based systems through illumination search}. In
  \bibinfo{booktitle}{\emph{Proceedings of the 30th ACM SIGSOFT International
  Symposium on Software Testing and Analysis}}. \bibinfo{pages}{79--90}.
\newblock


\end{thebibliography}

\end{document}